\documentclass[pdflatex]{sn-jnl}
\usepackage{graphicx}
\usepackage{hhline}
\usepackage{amsmath,amssymb,amsfonts}
\usepackage[title]{appendix}
\usepackage[compatibility=false]{caption}
\usepackage{subcaption}
\usepackage{subcaption}
\usepackage{booktabs}
\usepackage{appendix}
\usepackage{multirow}
\usepackage{float}
\usepackage{booktabs}
\usepackage{natbib}
\usepackage{longtable}
\usepackage{sidecap}
\usepackage{makecell}
\usepackage{tablefootnote}
\usepackage[compatibility=false]{caption}
\defcitealias{fazzari_2025}{F25}
\newcommand{\rthis}[1]{\textcolor{black}{#1}}
\usepackage{float}
\usepackage{comment}

\begin{document}

\title{Determination of the best dark matter profile for the Milky Way with Gaia DR3 using Bayesian Model Comparison}

\author[1]{\fnm{Aryan} \sur{Singh}}
 \email{ph25resch01005@iith.ac.in}
\author[1]{\fnm{Shantanu} \sur{Desai}}
 \email{shntn05@gmail.com}
\affil[1]{Department of Physics, IIT Hyderabad, Kandi, Telangana-502284, India}


\abstract{
We carry out a Bayesian model comparison analysis to determine the dark matter model that best describes the Milky Way rotation curve, using four recent compilations of rotation-curve velocities with the Gaia data. We compare these data using three different baryonic models along with seven different dark matter models.  \rthis{Within the adopted modeling framework and Gaia-based rotation curve datasets}, we find that the Einasto \rthis{profile provides the preferred phenomenological fit compared to the NFW profile} across most combinations of datasets and baryonic models. We also find that the cored dark profiles are preferred over the (cuspy) NFW profile. We also test MOND using three different interpolating functions and find that, 
\rthis{within the implementations considered, the models provide poorer fits than the preferred dark matter profiles.}
Finally, among the different baryonic models considered, none is decisively favored over the others.
}


\keywords{Galaxies, Dark Matter, Modified Gravity, Rotation Curve, Bayesian Analysis}
\maketitle

\section{Introduction}
\label{section:1}
Galaxy Rotation Curves (RCs) display the relationship between the circular velocity $v_c(r)$ and the radial distance $r$ of stars, gas, and other objects (assumed to be on circular orbits) measured with respect to the center of the galaxy. The study of RCs becomes important because the centripetal and gravitational forces are equal and opposite in circular orbits. If $M(r)$ is the mass enclosed within the orbit of radius $r$ in which the astrophysical object is rotating with the orbital speed $v_c(r)$, then
\begin{equation}\label{1}
    \frac{mv_c^2(r)}{r} = \frac{GmM(r)}{r^2}
\end{equation}
If the circular velocity $v_{obs}(r)$ is known, the mass distribution in the galaxy ($M_{est}(r)$) can be estimated. Conversely, if the mass distribution ($M_{obs}(r)$) is known, then the orbital speeds ($v_{est}(r)$) can be estimated. The $v_{est}(r)$ RC is sometimes referred to as the baryon curve, obtained from observations of the baryon components in the galaxy (bulge, stellar disk, dust, and gas).

Contrary to expectations, the two RCs $v_{est}(r)$ and $v_{obs}(r)$ were found to differ for a large number of spiral galaxies, especially within the radial distance from $r = 5$ kpc to $r = 30$ kpc. While $v_{obs}(r)$ ranges around $200$ km/s with some small variations (maximum of $v_{obs}(r)$ ranging from 230-250 km/s), $v_{est}(r)$ shows a steep decline from $v \approx 170$ km/s at $r = 5$ kpc to $v \approx 100$ km/s at $r = 30$ kpc, making our current understanding of physical laws inadequate \citep{Rubin70,Ostriker74,Rubin80,Bosma78}. The observed mass was found to be insufficient to account for such high speeds at large radii, thus establishing the problem of ``missing mass" in modern cosmology. It was posited that there exists unseen matter surrounding the galaxies, which was later referred to as dark matter~\citep{Faber79, Trimble87, Bosma23}. Subsequent observations at larger scales and other probes (such as the Cosmic Microwave background) in conjunction with results from cosmological simulations, confirmed that dark matter is cold (i.e., non-relativistic at the epoch of structure formation)~\citep{Kamionkowski96} and constitutes about 25\% of the total matter energy density budget of the universe \citep{Planck20}. The existence of cold dark matter is also construed as evidence of Physics beyond the Standard Model of Particle Physics~\citep{Hooper04}. As an alternative to postulating the existence of unseen cold matter, \citet{Milgrom} proposed a correction to Newton's laws of gravity in the low acceleration regime to explain the ``missing mass problem''. 

However, there are still some vexing issues with the prevailing Cold Dark Matter paradigm at small scales, such as the core-cusp and missing satellite problem~\citep{Bullock}, the radial acceleration relation in spiral galaxies with extremely small scatter \citet{Lelli16}, failure to detect any cold matter candidates in laboratory-based experiments~\citep{Merritt,Baudis}, etc. Therefore, alternatives to standard $\Lambda$CDM, such as Self-interacting Dark Matter (SIDM)~\citep{Arka} and other models~\citep{divalentino25}, have been proposed to alleviate these problems.

To resolve some of these issues using a data-driven approach, it is prudent to determine the best dark matter model using the latest available data. In this manuscript, we analyze the latest high-precision measurements of the Milky Way RC . The Milky Way RC has been notoriously difficult to measure due to our location within it. For a long time, there was considerable scatter in the data from the different methods used to measure the rotation curve, due to uncertainty in the distance to tracers and due to obscuration from the dense intergalactic dust and gas in our galaxy \citep{Sofue09,Mroz19,Russeil,Wang2020}. In recent years, high-precision measurements of the Milky Way's  RC  have become possible due to data from the Gaia satellite, which has enabled the measurement of 3-D velocities and distances of stars in the Milky Way.

In this work, we aim to determine the optimal dark matter and baryon mass models for the Milky Way, based on the latest high-precision compilations of the Galactic RC from GAIA DR3~\citep{GaiaDR3}, using Bayesian model comparison. This manuscript is structured as follows. In Section \ref{section:2}, we discuss the data used. In Section~\ref{section:3}, and \ref{section:4}, we discuss mass models for baryonic and dark matter, respectively.  In Section~\ref{section:5}, we discuss MOND models. In Section~\ref{sec:compare}, we give a brief description of the methodology of model comparison. In Section \ref{sec:results}, we discuss the results of our analysis. We conclude in Section~\ref{sec:conclusions}.

\section{Datasets used}
\label{section:2}
\begin{figure}[h]
    \centering
    \caption{Raw data plotted on velocity vs radial distance.}
    \includegraphics[width=1\textwidth]{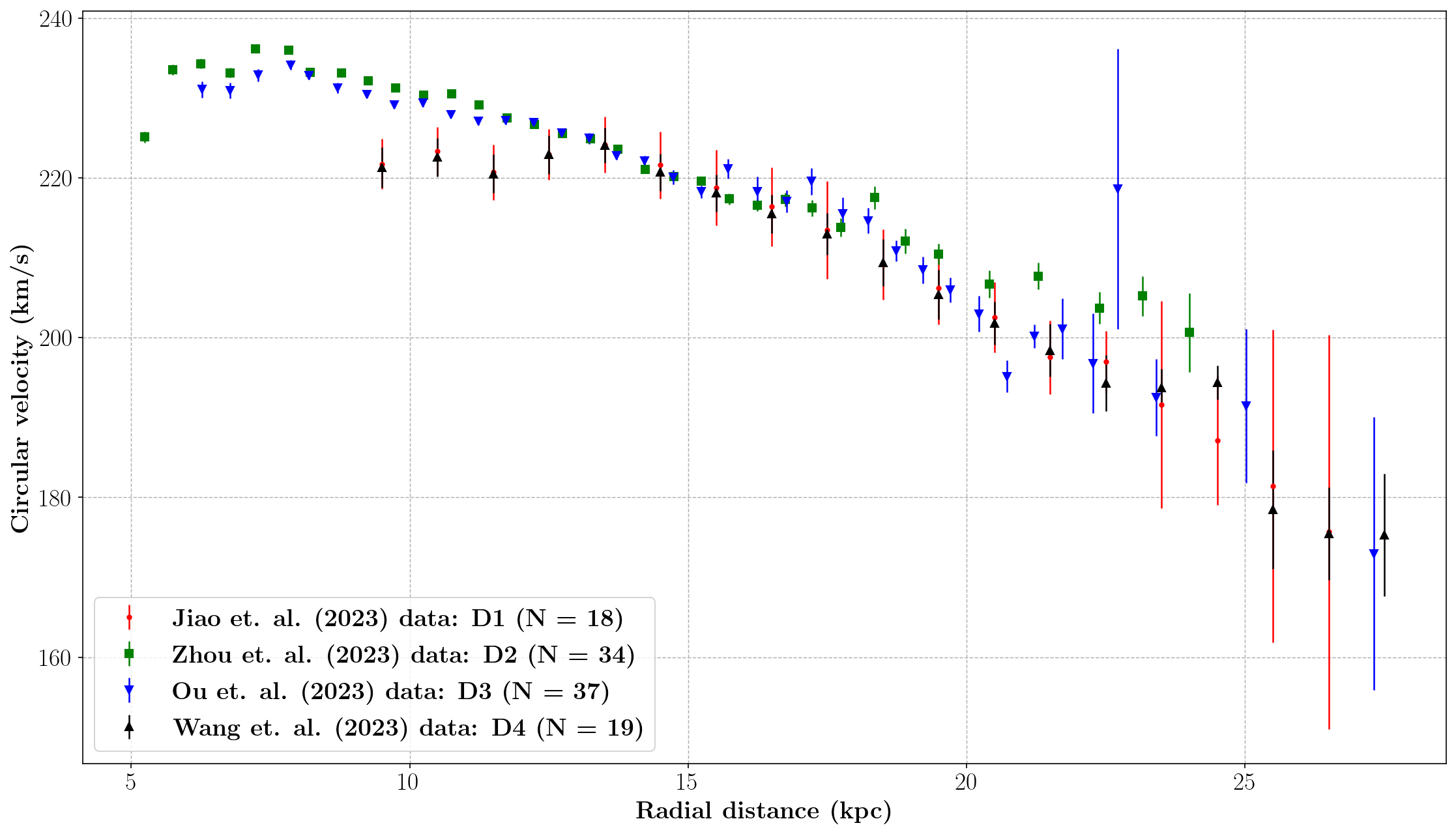}
    \label{fig:data}
\end{figure}

In this work, we utilize four RC datasets, each consisting of the radial distance $r$, the corresponding rotational velocity $v_{obs}(r)$, and the uncertainties in the rotational velocity estimates $\sigma_{i}$. These data contain rotational velocities ranging from $r \approx 5$ kpc to $r \approx 30$ kpc. We plot the raw data along with their $1\sigma$ uncertainties in Figure~\ref{fig:data}. All four sets of measurements used Gaia DR3~\citep {GaiaDR3}, which provided a robust RC up to 14 kpc using samples of OB and luminous red giant branch (RGB) stars. We briefly describe the datasets below and the ancillary observations used to construct these measurements. More details on the construction of these datasets and the tracers used can be found in the original works.
\begin{itemize}
    \item {\tt D1:} This consists of 18 data points ($N = 18$) from $r = 9.5$ kpc to $r = 26.5$ kpc from \citet[see their Table 3]{Jiao23}. The RCs were derived using both RGB and OB stars. The selected stars were within a scale height of $ |z| \leq 3$ kpc, which provides the most extended RC. 
    Based on this data, a decline in the RC beyond 19.5 kpc was observed, with the RC decreasing from $221.75$ km/s to about $175.68$ km/s. The errors in the dataset consist of both systematic and statistical uncertainties.  We note that asymmetric drift corrections have been applied in this work.
    \item {\tt D2:} This consists of 34 data points ($N = 34$) from $r = 5.24$ kpc to $r = 24.0$ kpc, which can be found in \citet[see their Table 4]{Zhou}. This work used RGB stars as tracers, obtained from APOGEE~\citep{Apogee} and LAMOST~\citep{LAMOST}, with distances derived by combining photometric and astrometric information from 2MASS~\citep{2MASS} and Gaia DR3 using supervised machine learning. The asymmetric drift corrections were included by directly solving the Jean's equation. The uncertainties were obtained from bootstrap resampling. The rotational velocity decays from $225.1$ km/s to $200.6$ km/s. To avoid contamination from the thick disk, only the stars within $ |z| \leq 0.5$ kpc are considered. As a result, the circular velocity for $r \sim 10$ kpc is higher compared to D1.
    \item {\tt D3:} This consists of 37 data points ($N = 37$) from $r = 6.27$ kpc to $r = 27.31$ kpc, and can be found in \citet[see their Table 1]{Eilers23}. Similar to D2, this study also used RGB stars as tracers, with spectroscopic observations from APOGEE DR17 and photometric measurements from Gaia DR3, WISE~\citep{WISE}, and 2MASS. The distances were obtained from the spectrophotometric parallaxes of RGB stars using the method described in \citet{Hogg19}.  
    The rotational velocity decays from $231.07$ km/s to $172.98$ km/s. Similar to D2, a restriction of $ |z| \leq 1$ kpc and $|v_z| \leq 100$ km/s on star selection was done to account for any flaring of the galactic disk. D3 consists of high circular velocities for $r \sim 10$ kpc compared to D1, but slightly lower compared to D2. Although no asymmetric drift corrections were applied, the selected stars were chosen with a metallicity cut of $[\alpha/\text{Fe}]< 0.12$ so that such corrections would be negligible. However, this work also estimated the systematic uncertainty due to the neglected drift corrections.
    \item {\tt D4:} This consists of 19 measurements ($N = 19$) from $r = 9.5$ kpc to $r = 27.5$ kpc, obtained from \citet[see their Table 1]{Wang23}. This RC was reconstructed using Lucy's inversion method from the 3-dimensional velocity components of sources from GAIA DR3 using both OB and RGB stars. Similar to {\tt D2}, the asymmetric drift corrections were obtained directly from the Jean's equation. The rotational velocity also declines from $221.3$ km/s to $175.3$ km/s.
\end{itemize}
A comparison of the RC for the four aforementioned datasets, along with the $1\sigma$ uncertainties, is presented in Figure \ref{fig:data}. We note that {\tt D1} and {\tt D4} correspond to measurements at the same radii, except for one extra measurement in D4 at $\sim 27.5$ kpc. Another difference between D1 and D4 is that systematic uncertainties have been included in D1, but not in D4. The datasets {\tt D1} and {\tt D3} agree for $r \geq 13$ kpc,  except for one point at 23 kpc. Finally, {\tt D2} shows larger velocities in the Milky Way outskirts compared to D1. More details on the comparison between the four datasets and the reasons for the discrepancies are discussed in \citet{Jiao23} (notably in their appendix A, where they depict the effect of selecting stars with a small, narrow range of scale height $z$ on the RC).

A detailed study of the systematic errors in the RCs reconstructed from the Gaia DR3 data can be found in ~\citet{Koop}. We briefly summarize their findings. The total systematic error up to $R \approx 18$ kpc is less than 5\%.
The first systematic effect arises from the scale length in the exponential function used to model the tracer population's number density in Jeans' equation. A change of  1~kpc can cause a change of up to 5\% in the outer regions and  2\% within $R < 14$ kpc.  If the number density is modelled as a power law~\citep{Eilers23}, the errors are much larger. Furthermore, the exponential function used to compute the radial velocity profile's derivative cannot fit the data well for $R > 12.5$ kpc.
If one fits the data in the region $11 < R < 22$ kpc, one obtains a larger scale height, which amounts to a systematic error of 2.5\% for $R < 13$ kpc and up to 5\% at large radii. If we assume all errors are independent and add them in quadrature, the total systematic error has been estimated to be 6\% for $R<14$ kpc and up to 15\% at larger radii~\citep{Koop}.  
However, these errors are not independent or Gaussian~\citep{Oman24}, so the above estimate is the worst case scenario.

\section{Baryon Mass Models}
\label{section:3}
We now discuss the baryon mass models used to account for the rotation curve. The baryon RC contribution, consisting of the superposition of the bulge ($v_b$) and disk ($v_d$) components of the baryon matter in a galaxy, can be represented as follows:
\begin{equation}\label{3}
    v_N^2(r) = v_b^2(r) + \sum_{d} v_d^2(r),
\end{equation}
where the sum is over different disk components, corresponding to the stars, dust, or gas, depending on the model. If either the bulge or the disk can be described by a cylindrical potential $\phi (r,z)$, then we can obtain the corresponding velocity ($v$) for that potential using:
\begin{equation}\label{4}
    v^2 (r) = r \left.\frac{\partial \phi (r,z)}{\partial r}\right|_{z = 0} 
\end{equation}
Here, $r$ and $z$ are the galactocentric coordinates, and the potential $\phi$ satisfies the 3-dimensional Poisson equation $\nabla^2 \phi = - 4 \pi G \rho$. In our study, we consider three baryon models for the spherical galactic bulge and axisymmetric galactic disks, which we enumerate below:
\begin{enumerate}
    \item {\tt B0:} This model has been used in some recent works to constrain mass models of the Milky Way~\citep{ManHoChan,Misiriotis}. For the bulge component of baryon matter, we use the Hernquist potential~\citep{deSalas19} described as follows:
    \begin{equation}
    \phi_{Hernquist}(r) = -\frac{G M_b}{r + r_b}
    \label{5}
    \end{equation}
    Here, $M_b = 1.55 \times 10^{10} \textup{M}_\odot$ and $r_b = 0.7$ kpc are constants whose values were obtained from \citet{deSalas19}. We then calculate $v_b$ using eq. \eqref{4}.
    \begin{equation}
    v_b^2 (r)  = \frac{GM_b r}{(r+r_b)^2}
    \label{6}
    \end{equation}
    The disk model used consists of five components: a stellar disk, warm and hot dust, HI gas, and $H_2$ gas. The surface mass densities for all the components can be represented by a flattened, thin disk depicted by an exponential profile as follows~\citep{Toomre}:
    \begin{equation}\label{7}
        \Sigma_d(r) = \Sigma_0 \exp\left[-\frac{r}{r_d}\right], \hspace{1cm} \Sigma_0 = \frac{M_d}{2 \pi r_d^2}, 
    \end{equation}
    where $M_d$ and $r_d$ are constant parameters of the components, whose values are constrained from the observations from COBE/DIRBE and COBE/FIRAS~\citep{Misiriotis} and summarized in Table~\ref{table1}. Among these, we consider the mass of the stellar disk as a free parameter.
    \begin{table}[h]
        \caption{\label{table1} Parameters of the components of the exponential thin disk used for the baryon model B0 obtained from ~\citet{deSalas19}.}
        \begin{tabular}{l|l|l}
        \hline
        Disk component & $r_d$ (Kpc) & $M_d$  ($10^9 \textup{M}_\odot$) \\
        \hline
        Stellar disk & 2.35 &  Free \\
        Cold disk & 5 & 0.07\\
        Warm disk & 3.3 & $2.2 \times 10^{-4}$\\
        $H_2$ gas disk & 2.57 & 1.3\\
        HI gas disk & 18.24 & 8.2\\
        \hline
        \end{tabular}
    \end{table}
    The RC for each disk component can be modeled by the following equation \citep{Freeman}.
    \begin{equation}\label{8}
        v_d^2(r) = \frac{G M_d r^2}{2 r_d^3} \left[I_0(x) K_0(x) - I_1(x) K_1(x)\right], \hspace{0.7cm} x = \frac{r}{2r_d}
    \end{equation}
    Here, $I_n$ and $K_n$ are the modified Bessel functions of the n$^{th}$ kind.  The total rotation curve is then given by: 
    \begin{equation}\label{eq:vnB0}
        v_N^2 (r) = \frac{G M_b r}{(r+r_b)^2} + \sum_d v_d^2
    \end{equation}
    \item {\tt B1:}  This model has previously been used in \citet{deSalas19} and has also been referred to as B1. For the bulge component of baryon matter, we choose to work with the Plummer potential described as follows \citep{Plummer}:
     \begin{equation}\label{eq:Plummer}
        \phi_{Plummer}(r) = - \frac{G M_b}{\sqrt{r^2 + r_b^2}},
    \end{equation}
    where $M_b = 1.067 \times 10^{10} \textup{M}_\odot$ and $r_b = 0.3$ kpc are constants obtained from \citet{deSalas19}. The disk components consist of the thin and thick disks, both described by the Miyamoto-Nagai potential~\citep{Miyamato} as follows:
    \begin{equation}\label{eq:MN}
        \phi_{MN,d}(r, z) = - \frac{G M_d}{\sqrt{r^2 + \left(r_d + \sqrt{z^2 + z_d^2}\right)^2}},
    \end{equation}
    where $M_d$, $r_d$, and $z_d$ are constant parameters for each of the disk components $d$. 
    We kept the mass of both thin and thick disks as free parameters.
    Their values are shown in Table \ref{table2}.
    \begin{table}[h]
        \caption{\label{table2}Parameters of the components of the Miyamoto-Nagai potential used in the baryonic model {\tt B1} obtained from \citet{deSalas19}.}
        \begin{tabular}{l|l|l|l}
        \hline
        Disk Component & $r_d$ (Kpc) & $z_d$ (kpc) & $M_d$  ($10^9 \textup{M}_\odot$) \\
        \hline
        Thick disk & 2.6 & 0.8 &  Free \\
        Thin disk & 5.3 & 0.25  & Free \\
        \hline
        \end{tabular}
    \end{table}
    The Miyamoto-Nagai potential for the galactic disk is axisymmetric and doesn't confine the baryon mass to a single plane, unlike the disk model on B0. However, it overestimates the baryon mass at larger galactic radii~\citep{deSalas19}. Once the potentials for the bulge and disk are specified, we can calculate ($v_b$) and ($v_d$) by applying Eq.~\eqref{4} to the Plummer and Miyamoto-Nagai potentials, respectively. The total velocity can be represented as follows:
    \begin{equation}\label{vnB1}
        v_N^2 (r) = \frac{G M_b r^2}{(r^2+r_b^2)^{3/2}} + \sum_d \frac{G M_d r^2}{(r^2 + (r_d + z_d)^2)^{3/2}},
    \end{equation}
    where the summation ($\sum_{d}$ is carried over the components mentioned in Table \ref{table2}).
    \item {\tt B2:} This model has also been used in \citet[also labeled as {\tt B2} there]{deSalas19} and \citet{Blanchard}. For this model, the bulge component consists of the Hernquist potential similar to that used in {\tt B0}. The disk model used consists of a double exponential profile as follows:
    \begin{equation}\label{9}
        \rho_d(r, z) = \rho_0  \exp\left[ -\frac{r}{r_d} - \frac{|z|}{z_d} \right]
    \end{equation}
    The model in Eq.~\eqref{9} describes the stellar disk, the cold and warm dust components, the atomic HI, and the molecular $H_2$ gas distribution. The normalization constant ($\rho_0$) can be described by two different parametric models: $\rho_{0d1}$ for stellar disk, warm and cold dust, $H_2$ gas, and $\rho_{0d2}$ for the atomic HI gas. These models are given by:  
    \begin{equation}\label{10}
        \rho_{0d1} = \frac{M_d}{4 \pi z_d r_d^2} \hspace{2cm} \rho_{0d2} = \frac{M_d e^{(r_t/r_d)}}{4 \pi z_d r_d(r_d + r_t)}
    \end{equation}
    The values of the parameters in Eq.~\eqref{9} and Eq.~\eqref{10} are tabulated in Table \ref{table3}. Similar to B0 and B1, we kept the mass of the stellar disk as a free parameter.
    \begin{table}[h]
        \caption{\label{table3}Parameters of the components of the exponential disk density profile corresponding to the baryonic model {\tt B2} obtained from \citet{deSalas19}.}
        \begin{tabular}{l|l|l|l|l}
        \hline
        Disk component & $z_d$ (kpc) & $r_d$ (Kpc) & $r_t$ (Kpc) & $M_d$  ($10^9 \textup{M}_\odot$) \\
        \hline
        Stellar disk & 0.14 & 2.35 & - & Free \\
        Cold disk & 0.1 & 5 & - & 0.07\\
        Warm disk & 0.09 & 3.3 & - & $2.2 \times 10^{-4}$\\
        $H_2$ gas disk & 0.08 & 2.57 & - & 1.2\\
        HI gas disk & 0.52 & 18.24 & 2.75 & 8.2\\
        \hline
        \end{tabular}
    \end{table}
    The contribution of the disk density profile mentioned above in the Eq.~\eqref{9} and \eqref{10} to the rotation curve can be calculated by the potential described in \citet[see Appendix A]{Gilmore}. This potential can be written as follows:
    \begin{equation}\label{11}
        \phi_d(r,z) = \frac{-4 \pi G \rho_{0d}}{r_d} \int_0^{\infty} dk \frac{J_0(kr) \hspace{0.2cm} \beta_d e^{-k|z|} - k e^{-\beta_d|z|}}{(\beta_d^2 - k^2) (k^2 + \alpha_d^2)^{-3/2}}, \left[\alpha_d = \frac{1}{r_d}, \beta_d = \frac{1}{z_d}\right]
    \end{equation}
    We can then calculate $v_d$ by applying Eq.~\eqref{4} to the potential in Eq.~\eqref{11}. The total velocity can then be represented as follows:
    \begin{equation}\label{vnB2}
    v_N^2 (r) = \frac{G M_b r}{(r+r_b)^2} + \sum_d \frac{4 \pi G \rho_{0d} r}{r_d} \int_0^{\infty} dk \hspace{0.2cm} k J_1(kr) \frac{(k^2 + \alpha_d^2)^{3/2}}{(\beta_d + k)},
    \end{equation}
    where the summation is over different disk components.
\end{enumerate}
We note that the above sets of models are not exhaustive. Other models, such as the de Vaucouleur profile~\citep{Jain}, exponential and Gaussian profiles~\citep{Jiao23}, have also been used to model the bulge. However, the de Vaucouleurs profile fails to fit the inner MW rotation curve, and the exponential and Gaussian profiles exhibit angular dependence and are difficult to evaluate. Similarly, more complex models have also been used for the disk (\citealt{Lopez,Pouli}). However, these are similar to the {\tt B2} model in \citet{Blanchard}, and hence we do not consider them further.

\section{Dark matter mass models}
\label{section:4}

We now discuss the dark matter profiles used for this analysis.
\begin{enumerate}
    \item {\tt Navarro Frenk White (NFW) profile}~\citep{NFW97}: This is one of the most widely used dark matter profiles, which was found to fit the dark matter halos identified in N-body simulations involving only gravity. In this profile,   a cuspy halo was favored with $\rho \propto r^{-1}$ towards the center, whereas it scales according to $\rho \propto r^{-3}$ in the outer parts. This profile was found to be universal across cosmological models and simulations. The NFW density profile is given by:
    \begin{equation}\label{16}
        \rho_{NFW}(r) =\frac{\rho_0}{(r/r_s)(1+r/r_s)^2},
    \end{equation}
    where  $\rho_0$ is a normalization constant, and $r_s$ is a scale radius at which the profile begins to change its shape.
    \item {\tt Isothermal profile} \citep{Kormendy14}: It was used to explain the structure of dark matter halos based on earlier simulations with poor resolution. It described the DM density profile with a $r^{-2}$ fall-off to achieve a constant circular velocity in the galaxy's outer regions. The density profile is given by:
    \begin{equation}\label{17}
        \rho_{isothermal}(r) =\frac{\rho_0}{[1+(r/r_c)^2]}, 
    \end{equation}
    where $\rho_0$ and $r_c$ are the central density and core radius of the DM halo. The subsequent mass distribution has a divergence $\propto r$.
    \item {\tt Burkert profile}~\citep{Burkert95}: This is also a cored dark matter profile, which shows $\propto r^{-3}$ decay, unlike the isothermal profile, and may describe a DM halo with decaying mass in outer regions. The density profile is given by:
    \begin{equation}\label{18}
        \rho_{Burkert}(r) =\frac{\rho_0}{[1 + r/r_s][1 + (r/r_s)^2]},
    \end{equation}
    where, $\rho_s$ is the central density and $r_s$ is the scale radius. The Burkert profile resembles the isothermal profile in the inner region $(r \ll r_s)$ of the DM halo, where $r_s$ of the Burkert profile resembles the core radius $r_c$ of the Isothermal profile. 
    \item {\tt gNFW profile} \citep{Zhao96}: This profile generalizes the NFW profile and modifies its cuspy inner region to describe an arbitrary power-law shaped cored center $\rho(r) \propto r^{-\alpha}$ and an outer region similar to the NFW profile. It has also been referred to as $\alpha\beta\gamma$ profile in literature~\citep{Straight}. This gNFW model has also been shown to fit baryonified dark matter haloes \citep{Gopika}. The density profile is given by:
    \begin{equation}\label{eq:gNFW}
        \rho_{gNFW}(r) =\frac{\rho_0}{(r/r_c)^\alpha(1+r/r_c)^{3-\alpha}},
    \end{equation}
    \item  {\tt $\beta$ profile} \citep{Spano}: This is a variant of the isothermal profile which has a constant density in the core region of the halo, but decays faster than the isothermal profile given by $\propto r^{-3}$. The general form of the profile is described in \citet[see section 3.2]{NFWX}. It was first used to model spiral and irregular galaxies in the GHASP $H\alpha$ kinematic survey~\citep{GHASP}. The $\beta$ density profile is given by:
    \begin{equation}\label{20}
        \rho_{\beta}(r) =\frac{\rho_0}{[1 + (r/r_c)^2]^{3/2}}
    \end{equation}
    where, $r_c$ is the core radius and $\rho_0$ is the central density of the halo.
    \item  {\tt Hernquist profile} \citep{Hernquist90}: This profile, which was originally proposed to model the spherical bulge of elliptical galaxies, closely resembles the empirical de Vaucouleurs' $R^{1/4}$ law. It is similar to the NFW profile, but with a faster decay at large radii. It has been used to model dark matter profiles in N-body simulations \citep{Tormen97} and the dark matter halos of our galaxy and M31 \citep{Jain}. The Hernquist density profile can be written as:
    \begin{equation}\label{21}
        \rho_{Hernquist}(r) =\frac{\rho_0}{(r/r_c)(1+r/r_c)^3},
    \end{equation}
    where $\rho_0$ is a normalization constant and $r_c$ is the core radius. Both $\rho_0$ and $r_c$ are kept as free parameters.
    \item  {\tt Einasto profile} \citep{Einasto65} assumes a power law slope for the logarithmic density profile. Various studies have found that the Einasto profile is a better fit for cosmological N-body simulations than the NFW profile in high-density regions~\citep[and references therein]{Ludlow13}.  It can be used to describe both a core and a cusp. This profile has also been used to fit SIDM simulations~\citep{Straight,Dalui} and to obtain observational constraints on SIDM parameters~\citep{Gopikagroup}.
    The Einasto density profile can be written as:
    \begin{equation}\label{eq:einasto}
        \rho_{Einasto}(r) = \rho_0 \hspace{0.1cm} \exp\left[ - \frac{2}{\alpha} \left( \left(\frac{r}{r_c}\right)^{\alpha} - 1 \right) \right]
    \end{equation}
    Here, $\rho_0$ and $r_c$ are the scale parameters, at which $ \rho(r) \propto r^{-2}$, and $\alpha$ is the Einasto index, characterizing the degree of curvature of the profile. 
\end{enumerate}
The contribution of the halo component ($v_{DM}^2(r)$) is calculated from Eq.~\eqref{23}, and can be  added to $v_N^2(r)$ in Eq.~\ref{3} to get DM corrected models of rotation curve as follows:
\begin{equation}\label{23}
    v_{DM}^2(r) = \frac{4 \pi G}{r} \int_0^r  dr' r'^2 \rho_{DM}(r')
\end{equation}
The total velocity, which includes the baryon contributions along with dark matter, is given as follows:
\begin{equation}\label{24}
    v_{total}^2(r) = v_N^2(r) + v_{DM}^2,
\end{equation}
where $v_N^2(r)$ is described by Eqns.~\eqref{eq:vnB0}, \eqref{vnB1}, \eqref{vnB2} for {\tt B0, B1, and B2}, respectively.

\section{MOND  as an alternative to dark matter}
\label{section:5}
An alternative model to dispense with dark matter is to replace it with modified gravity. The most widely studied model for the same is Milgrom's law for MOdified Newtonian Dynamics (MOND). It states that for accelerations smaller than a constant value, denoted as $a_0$, depart from the standard Newton's gravitational law in Eq.~\eqref{1}, requiring a modification to the gravitational force law \citep{Milgrom}. More details on the successes and failures of MOND are reviewed elsewhere~\citep{Desmond}. The modified dynamical acceleration $a_{dyn}$ is related to the standard Newtonian acceleration $a_{N}$ by introducing an interpolating function as follows:
\begin{equation}\label{25}
    a_{dyn} = \nu\left(\frac{a_0}{a_N}\right) a_{N}
\end{equation}
The interpolating functions provide the transition between Newtonian and MOND regimes. The interpolating functions that we use in this work are enumerated below:
\begin{enumerate}
    \item {\tt Standard interpolating function}\citep{Famaey11}: 
    \begin{equation}\label{26}
        \nu\left(\frac{a_0}{a_N}\right) = \sqrt{\frac{1}{2} + \sqrt{\frac{1}{4} + \left(\frac{a_0}{a_N}\right)^2}}
    \end{equation}
    \item {\tt Simple interpolating function}~\citep{Famaey11}:
    \begin{equation}\label{27}
        \nu\left(\frac{a_0}{a_N}\right) = \frac{1}{2} + \sqrt{\frac{1}{4} + \left(\frac{a_0}{a_N}\right)}
    \end{equation}
    The simple interpolating function decays more slowly than the standard interpolating function; hence, it may add more constant acceleration than required, even in some high-acceleration environments.
    \item {\tt Radial acceleration relation (RAR)}~\citep{Lelli16}: 
    \begin{equation}\label{28}
        \nu\left(\frac{a_0}{a_N}\right) = \left[1 - \exp(-\sqrt{\frac{a_N}{a_0}})\right]^{-1}
    \end{equation}
    The RAR has been shown to be a very good fit to the rotation curves of spiral galaxies~\citep{Lelli16} and is the most widely used.
\end{enumerate}
The interpolating functions (IFs) described above are conditioned to equate the two accelerations $a_{dyn}$ and $a_N$ when $a_N \gg a_0$. Once we specify $a_N$, we can calculate $a_{dyn}$, hence $v_{dyn}$. \rthis{We note however, that this analysis does not include the external field effect~\citep{Famaey11} or non-spherical baryonic structure.}
\begin{equation}\label{29}
    a_N(r) = \frac{v_N^2(r)}{r} = \frac{GM}{r^2} \rightarrow a_{dyn} = \nu\left(\frac{a_0}{a_N}\right) \frac{GM}{r^2}
\end{equation}
\begin{equation}\label{30}
    v_{dyn}^2(r) = a_{dyn}(r) \times r = v_N^2(r) \times \nu\left(\frac{a_0}{a_N}\right) = \nu\left(\frac{a_0}{a_N}\right) \frac{GM}{r}
\end{equation}
For low accelerations $a_N \ll a_0$, and
\begin{equation}\label{31}
    v_{dyn}^2 (r)  \rightarrow \sqrt{G M a_0}
\end{equation}.

\section{Bayesian Model Comparison}
\label{sec:compare}
We provide a brief overview of Bayesian model comparison and refer the reader elsewhere for more details~\citep{Sanjib,Weller,Krishakeotwash}. Bayesian model comparison is the most robust among all model comparison techniques used in the literature~\citep{Sanjib,Liddle07}. To compare a model ($M_2$) with another model ($M_1$), we define the Bayes factor ($B_{21}$) given by:
\begin{equation}\label{32}
    B_{21}= \frac{\int P(D|M_2, \theta_2)P(\theta_2|M_2) \, d\theta_2}{\int P(D|M_1, \theta_1)P(\theta_1|M_1) \, d\theta_1}
\end{equation}
where $P(D|M_2,\theta_2)$ is the likelihood for the model $M_2$ given the data $D$, and $P(\theta_2|M_2)$ is the prior for the parameter vector $\theta_2$ of the model $M_2$. The denominator refers to the same, but for model $M_1$. If $B_{21}$ is greater than 1.0, then model $M_2$ is preferred over $M_1$ and vice-versa. The significance can be evaluated using Jeffreys' scale. If a model $M_2$ is decisively favored compared to model $M_1$, its Bayes factor has to be greater than 100 according to Jeffreys' scale~\citep{Weller}.

For our analysis, $M_1$ consists of the hypothesis that the rotation curve can be described by the NFW dark matter profile and a given baryon model. Therefore, by definition, this model has a Bayes factor of 1.0. $M_2$ refers to all the other dark matter or MOND models for the same baryon model. The detailed list of models along with the free parameters for each of these models can be found in Table \ref{tab:models}.
\begin{longtable}[c]{|l|c|l|}
    \caption{\label{tab:models}Summary of the combinations of the models used for our analysis along with priors used. The first seven models correspond to the {\tt baryon + DM profile} models. The next three are the MOND-corrected models, with $a_0$ kept as a free parameter. The last three are the MOND-corrected models in the standard paradigm, where $a_0 = 1.2 \times 10^{-10} m/s^2$. For B0 and B2, $M_{d}$ refers to the stellar disk. For B1, $M_{d}$ refers to both thin and thick disks.}\\
    \hline
    \textbf{\multirow{2}{8em}{Model M}} & \textbf{\multirow{2}{9em}{Parameters $\theta$ [Units]}} & \textbf{\multirow{2}{8em}{Bayesian priors}} \\[2pt]
    & & \\[2pt]
    \hline
    \endfirsthead
    \hline
    \textbf{\multirow{2}{8em}{Model M}} & \textbf{\multirow{2}{9em}{Parameters $\theta$ [Units]}} & \textbf{\multirow{2}{8em}{Bayesian priors}} \\
    [2pt]
    & & \\[2pt]
    \hline
    \endhead
    \hline\hline
    \endlastfoot
    \hline
    \multicolumn{3}{|c|}{dark matter profiles} \\[2pt]
    \multirow{3}{14em}{(B0, B1 or B2) + NFW profile}        & $M_{d}$ [$10^9 \times M_{\odot}$]        & $\mathcal{U}(10^{-3}$, $100)$ \\[2pt]
                                                            & $\rho_0$ [$10^7 \times M_{\odot} \text{kpc}^{-3}]$ & $\mathcal{U}(10^{-2}$, $100)$ \\[2pt]
                                                            & $r_s$ [kpc]                                        & $\mathcal{U}(10^{-2}$, $50)$  \\[2pt]
    \multirow{3}{14em}{(B0, B1 or B2) + Isothermal profile} & $M_{d}$ [$10^9 \times M_{\odot}$]        & $\mathcal{U}(10^{-3}$, $100)$ \\[2pt]
                                                            & $\rho_0$ [$10^7 \times M_{\odot} \text{kpc}^{-3}]$ & $\mathcal{U}(10^{-2}$, $1000)$ \\[2pt]
                                                            & $r_s$ [kpc]                                        & $\mathcal{U}(10^{-2}$, $100)$  \\[2pt]
    \multirow{3}{14em}{(B0, B1 or B2) + Burkert profile}    & $M_{d}$ [$10^9 \times M_{\odot}$]        & $\mathcal{U}(10^{-3}$, $100)$ \\[2pt]
                                                            & $\rho_0$ [$10^7 \times M_{\odot} \text{kpc}^{-3}]$ & $\mathcal{U}(10^{-2}$, $100)$ \\[2pt]
                                                            & $r_s$ [kpc]                                        & $\mathcal{U}(10^{-2}$, $50)$  \\[2pt]
    \multirow{4}{14em}{(B0, B1 or B2) + gNFW profile}       & $M_{d}$ [$10^9 \times M_{\odot}$]        & $\mathcal{U}(10^{-3}$, $100)$ \\[2pt]
                                                            & $\rho_0$ [$10^7 \times M_{\odot} \text{kpc}^{-3}]$ & $\mathcal{U}(10^{-2}$, $1000)$ \\[2pt]
                                                            & $r_s$ [kpc]                                        & $\mathcal{U}(10^{-2}$, $100)$  \\[2pt]
                                                            & $\alpha$                                           & $\mathcal{U}(0$, $1.5)$  \\[2pt]
    \multirow{3}{14em}{(B0, B1 or B2) + $\beta$ profile}    & $M_{d}$ [$10^9 \times M_{\odot}$]        & $\mathcal{U}(10^{-3}$, $100)$ \\[2pt]
                                                            & $\rho_0$ [$10^7 \times M_{\odot} \text{kpc}^{-3}]$ & $\mathcal{U}(10^{-2}$, $100)$ \\[2pt]
                                                            & $r_s$ [kpc]                                        & $\mathcal{U}(10^{-2}$, $50)$  \\[2pt]
    \multirow{3}{14em}{(B0, B1 or B2) + Hernquist profile}  & $M_{d}$ [$10^9 \times M_{\odot}$]        & $\mathcal{U}(10^{-3}$, $100)$ \\[2pt]
                                                            & $\rho_0$ [$10^7 \times M_{\odot} \text{kpc}^{-3}]$ & $\mathcal{U}(10^{-2}$, $100)$ \\[2pt]
                                                            & $r_s$ [kpc]                                        & $\mathcal{U}(10^{-2}$, $50)$  \\[2pt]
    \multirow{4}{14em}{(B0, B1 or B2) + Einasto profile}    & $M_{d}$ [$10^9 \times M_{\odot}$]        & $\mathcal{U}(10^{-3}$, $100)$ \\[2pt]
                                                            & $\rho_0$ [$10^7 \times M_{\odot} \text{kpc}^{-3}]$ & $\mathcal{U}(10^{-2}$, $100)$ \\[2pt]
                                                            & $r_s$ [kpc]                                        & $\mathcal{U}(10^{-2}$, $100)$  \\[2pt]
                                                            & $\alpha$                                           & $\mathcal{U}(0$, $7)$  \\[2pt]
    \multicolumn{3}{|c|}{$a_0$ as free parameter} \\[2pt]
    \multirow{2}{14em}{(B0, B1 or B2) + standard IF}        & $M_{d}$ [$10^9 \times M_{\odot}$]   & $\mathcal{U}(10^{-3}$, $100)$ \\[2pt]
                                                            & $a_0$ [$10^{-10}$ m/s$^{2}$]           & $\mathcal{U}(0$, $10^4)$ \\[2pt]
    \multirow{2}{14em}{(B0, B1 or B2) + simple IF}          & $M_{d}$ [$10^9 \times M_{\odot}$]   & $\mathcal{U}(10^{-3}$, $100)$ \\[2pt]
                                                            & $a_0$ [$10^{-10}$ m/s$^{2}$]           & $\mathcal{U}(0$, $10^4)$ \\[2pt]
    \multirow{2}{14em}{(B0, B1 or B2) + RAR}                & $M_{d}$ [$10^9 \times M_{\odot}$]   & $\mathcal{U}(10^{-3}$, $100)$ \\[2pt]
                                                            & $a_0$ [$10^{-10}$ m/s$^{2}$]           & $\mathcal{U}(0$, $10^4)$ \\[2pt]
    \multicolumn{3}{|c|}{standard paradigm} \\[2pt]
    (B0, B1 or B2) + standard IF         & $M_{d}$ [$10^9 \times M_{\odot}$]   & $\mathcal{U}(10^{-3}$, $100)$ \\[2pt]
    (B0, B1 or B2) + simple IF           & $M_{d}$ [$10^9 \times M_{\odot}$]   & $\mathcal{U}(10^{-3}$, $100)$ \\[2pt]
    (B0, B1 or B2) + RAR                 & $M_{d}$ [$10^9 \times M_{\odot}$]   & $\mathcal{U}(10^{-3}$, $100)$ \\[2pt]
\end{longtable}
Therefore, the Bayes factors are computed with respect to the same baryon model along with the NFW profile. The first step in calculating Bayesian evidence is determining the likelihood. We assume a Gaussian likelihood given by $P(D|M, \theta)\equiv -0.5 \chi^2$ where $\chi^2$ is given by:
\begin{equation}\label{33}
    \chi^2(\theta) = \sum_{i = 1}^{n} \frac{(v_{tot} - v_{obs})^2}{\sigma_i^2}
\end{equation}
where $v_{obs}$ is the observed circular velocity and $\sigma_i$ is the Gaussian-distributed uncertainty associated with each data point.
In Eq.~\eqref{33}, $v_{tot}$ can be obtained from Eq.~\eqref{24} and Eq.~\eqref{30} for dark matter and MOND models, respectively.  The priors for each of the model parameters can be found in Table~\ref{tab:models}.
We use dynamical nested sampling, implemented in the {\tt dynesty} package \citep{Speagle}, to calculate the Bayesian evidence for each model. The nested sampler also returns the posteriors, which are shown in Fig.~\ref{fig:placeholder} of Appendix \ref{appendixb}. The parameter estimates were calculated using the median of the posterior chain generated in the analysis and tabulated in Table~\ref{tab:param1} in  Appendix \ref{appendixa}. 

\section{Results and discussion}
\label{sec:results}
Our results for the $\ln$ (evidence) for the aforementioned models, using all four datasets and combinations of baryon + dark matter/MOND interpolating functions, are presented in Table \ref{tab:lnZ}. We first compare the evidence against the {\tt baryon + NFW profile}, which is chosen as the null hypothesis, and then select the best-suited DM model or MOND IF. We then report the corresponding Bayes factors for the different dark matter and baryon models in Table \ref{tab:baye}. Finally, we also compare the different baryonic models in conjunction with the Einasto dark matter model against each other, and calculate the Bayes factor by choosing the B0 baryonic model as the null hypothesis. This table of Bayes factors for the baryonic models can be found in Table~\ref{tab:baryon}.

\begin{longtable}[t]{|l|c|c|c|c|}
\caption{\label{tab:lnZ} Bayesian evidence (on log scale) for all the models.}\\
\hline
    \multirow{2}{4em}{Model} & \multicolumn{4}{c|}{$\ln$ (Evidence)} \\
    & D1 (N = 18) & D2 (N = 34) & D3 (N = 37) & D4 (N = 19)\\[3pt]
    \hline
    \endfirsthead
    \hline
    \multirow{2}{4em}{Model} & \multicolumn{4}{c|}{$\ln$ (Evidence) } \\
    & D1 (N = 18) & D2 (N = 34) & D3 (N = 37) & D4 (N = 19)\\[3pt]
    \hline
    \endhead
    \hline \hline
    \endlastfoot
    B0 + NFW profile        & $-19.1 \pm 0.07$ & $-238.3 \pm 0.09$ & $-225.1 \pm 0.10$ & $-40.6 \pm 0.08$ \\[3pt]
    B0 + Isothermal profile & $-26.5 \pm 0.07$ & $-374.6 \pm 0.11$ & $-477.1 \pm 0.10$ & $-65.3 \pm 0.08$ \\[3pt]
    B0 + Burkert profile    & $-17.5 \pm 0.07$ & $-121.4 \pm 0.09$ & $-159.8 \pm 0.09$ & $-36.7 \pm 0.07$ \\[3pt]
    B0 + gNFW profile       & $-22.5 \pm 0.09$ & $-175.9 \pm 0.12$ & $-203.5 \pm 0.13$ & $-44.9 \pm 0.10$ \\[3pt]
    B0 + $\beta$ profile    & $-17.3 \pm 0.07$ & $-124.8 \pm 0.10$ & $-151.0\pm 0.10$ & $-35.1 \pm 0.08$ \\[3pt]
    B0 + Hernquist profile  & $-18.1 \pm 0.08$ & $-163.1 \pm 0.09$ & $-153.2 \pm 0.11$ & $-34.9 \pm 0.08$ \\[3pt]
    B0 + Einasto profile    & $-17.1 \pm 0.09$ & $-141.5 \pm 0.12$ & $-97.0 \pm 0.12$ & $-24.1 \pm 0.10$ \\[3pt]
    \multicolumn{5}{|c|}{$a_0$ as free parameter}\\
    B0 + standard IF    & $-28.38 \pm 0.08$ & $-304.86 \pm 0.09$ & $-403.35 \pm 0.09$ & $-69.59 \pm 0.08$ \\[3pt]
    B0 + simple IF      & $-28.52 \pm 0.08$ & $-283.26 \pm 0.09$ & $-398.95 \pm 0.09$ & $-68.95 \pm 0.08$ \\[3pt]
    B0 + RAR            & $-28.22 \pm 0.08$ & $-261.55 \pm 0.09$ & $-383.67 \pm 0.09$ & $-68.07 \pm 0.08$ \\[3pt]
    \multicolumn{5}{|c|}{constant $a_0$; standard MOND paradigm}\\
    B0 + standard IF    & $-20.80 \pm 0.04$ & $-3833.06$ & $-1891.96$ & $-59.04 \pm 0.04$ \\[3pt]
    B0 + simple IF      & $-19.63 \pm 0.03$ & $-2614.15$ & $-1414.74$ & $-58.61 \pm 0.04$ \\[3pt]
    B0 + RAR            & $-18.78 \pm 0.03$ & $-2055.74$ & $-1155.07$ & $-58.35 \pm 0.01$ \\[3pt]
    \hline
    B1 + NFW profile         & $-17.9 \pm 0.07$ & $-199.8 \pm 0.09$ & $-207.3 \pm 0.09$ & $-37.9 \pm 0.07$ \\[3pt]
    B1 + Isothermal profile  & $-22.6 \pm 0.08$ & $-217.1 \pm 0.11$ & $-292.1 \pm 0.10$ & $-49.8 \pm 0.09$ \\[3pt]
    B1 + Burkert profile     & $-16.4 \pm 0.07$ & $-124.3 \pm 0.09$ & $-148.9 \pm 0.09$ & $-33.9 \pm 0.07$ \\[3pt]
    B1 + gNFW profile        & $-21.4 \pm 0.08$ & $-153.6 \pm 0.12$ & $-184.7 \pm 0.12$ & $-41.6 \pm 0.09$ \\[3pt]
    B1 + $\beta$ profile     & $-16.3 \pm 0.07$ & $-128.6 \pm 0.09$ & $-145.2 \pm 0.09$ & $-32.2 \pm 0.07$ \\[3pt]
    B1 + Hernquist profile   & $-16.9 \pm 0.07$ & $-158.2 \pm 0.09$ & $-146.8 \pm 0.09$ & $-33.0 \pm 0.07$ \\[3pt]
    B1 + Einasto profile     & $-16.6 \pm 0.09$ & $-145.1 \pm 0.12$ & $-96.6 \pm 0.12$ & $-24.1 \pm 0.10$ \\[3pt]
    \multicolumn{5}{|c|}{$a_0$ as free parameter}\\
    B1 + standard IF    & $-27.88 \pm 0.08$ & $-423.28 \pm 0.09$ & $-407.11 \pm 0.1$ & $-62.86 \pm 0.08$ \\[3pt]
    B1 + simple IF      & $-26.07 \pm 0.08$ & $-285.45 \pm 0.09$ & $-339.01 \pm 0.09$ & $-55.90 \pm 0.08$ \\[3pt]
    B1 + RAR            & $-26.17 \pm 0.08$ & $-273.71 \pm 0.09$ & $-331.24 \pm 0.09$ & $-55.99 \pm 0.08$ \\[3pt]
    \multicolumn{5}{|c|}{constant $a_0$; standard MOND paradigm}\\
    B1 + standard IF    & $-21.97 \pm 0.04$ & $-410.77 \pm 0.05$ & $-408.85 \pm 0.05$ & $-83.50 \pm 0.04$ \\[3pt]
    B1 + simple IF      & $-23.96 \pm 0.04$ & $-437.05 \pm 0.05$ & $-474.67 \pm 0.05$ & $-92.71 \pm 0.04$ \\[3pt]
    B1 + RAR            & $-25.78 \pm 0.04$ & $-596.55 \pm 0.05$ & $-573.27 \pm 0.05$ & $-99.11 \pm 0.04$ \\[3pt]
    \hline
    B2 + NFW profile         & $-18.9 \pm 0.07$ & $-228.0 \pm 0.09$ & $-224.5 \pm 0.10$ & $-40.5 \pm 0.08$ \\[3pt]
    B2 + Isothermal profile  & $-26.1 \pm 0.07$ & $-336.1 \pm 0.10$ & $-453.8 \pm 0.10$ & $-64.3 \pm 0.07$ \\[3pt]
    B2 + Burkert profile     & $-17.3 \pm 0.07$ & $-121.1 \pm 0.09$ & $-159.4 \pm 0.09$ & $-36.5 \pm 0.07$ \\[3pt]
    B2 + gNFW profile        & $-22.42 \pm 0.08$ & $-167.98 \pm 0.12$ & $-203.06 \pm 0.12$ & $-44.58 \pm 0.10$ \\[3pt]
    B2 + $\beta$ profile     & $-17.16 \pm 0.07$ & $-125.15 \pm 0.10$ & $-151.02 \pm 0.10$ & $-35.05 \pm 0.07$ \\[3pt]
    B2 + Hernquist profile   & $-17.95 \pm 0.07$ & $-159.74 \pm 0.09$ & $-163.35 \pm 0.10$ & $-34.91 \pm 0.08$ \\[3pt]
    B2 + Einasto profile     & $-16.87 \pm 0.09$ & $-138.60 \pm 0.12$ & $-100.88 \pm 0.12$ & $-24.02 \pm 0.06$ \\[3pt]
    \multicolumn{5}{|c|}{$a_0$ as free parameter}\\
    B2 + standard IF   & $-28.02 \pm 0.08$ & $-279.42 \pm 0.09$ & $-384.12 \pm 0.09$ & $-68.64 \pm 0.08$ \\[3pt]
    B2 + simple IF     & $-28.13 \pm 0.08$ & $-257.69 \pm 0.09$ & $-379.52 \pm 0.09$ & $-67.80 \pm 0.08$ \\[3pt]
    B2 + RAR           & $-27.77 \pm 0.08$ & $-238.01 \pm 0.09$ & $-364.92 \pm 0.09$ & $-66.93 \pm 0.08$ \\[3pt]
    \multicolumn{5}{|c|}{constant $a_0$; standard MOND paradigm}\\
    B2 + standard IF   & $-20.11 \pm 0.04$ & $-3265.88$ & $-1651.68$ & $-57.93 \pm 0.04$ \\[3pt]
    B2 + simple IF     & $-18.94 \pm 0.04$ & $-2160.42$ & $-1214.28$ & $-57.61 \pm 0.04$ \\[3pt]
    B2 + RAR           & $-18.23 \pm 0.03$ & $-1661.44$ & $-980.34$ & $-57.66 \pm 0.04$ \\[3pt]
    \hline
\end{longtable}

\begin{longtable}[l]{|l|c|c|c|c|}
\caption{\label{tab:baye} Bayes factors for all the models used with the null hypothesis consisting of {\tt baryon + NFW profile}. The table is divided into three sections, each section comparing DM models or MOND IFs against the null hypothesis with the same baryon model. {\tt B0 + NFW profile}, {\tt B1 + NFW profile}, and {\tt B2 + NFW profile} are assumed as the null hypothesis for the three sections, respectively. Each section is further subdivided into three parts. Part one compares {\tt baryon + DM profiles} against the null hypothesis. Part two compares the {\tt baryon + MOND IFs}, with $a_0$ as a free parameter, against the null hypothesis. Part three compares the {\tt baryon + MOND IFs}, with $a_0 = 1.2 \times 10^{-10} m/s^{-2}$, against the null hypothesis.}\\
\hline
    \multirow{2}{6em}{Model} & \multicolumn{4}{c|}{Bayes factor for the models} \\
    & D1 (N = 18) & D2 (N = 34) & D3 (N = 37) & D4 (N = 19)\\[3pt]
    \hline
    \endfirsthead
    \hline
    \multirow{2}{6em}{Model} & \multicolumn{4}{c|}{Bayes factor for the models} \\
    & D1 (N = 18) & D2 (N = 34) & D3 (N = 37) & D4 (N = 19)\\[3pt]
    \hline
    \endhead
    \hline \hline
    \endlastfoot
    \rule{0pt}{3ex}\bf{B0 + NFW profile}    & $1.0$ & $1.0$ & $1.0$ & $1.0$ \\[3pt]
    B0 + Isothermal profile  & $6.03 \times 10^{-4}$ & $6.074 \times 10^{-60}$ & $3.44 \times 10^{-110}$ & $1.95 \times 10^{-11}$ \\[3pt]
    B0 + Burkert profile     & $4.85$ & $5.67 \times 10^{50}$ & $2.06 \times 10^{28}$ & $49.21$ \\[3pt]
    B0 + gNFW profile        & $0.03$ & $1.19 \times 10^{27}$ & $2.19 \times 10^9$ & $0.014$ \\[3pt]
    B0 + $\beta$ profile     & $5.94$ & $1.86 \times 10^{49}$ & $1.47 \times 10^{32}$ & $264.37$ \\[3pt]
    B0 + Hernquist profile   & $2.88$ & $4.53 \times 10^{32}$ & $1.56 \times 10^{31}$ & $327.25$ \\[3pt]
    B0 + Einasto profile     & $8.05$ & $1.02 \times 10^{42}$ & $4.11 \times 10^{55}$ & $1.47 \times 10^{7}$ \\[3pt]
    \multicolumn{5}{|c|}{$a_0$ as free parameter}\\
    B0 + standard IF         & $9.91 \times 10^{-5}$ & $1.22 \times 10^{-29}$ & $3.77 \times 10^{-78}$ & $2.75 \times 10^{-13}$ \\[3pt]
    B0 + simple IF           & $8.64 \times 10^{-5}$ & $2.93 \times 10^{-20}$ & $3.06 \times 10^{-76}$ & $5.23 \times 10^{-13}$ \\[3pt]
    B0 + RAR                 & $1.16 \times 10^{-4}$ & $7.82 \times 10^{-11}$ & $1.32 \times 10^{-69}$ & $1.26 \times 10^{-12}$ \\[3pt] 
    \multicolumn{5}{|c|}{constant $a_0$; standard MOND paradigm}\\
    B0 + standard IF         & $0.194$ & $<10^{-308}$ & $<10^{-308}$ & $1.05 \times 10^{-8}$ \\[3pt]
    B0 + simple IF           & $0.62$ & $<10^{-308}$ & $<10^{-308}$ & $1.62 \times 10^{-8}$ \\[3pt]
    B0 + RAR                 & $1.47$ & $<10^{-308}$ & $<10^{-308}$ & $2.10 \times 10^{-8}$ \\[3pt]
    \hline
    \rule{0pt}{3ex}\bf{B1 + NFW profile}    & $1.0$ & $1.0$ & $1.0$ & $1.0$ \\[3pt]
    B1 + Isothermal profile  & $9.35 \times 10^{-3}$ & $3.05 \times 10^{-8}$ & $1.61 \times 10^{-37}$ & $7.12 \times 10^{-6}$ \\[3pt]
    B1 + Burkert profile     & $4.76$ & $6.33 \times 10^{32}$ & $2.30 \times 10^{25}$ & $54.99$ \\[3pt]
    B1 + gNFW profile        & $0.032$ & $1.08 \times 10^{20}$ & $6.66 \times 10^{9}$ & $0.027$ \\[3pt]
    B1 + $\beta$ profile     & $5.55$ & $7.89 \times 10^{30}$ & $10^{27}$ & $326.73$ \\[3pt]
    B1 + Hernquist profile   & $2.74$ & $1.12 \times 10^{18}$ & $1.99 \times 10^{26}$ & $139.84$ \\[3pt]
    B1 + Einasto profile     & $3.91$ & $5.53 \times 10^{23}$ & $1.17 \times 10^{48}$ & $1.03 \times 10^{6}$ \\[3pt]
    \multicolumn{5}{|c|}{$a_0$ as free parameter}\\
    B1 + standard IF         & $5.09 \times 10^{-5}$ & $8.90 \times 10^{-98}$ & $1.76 \times 10^{-87}$ & $1.58 \times 10^{-11}$ \\[3pt]
    B1 + simple IF           & $3.11 \times 10^{-4}$ & $6.41 \times 10^{-38}$ & $6.71 \times 10^{-58}$ & $1.66 \times 10^{-8}$ \\[3pt]
    B1 + RAR                 & $2.81 \times 10^{-4}$ & $8.02 \times 10^{-33}$ & $1.58 \times 10^{-54}$ & $1.51 \times 10^{-8}$ \\[3pt]
    \multicolumn{5}{|c|}{constant $a_0$; standard MOND paradigm}\\
    B1 + standard IF         & $1.86 \times 10^{-2}$ & $2.41 \times 10^{-92}$ & $3.10 \times 10^{-88}$ & $1.71 \times 10^{-20}$ \\[3pt]
    B1 + simple IF           & $2.56 \times 10^{-3}$ & $9.26 \times 10^{-104}$ & $8.13 \times 10^{-117}$ & $1.72 \times 10^{-24}$ \\[3pt]
    B1 + RAR                 & $4.13 \times 10^{-4}$ & $4.99 \times 10^{-173}$ & $1.21 \times 10^{-159}$ & $2.86 \times 10^{-27}$ \\[3pt]
    \hline
    \rule{0pt}{3ex}\bf{B2 + NFW profile}    & $1.0$ & $1.0$ & $1.0$ & $1.0$ \\[3pt]
    B2 + Isothermal profile  & $7.19 \times 10^{-4}$ & $1.15 \times 10^{-47}$ & $2.62 \times 10^{-100}$ & $5.01 \times 10^{-11}$ \\[3pt]
    B2 + Burkert profile     & $4.7$ & $2.69 \times 10^{46}$ & $1.77 \times 10^{28}$ & $61.08$ \\[3pt]
    B2 + gNFW profile        & $2.96 \times 10^{-2}$ & $1.21 \times 10^{26}$ & $2.067 \times 10^{9}$ & $1.82 \times 10^{-2}$ \\[3pt]
    B2 + $\beta$ profile     & $5.66$ & $4.85 \times 10^{44}$ & $8.27 \times 10^{31}$ & $251.9$ \\[3pt]
    B2 + Hernquist profile   & $2.58$ & $4.59 \times 10^{29}$ & $3.67 \times 10^{26}$ & $290.83$ \\[3pt]
    B2 + Einasto profile     & $7.62$ & $6.99 \times 10^{38}$ & $4.93 \times 10^{53}$ & $1.09 \times 10^{7}$ \\[3pt]
    \multicolumn{5}{|c|}{$a_0$ as free parameter}\\
    B2 + standard IF         & $1.08 \times 10^{-4}$ & $4.87 \times 10^{-23}$ & $4.82 \times 10^{-70}$ & $6.5 \times 10^{-13}$ \\[3pt]
    B2 + simple IF           & $9.82 \times 10^{-5}$ & $1.32 \times 10^{-13}$ & $4.81 \times 10^{-68}$ & $1.50 \times 10^{-12}$ \\[3pt]
    B2 + RAR                 & $1.39 \times 10^{-4}$ & $4.711 \times 10^{-5}$ & $1.056 \times 10^{-61}$ & $3.58 \times 10^{-12}$ \\[3pt]
    \multicolumn{5}{|c|}{constant $a_0$; standard MOND paradigm}\\
    B2 + standard IF         & $0.29$ & $<10^{-308}$ & $<10^{-308}$ & $2.91 \times 10^{-8}$ \\[3pt]
    B2 + simple IF           & $0.95$ & $<10^{-308}$ & $<10^{-308}$ & $3.99 \times 10^{-8}$ \\[3pt]
    B2 + RAR                 & $1.94$ & $<10^{-308}$ & $<10^{-308}$ & $3.79 \times 10^{-8}$ \\[3pt]
    \hline
\end{longtable}

\begin{figure}[h]
    \centering
    \caption{Rotation curve of the Milky Way constructed from the best-fit parameters {\tt baryon + Einasto profile} for the {\tt D3} dataset, which has the strongest Bayesian evidence.}
    \includegraphics[width=1\textwidth]{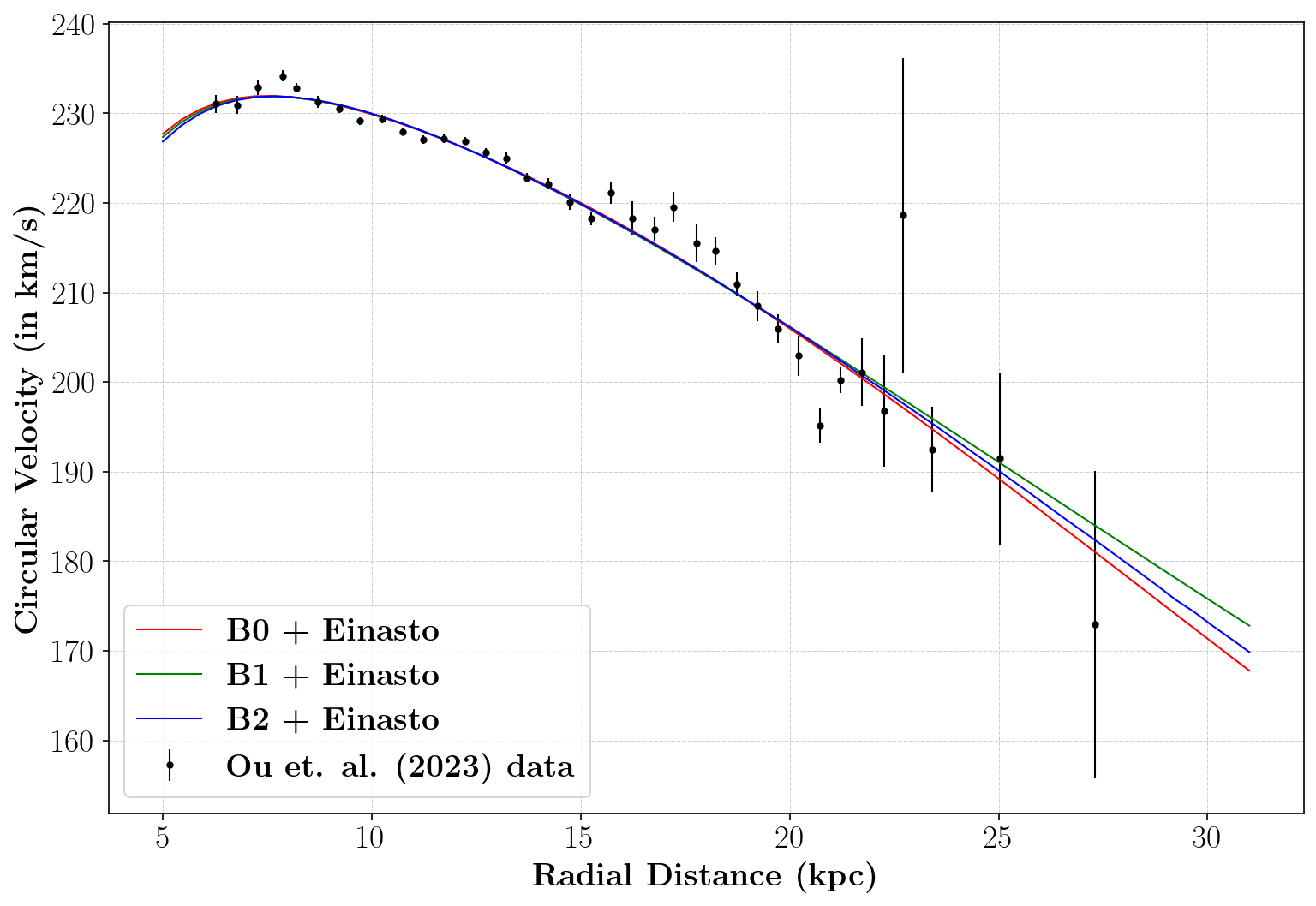}
    \label{rc}
\end{figure}

 We now present several salient features from these results:
\begin{itemize}
    \item  Among D3 and D4 datasets, the Einasto dark matter profile has the highest  Bayes factors  ($> 100$) for all baryonic models. We also show the best-fit Einasto profile along with all three baryon models for the {\tt D3} dataset in Fig.~\ref{rc}. We also show the best-fit Einasto profile along with all three baryon models for the {\tt D3} dataset in Fig.~\ref{rc}. 
    \item For the D1 dataset, the Bayes factor for the Einasto model for all baryonic models is less than 10, indicating only substantial evidence. For the B1 baryonic model, the Bayes factor for Burkert and $\beta$-profile is marginally larger compared to the Einasto profile.
    \item Only the {\tt D2} data show a higher Bayes factor for the  $\beta$ profile and the Burkert profile, as compared to that of the Einasto profile for all baryonic models. However, even for {\tt D2}, the Einasto profile yields a very high Bayes factor ($>100$) for all the baryonic models, indicating that it is decisively favored over the NFW profile.
    \item Among the cored profiles, both the Burkert and $\beta$ profile have Bayes factors $>100$ for {\tt D2}, and {\tt D3} datasets (for all baryonic models). Therefore, both these cored profiles provide a much better fit than the NFW model. In addition, the $\beta$ profile has a Bayes factor of $>100$ for {\tt D4} dataset.
    \item We note that the best-fit value of $\alpha$ for the Einasto profile is \rthis{positive} for all baryonic-dataset combinations, indicating it is closer to a cored profile.
    \item We find that the gNFW profile is decisively favored (Bayes factor $> 100$) over the NFW profile for the {\tt D2} and {\tt D3} datasets. However, for {\tt D1} and {\tt D4}, the Bayes factors are $<0.1$, indicating that the NFW model is moderately favored over gNFW.
    \item We note that the Hernquist profile is favored against the NFW profile across all datasets, especially for the {\tt D2}, {\tt D3}, and {\tt D4} datasets.
    \item The isothermal profile yields a very small Bayes factor across all the datasets for all baryonic models.
    \item The standard MOND paradigm ($a_0 = 1.2 \times 10^{-10} m/s^2$) is strongly disfavored \rthis{compared to NFW profile} across almost all datasets \rthis{based on the assumptions in this work.} This arises from the nature of the interpolating functions, which produce constant velocities at large radii, in contradiction with the declining trend in the rotation curve (see equation \eqref{31} and Figure \ref{fig:data}). This is also in accord with the conclusions from recent works comparing MOND models with Gaia-based RC~\citep{Starkman,ManHoChan,Blanchard}. Even when $a_0$ is kept as a free parameter, the Bayes factors are very small and insignificant against the NFW model. \rthis{The MOND fits for most of the combinations of baryonic models, and interpolating functions are also very bad. We also note that for B1+RAR,  when $a_0$ is a free parameter, the best-fit values of $a_0$ are about an order of magnitude smaller. }
   \item When we compare the baryonic models with each other, keeping the {B0 + Einasto} model as the null hypothesis, we find the Bayes factors for the baryonic models across most of the datasets to be less than 10, implying that there is no strong preference for any one baryonic model. The highest Bayes factor is obtained for the B2 model when considering the D2 dataset, but even here, the value is around 20, and hence it is not decisively favored compared to B0. 
    
\end{itemize}
Therefore, the main conclusion from our  Bayesian inference is that \rthis{within the adopted modeling framework and Gaia-based rotation-curve datasets, the Einasto profile provides the most preferred phenomenological fit}  across most combinations of datasets and baryonic models.  \rthis{We note, however, that some of the posteriors for $M_d$ remain very broad, especially for D1-based fits where the halo dominates the outer rotation curve. This suggests that some of the RC data do not strongly constrain the baryonic contribution in those cases, and therefore the Bayesian evidence may still retain some sensitivity to the adopted prior volume for $M_d$. Nevertheless, this dependence on $M_d$ would be there for all dark matter models for a given fixed baryonic model. Therefore, we do not expect it to make a big difference while comparing the  BFs for two dark matter profiles, as long as both are used in conjunction with the same baryonic profile. }


\begin{longtable}[l]{|l|c|c|c|c|}
\caption{\label{tab:baryon} Bayes factors for the different combinations of baryonic models + Einasto profile, with the null hypothesis consisting of {\tt B0 + Einasto profile}.}\\
\hline
    \multirow{2}{6em}{Model} & \multicolumn{4}{c|}{Bayes factor for the Einasto models} \\
    & D1 (N = 18) & D2 (N = 34) & D3 (N = 37) & D4 (N = 19)\\[3pt]
    \hline
    \endfirsthead
    \hline
    \multirow{2}{6em}{Model} & \multicolumn{4}{c|}{Bayes factor for the Einasto models} \\
    & D1 (N = 18) & D2 (N = 34) & D3 (N = 37) & D4 (N = 19)\\[3pt]
    \hline
    \endhead
    \hline \hline
    \endlastfoot
    \rule{0pt}{3ex}\bf{B0 + Einasto profile}    & $1.0$ & $1.0$ & $1.0$ & $1.0$ \\[3pt]
    B1 + Einasto profile     & $1.56$ & $0.02$ & $1.41$ & $1.02$ \\[3pt]
    B2 + Einasto profile     & $1.23$ & $19.27$ & $0.02$ & $0.81$ \\[3pt]
    \hline
\end{longtable}

\subsection{Comparison  with  recent works}
\rthis{We now compare our results for the best-fit halo model parameters (cf. Table~\ref{tab:param1}) with some recent works which have estimated the same using  using Milky Way RC measurements as well as cosmological simulations, focusing on the Einasto profile. }

\rthis{The best-fit values for the Einasto profile parameters using the pre-Gaia Milky RC data obtained in ~\citet{Jain} are $\alpha=0.041^{+0.471}_{-0.109}$,  and $r_c=8.745^{+24.138}_{-5.923}$ kpc. 
These parameters are in agreement with our best-fit values within $1\sigma$. However, we note that this work also uses different baryonic models. \citet{deSalas19} used the Gaia DR2 based RC curve~\citep{Eilers19} and obtained Einasto profile parameter values of $r_c=9.2^{5.3}_{-2.7}$ kpc and $\alpha=0.384\pm 0.03$ using the B1 baryonic model. These agree with our best-fit values within $1\sigma$. Although no estimate of $\rho_0$ was obtained in  \citet{deSalas19}, the inferred value of the  dark matter density at the location of the Sun using the Einasto profile was $(8.17 \pm 0.71) \times 10^6 M_{\odot}/kpc^3$, which is roughly of the same order of magnitude as our best-fit value of $\rho_0$. The Einasto profile parameters were also obtained in \cite{Jiao23} using D1 and D3 rotation curves for multiple baryonic models. We find that the posteriors for $\alpha$ and $r_s$ (cf. Fig.~6 of \cite{Jiao23}) are in agreement with our best-fit values. The best-fit value for $\alpha$ using the D3 dataset  estimated in ~\cite{Eilers23} ($\alpha=0.91^{+0.04}_{-0.05}$) agrees with our best-fit estimates within $1\sigma$. We note that some of the aforementioned works used a slightly different parametric form for the Einasto profile~\citep{Jain,Eilers23} and hence $\rho_0$ cannot be directly compared with our fits. Therefore, all other estimates of Einasto $\alpha$ using Milky way RC are consistent with a core  in accord with our findings.} 

\rthis{A comparative study of dark matter models using pre-GAIA Milky Way RC data~\citep{Huang16} was also carried out in ~\citet{Lin19}. This work did model comparison of four dark matter models (NFW, Burkert, isothermal, core-modified) and 56 baryonic models, with the best-fit model selected based on the minimum reduced $\chi^2$. This work found that for 32 of the 56 baryon models, the NFW profile fits the data the best, whereas for the remaining, the Burkert profile provides the best fit~\citep{Lin19}.}

\rthis{Although a large number of cosmology simulations of Milky Way haloes have been generated in recent years~\citep{Lazar,Grand,Smirnov25,Sanderson,Ngu,Pillepich} (and references therein), we are aware of a handful of these~\citep{Lazar,Smirnov25} which have done a model comparison of dark matter profiles to fit the synthetic haloes. ~\citet{Smirnov25} found that  Milky haloes contain a dark matter core with a scale of $\sim 1$ kpc. ~\citet{Smirnov25} also applied the Bayesian Information criterion to compare between gNFW, isothermal, Burkert, and modified core profiles and found that the isothermal profile is the most preferred for the synthetic haloes. \cite{Lazar} fitted dark matter profiles for Milky way haloes from the FIRE-2 simulations~\citep{FIRE2} with stellar masses ranging from $10^{10-11} M_{\odot}$ to the core-Einasto profile (keeping $\alpha$ as fixed) as well as gNFW profile (with a slightly different parameterization compared to that in  Eq.~\ref{eq:gNFW}). The core-Einasto profile is an extension of the Einasto profile in which an extra parameter is added to the numerator inside the exponent in Eq.~\ref{eq:einasto}, so that the density contains a core at the center. This work found that the core-Einasto profile is a better fit than core NFW as well as some other profiles~\citep{Lazar}. The best-fit values of $r_c$ for gNFW profile for 12 Milky Way haloes range from 2.21-5.46 kpc~\citep{Lazar}, which are in agreement with our best-fit values of $r_c$ for gNFW profile. Similarly, the best-fit values of $r_c$ for the core-Einasto profile for haloes with good fits were found to be between 0.41 and 2.64, which are consistent with our estimates. Therefore, our estimated values for Einasto and gNFW profile parameters are in agreement with recent  results from cosmological simulations.}

\subsection{Sensitivity to prior choices}
\rthis{As we can see from Table~\ref{tab:models}, we have chosen  very broad priors for disk mass and halo parameters. These broad priors were chosen in order to get closed posteriors across all combinations of datasets and models used. To test the sensitvity to priors, we recalculated the Bayesian evidence for the Einasto model for the D3 dataset using a narrow range of priors as: $\alpha \in \mathcal{U}[10^{-2},3]$ and $\rho \in \mathcal{U} [10^{-2},50] \times 10^7 M_{\odot}/kpc^3$ and $r_s \in [10^{-2},100]$ kpc. We find the new values of $\ln$ (evidence) to be $-95.4 \pm 0.11$ (B0), $-95.05 \pm 0.09$ (B1), and $-94.99 \pm 0.11$ (B2). When we choose $\alpha \in \mathcal{U}[10^{-2},3]$, $\rho \in \mathcal{U} [10^{-2},100] \times 10^7 M_{\odot}/kpc^3$ and $r_s \in [10^{-2},50]$ kpc, we get $\ln Z = -95.99 \pm 0.11$ (B0), $-95.4 \pm 0.11$ (B1), and  $-95.13 \pm 0.11$  (B2)
Therefore, in both cases, $\ln$ (evidence) is larger,  with additive offsets of  about 1.0 (B0), 1.2-1.5 (B1), and 5.6-5.9 (B2) than those in Table ~\ref{tab:baye}.  Similarly, when we recalculate the Bayes factor  for the Burkert profile using conservative priors: $\rho \in \mathcal{U} [10^{-2},50] \times 10^7 M_{\odot}/kpc^3$, we find the values of $\ln$ (evidence) to be $-157.9 \pm 0.09$, $-147.3 \pm 0.08$, and $-157.7 \pm 0.09$ for B0,  B1, and B2. Although these are larger than those in Table~\ref{tab:baye} by additive offsets of 1.9 (B0), 1.6 (B1), 1.7 (B2), they are still not comparable with the evidences for the Einasto profile. Therefore, our conclusions regarding the efficacy of the Einasto profile compared to other profiles are robust with respect to the choice of priors. 
}

\section{Conclusions}
\label{sec:conclusions}
In this work, we conducted a comprehensive model comparison study of dynamical models for the Milky Way using four different compilations of RC data from Gaia DR3 (in conjunction with other data). These compilations using the latest Gaia data have found a decline in the Milky Way's rotation curve at distances beyond about 20 kpc.

We compare these data using three different baryon potentials, seven different dark models, and MOND models (for three interpolating functions). For each of the three baryonic models, we considered the disk mass to be a free parameter.
We perform a model selection analysis by calculating the Bayes factor with respect to the null hypothesis, which consists of the NFW model in conjunction with the Hernquist model and the Plummer model for the bulge, and the exponential disk model, the Miyamato-Nagai disk model, and the double exponential disk model for the baryon components. In all, we consider 13 dynamical models for each dataset. The Bayes factor relative to the widely used NFW model, used as the null hypothesis, is shown in Table~\ref{tab:baye} and a similar table comparing the baryonic models in Table~\ref{tab:baryon}.
We find that for almost all datasets, the Einasto profile is the \rthis{preferred phenomenological fit} over the NFW profile for \rthis{2-3} of the four datasets. The only exception is the {\tt D2} dataset, for which the $\beta$ \rthis{or  Burkert}  profile is the most preferred  over the NFW profile \rthis{for all baryonic models}. However, even for this profile, the {\tt D2}, Einasto profile shows very high Bayes factors, pointing to decisive evidence. \rthis{Our best-fit values for the Einasto profile are comparable with other recent estimates using the Milky Way RC as well as estimates from cosmological simulations.}
However, we note that all four datasets are anchored to Gaia DR3, and three of them cover nearly identical radial ranges. They share common systematic uncertainties, including parallax zero-point offsets and assumptions on $R_0$ and $V_0$. Therefore, these datasets are not truly independent. 

We also note that \rthis{within the MOND implementations considered here, the models provide poorer fits than the preferred dark matter profiles}, in accordance with other recent results that have tested the viability of MOND models in our galaxy. Finally, when we compare the evidence for the baryonic models, we do not find any one baryonic model decisively favored over the others.

More insights into dark matter should be found in the coming years with the upcoming Vera Rubin LSST, Square Kilometer Array, Thirty Meter Telescope, and Extremely Large Telescope.
\section*{Acknowledgments}
AS would like to extend his gratitude to the Council of Scientific and Industrial Research (CSIR), Govt. of India, for their continuous support through the Junior Research Fellowship, which has played a crucial role in the successful completion of our research. The computational work used for this analysis was supported by the National Supercomputing Mission (NSM), Government of India, through access to the ``PARAM SEVA'' facility at IIT Hyderabad. The NSM is implemented by the Centre for Development of Advanced Computing (C-DAC) with funding from the Ministry of Electronics and Information Technology (MeitY) and the Department of Science and Technology (DST). We are also grateful to the anonymous referee for many useful comments and constructive feedback on this manuscript.

\section*{Statements and Declarations}
AS was supported by CSIR, Govt. of India. The authors have no competing interests to declare that are relevant to the content of this article.

\bibliographystyle{sn-basic}
\bibliography{references}

\newpage
\appendix
\renewcommand{\thetable}{A\arabic{table}}
\setcounter{table}{0}
\setcounter{figure}{0}
\section{Appendix}
\subsection{Parameter estimates of the DM profiles or MOND IFs coupled with baryon models}\label{appendixa}

\begin{longtable}[c]{|l|c|c|c|c|c|}
    \caption{\label{tab:param1}Parameter estimates for baryon models + DM profiles or MOND IFs along with $1\sigma$ uncertainties. The disk mass $M_d$ is reported in units of $10^9 \times M_{\odot}$.  \rthis{Note that $M_d$ corresponds to mass of stellar disk for B0 and B2 profiles, and  the mass of both the thin and thick disk for B1 profile.}     The DM density parameter $\rho_0$ is reported in units of $10^7 \times M_{\odot} \text{kpc}^{-3}$ and the radius $r_c$ or $r_s$ is reported in $kpc$. The MOND parameter ($a_0$) is reported in units of $10^{-10} m/s^2$.}\\
    \hline
    \rule{0pt}{3ex}\multirow{2}{6em}{Model} & \rule{0pt}{3ex}\multirow{2}{1em}{$\theta$} & \multicolumn{4}{c|}{Parameters for different datasets} \\[3pt]
    & & D1 (N = 18) & D2 (N = 33) & D3 (N = 37) & D4 (N = 19) \\[3pt]
    \hline
    \endfirsthead
    \hline
    \rule{0pt}{3ex}\multirow{2}{6em}{Model} & \rule{0pt}{3ex}\multirow{2}{1em}{$\theta$} & \multicolumn{4}{c|}{Parameters for different datasets} \\[3pt]
    & & D1 (N = 18) & D2 (N = 33) & D3 (N = 37) & D4 (N = 19) \\[3pt]
    \hline
    \endhead
    \hline\hline
    \endlastfoot
    \multicolumn{6}{|c|}{Dark matter profiles}\\[2pt]
    \rule{0pt}{3ex}\multirow{3}{8em}{B0 + Burkert profile}
        & $M_d$ & $8.43^{+10.73}_{-6.08}$ & $16.66^{+4.94}_{-4.89}$ & $1.73^{+2.3}_{-1.26}$ & $4.76^{+6.62}_{-3.51}$ \\[4pt]
        & $\rho_0$ & $28.96^{+12.86}_{-12.28}$ & $22.05^{+6.46}_{-5.22}$ & $44.08^{+2.81}_{-4.68}$ & $42.59^{+10.91}_{-12.16}$ \\[4pt]
        & $r_c$ & $3.37^{+0.8}_{-0.47}$ & $3.79^{+0.43}_{-0.36}$ & $2.91^{+0.12}_{-0.07}$ & $2.87^{+0.39}_{-0.25}$ \\[4pt]
        \hline
    \rule{0pt}{3ex}\multirow{3}{8em}{B1 + Burkert profile}
        & $M_d$ & $17.08^{+14.87}_{-11.61}$ & $27.21^{+8.98}_{-11.76}$ & $20.39^{+5.99}_{-6.31}$ & $19.89^{+10.71}_{-11.3}$ \\[4pt]
        & $\rho_0$ & $21.51^{+17.51}_{-12.8}$ & $15.24^{+13.6}_{-6.99}$ & $20.7^{+7.89}_{-6.05}$ & $22.06^{+18.2}_{-10.61}$ \\[4pt]
        & $r_c$ & $3.69^{+1.38}_{-0.7}$ & $4.13^{+1.05}_{-0.8}$ & $3.74^{+0.48}_{-0.38}$ & $3.56^{+0.91}_{-0.65}$ \\[4pt]
        \hline
    \rule{0pt}{3ex}\multirow{3}{8em}{B2 + Burkert profile}
        & $M_d$ & $9.09^{+11.24}_{-6.63}$ & $17.48^{+7.13}_{-5.87}$ & $2.12^{+2.71}_{-1.52}$ & $5.36^{+7.0}_{-3.91}$ \\[4pt]
        & $\rho_0$ & $28.33^{+13.2}_{-12.23}$ & $21.84^{+7.57}_{-6.99}$ & $43.5^{+3.23}_{-5.19}$ & $41.71^{+11.43}_{-12.45}$ \\[4pt]
        & $r_c$ & $3.4^{+0.83}_{-0.49}$ & $3.8^{+0.63}_{-0.41}$ & $2.93^{+0.14}_{-0.08}$ & $2.89^{+0.41}_{-0.27}$ \\[4pt]
        \hline
    \rule{0pt}{3ex}\multirow{3}{8em}{B0 + Isothermal profile}
        & $M_d$ & $27.32^{+5.23}_{-4.16}$ & $34.21^{+0.84}_{-0.85}$ & $31.05^{+1.19}_{-1.21}$ & $33.77^{+3.36}_{-2.67}$ \\[4pt]
        & $\rho_0$ & $319.05^{+415.95}_{-257.71}$ & $10.81^{+1.04}_{-0.91}$ & $16.76^{+2.76}_{-2.28}$ & $375.16^{+390.71}_{-274.87}$ \\[4pt]
        & $r_s$ & $0.43^{+0.56}_{-0.15}$ & $2.67^{+0.13}_{-0.13}$ & $2.08^{+0.17}_{-0.16}$ & $0.38^{+0.36}_{-0.11}$ \\[4pt]
        \hline
    \rule{0pt}{3ex}\multirow{3}{8em}{B1 + Isothermal profile}
        & $M_d$ & $33.53^{+6.96}_{-5.02}$ & $42.96^{+0.87}_{-0.87}$ & $42.6^{+1.02}_{-1.08}$ & $42.27^{+6.65}_{-5.02}$ \\[4pt]
        & $\rho_0$ & $168.89^{+459.89}_{-154.74}$ & $5.58^{+0.71}_{-0.61}$ & $5.75^{+0.92}_{-0.76}$ & $26.29^{+302.33}_{-21.24}$ \\[4pt]
        & $r_s$ & $0.54^{+1.3}_{-0.26}$ & $3.26^{+0.2}_{-0.19}$ & $3.16^{+0.24}_{-0.23}$ & $1.27^{+1.63}_{-0.91}$ \\[4pt]
        \hline
    \rule{0pt}{3ex}\multirow{3}{8em}{B2 + Isothermal profile}
        & $M_d$ & $28.12^{+5.27}_{-4.24}$ & $36.21^{+0.91}_{-0.9}$ & $33.1^{+1.23}_{-1.24}$ & $34.73^{+3.33}_{-2.75}$ \\[4pt]
        & $\rho_0$ & $310.9^{+430.56}_{-254.41}$ & $9.87^{+0.99}_{-0.87}$ & $14.75^{+2.36}_{-1.96}$ & $340.47^{+406.9}_{-255.44}$ \\[4pt]
        & $r_s$ & $0.43^{+0.59}_{-0.15}$ & $2.78^{+0.14}_{-0.14}$ & $2.21^{+0.17}_{-0.16}$ & $0.4^{+0.4}_{-0.13}$ \\[4pt]
        \hline
    \rule{0pt}{3ex}\multirow{3}{8em}{B0 + NFW profile}
        & $M_d$ & $5.97^{+8.5}_{-4.33}$ & $11.59^{+1.59}_{-1.59}$ & $0.68^{+0.96}_{-0.49}$ & $3.13^{+4.36}_{-2.27}$ \\[4pt]
        & $\rho_0$ & $20.88^{+10.8}_{-9.31}$ & $11.23^{+1.4}_{-1.26}$ & $24.34^{+1.02}_{-1.51}$ & $35.55^{+9.56}_{-9.65}$ \\[4pt]
        & $r_s$ & $4.03^{+1.11}_{-0.65}$ & $5.45^{+0.29}_{-0.26}$ & $3.93^{+0.1}_{-0.07}$ & $3.21^{+0.44}_{-0.3}$ \\[4pt]
        \hline
    \rule{0pt}{3ex}\multirow{3}{8em}{B1 + NFW profile}
        & $M_d$ & $13.34^{+13.68}_{-9.3}$ & $26.2^{+1.73}_{-1.84}$ & $15.99^{+2.75}_{-2.79}$ & $13.6^{+10.46}_{-8.42}$ \\[4pt]
        & $\rho_0$ & $15.64^{+13.58}_{-10.13}$ & $5.11^{+0.86}_{-0.74}$ & $11.13^{+2.26}_{-1.97}$ & $21.03^{+15.3}_{-11.13}$ \\[4pt]
        & $r_s$ & $4.46^{+2.15}_{-0.94}$ & $7.21^{+0.49}_{-0.44}$ & $5.27^{+0.41}_{-0.36}$ & $3.87^{+1.24}_{-0.69}$ \\[4pt]
        \hline
    \rule{0pt}{3ex}\multirow{3}{8em}{B2 + NFW profile}
        & $M_d$ & $6.45^{+8.81}_{-4.64}$ & $14.2^{+1.58}_{-1.66}$ & $0.98^{+1.27}_{-0.7}$ & $3.3^{+4.77}_{-2.42}$ \\[4pt]
        & $\rho_0$ & $20.61^{+11.11}_{-9.44}$ & $9.61^{+1.25}_{-1.09}$ & $23.98^{+1.22}_{-1.91}$ & $35.29^{+9.97}_{-9.95}$ \\[4pt]
        & $r_s$ & $4.06^{+1.16}_{-0.69}$ & $5.81^{+0.31}_{-0.3}$ & $3.96^{+0.13}_{-0.08}$ & $3.22^{+0.45}_{-0.32}$ \\[4pt]
        \hline
    \rule{0pt}{3ex}\multirow{3}{8em}{B0 + gNFW profile}
        & $M_d$ & $7.53^{+9.33}_{-5.43}$ & $15.78^{+1.91}_{-1.9}$ & $1.3^{+1.65}_{-0.94}$ & $3.64^{+5.14}_{-2.69}$ \\[4pt]
        & $\rho_0$ & $213.82^{+264.4}_{-108.44}$ & $67.68^{+8.84}_{-8.14}$ & $162.09^{+9.5}_{-15.24}$ & $251.12^{+113.04}_{-77.52}$ \\[4pt]
        & $r_c$ & $1.5^{+0.64}_{-0.6}$ & $2.87^{+0.16}_{-0.14}$ & $2.01^{+0.08}_{-0.05}$ & $1.55^{+0.27}_{-0.29}$ \\[4pt]
        & $\alpha$ & $0.1^{+0.1}_{-0.07}$ & $0.0^{+0.0}_{-0.0}$ & $0.0^{+0.0}_{-0.0}$ & $0.03^{+0.04}_{-0.02}$ \\[4pt]
        \hline
    \rule{0pt}{3ex}\multirow{3}{8em}{B1 + gNFW profile}
        & $M_d$ & $18.86^{+13.6}_{-11.96}$ & $29.1^{+2.16}_{-2.29}$ & $19.57^{+3.11}_{-3.21}$ & $18.01^{+10.55}_{-10.53}$ \\[4pt]
        & $\rho_0$ & $149.55^{+246.44}_{-101.17}$ & $36.62^{+7.0}_{-5.81}$ & $67.66^{+15.12}_{-12.58}$ & $128.81^{+123.26}_{-70.85}$ \\[4pt]
        & $r_c$ & $1.65^{+0.97}_{-0.77}$ & $3.45^{+0.25}_{-0.23}$ & $2.74^{+0.22}_{-0.2}$ & $1.92^{+0.68}_{-0.45}$ \\[4pt]
        & $\alpha$ & $0.12^{+0.13}_{-0.09}$ & $0.01^{+0.01}_{-0.0}$ & $0.0^{+0.01}_{-0.0}$ & $0.04^{+0.06}_{-0.03}$ \\[4pt]
        \hline
    \rule{0pt}{3ex}\multirow{3}{8em}{B2 + gNFW profile}
        & $M_d$ & $8.47^{+10.0}_{-6.12}$ & $18.29^{+1.96}_{-2.01}$ & $2.01^{+2.18}_{-1.4}$ & $3.98^{+5.65}_{-2.92}$ \\[4pt]
        & $\rho_0$ & $209.83^{+269.77}_{-108.85}$ & $60.01^{+8.32}_{-7.39}$ & $156.38^{+13.08}_{-18.76}$ & $247.42^{+111.51}_{-80.38}$ \\[4pt]
        & $r_c$ & $1.51^{+0.65}_{-0.62}$ & $3.01^{+0.18}_{-0.16}$ & $2.04^{+0.11}_{-0.06}$ & $1.56^{+0.28}_{-0.3}$ \\[4pt]
        & $\alpha$ & $0.1^{+0.1}_{-0.07}$ & $0.0^{+0.01}_{-0.0}$ & $0.0^{+0.0}_{-0.0}$ & $0.03^{+0.05}_{-0.02}$ \\[4pt]
        \hline
    \rule{0pt}{3ex}\multirow{3}{8em}{B0 + $\beta$ profile}
        & $M_d$ & $9.38^{+11.25}_{-6.78}$ & $32.21^{+1.47}_{-2.11}$ & $1.07^{+1.72}_{-0.79}$ & $5.59^{+7.18}_{-4.11}$ \\[4pt]
        & $\rho_0$ & $15.51^{+6.32}_{-6.33}$ & $6.18^{+0.78}_{-0.5}$ & $26.41^{+0.98}_{-1.87}$ & $21.33^{+5.25}_{-5.99}$ \\[4pt]
        & $r_c$ & $3.95^{+0.86}_{-0.5}$ & $5.69^{+0.21}_{-0.28}$ & $3.26^{+0.09}_{-0.04}$ & $3.46^{+0.44}_{-0.28}$ \\[4pt]
        \hline
    \rule{0pt}{3ex}\multirow{3}{8em}{B1 + $\beta$ profile}
        & $M_d$ & $17.63^{+14.83}_{-11.89}$ & $39.69^{+3.26}_{-38.99}$ & $32.85^{+4.41}_{-26.88}$ & $19.38^{+11.05}_{-11.11}$ \\[4pt]
        & $\rho_0$ & $11.89^{+8.62}_{-6.63}$ & $4.34^{+27.09}_{-1.01}$ & $6.35^{+17.84}_{-1.65}$ & $12.4^{+8.75}_{-5.57}$ \\[4pt]
        & $r_c$ & $4.25^{+1.34}_{-0.7}$ & $5.95^{+0.66}_{-2.86}$ & $5.21^{+0.61}_{-1.85}$ & $4.1^{+0.9}_{-0.65}$ \\[4pt]
        \hline
    \rule{0pt}{3ex}\multirow{3}{8em}{B2 + $\beta$ profile}
        & $M_d$ & $10.14^{+12.01}_{-7.34}$ & $35.78^{+1.35}_{-1.86}$ & $1.11^{+1.75}_{-0.81}$ & $6.12^{+7.84}_{-4.41}$ \\[4pt]
        & $\rho_0$ & $15.19^{+6.48}_{-6.42}$ & $5.34^{+0.59}_{-0.4}$ & $26.43^{+0.97}_{-1.87}$ & $21.11^{+5.3}_{-6.31}$ \\[4pt]
        & $r_c$ & $3.98^{+0.91}_{-0.52}$ & $6.03^{+0.21}_{-0.27}$ & $3.26^{+0.09}_{-0.04}$ & $3.47^{+0.48}_{-0.29}$ \\[4pt]
        \hline
    \rule{0pt}{3ex}\multirow{3}{8em}{B0 + Hernquist profile}
        & $M_d$ & $5.83^{+7.92}_{-4.24}$ & $6.3^{+2.19}_{-2.35}$ & $0.46^{+0.71}_{-0.33}$ & $2.57^{+4.06}_{-1.89}$ \\[4pt]
        & $\rho_0$ & $5.63^{+1.74}_{-1.9}$ & $5.94^{+0.79}_{-0.69}$ & $7.71^{+0.19}_{-0.27}$ & $7.73^{+1.25}_{-1.37}$ \\[4pt]
        & $r_c$ & $10.2^{+1.77}_{-1.09}$ & $10.18^{+0.5}_{-0.48}$ & $9.2^{+0.13}_{-0.1}$ & $8.94^{+0.69}_{-0.53}$ \\[4pt]
        \hline
    \rule{0pt}{3ex}\multirow{3}{8em}{B1 + Hernquist profile}
        & $M_d$ & $11.05^{+12.4}_{-7.85}$ & $19.27^{+2.72}_{-2.9}$ & $2.14^{+2.74}_{-1.55}$ & $6.96^{+8.37}_{-4.95}$ \\[4pt]
        & $\rho_0$ & $4.99^{+2.37}_{-2.46}$ & $3.56^{+0.66}_{-0.56}$ & $7.93^{+0.52}_{-0.86}$ & $7.06^{+1.78}_{-2.28}$ \\[4pt]
        & $r_c$ & $10.59^{+2.75}_{-1.4}$ & $12.03^{+0.77}_{-0.7}$ & $9.19^{+0.34}_{-0.19}$ & $9.26^{+1.24}_{-0.7}$ \\[4pt]
        \hline
    \rule{0pt}{3ex}\multirow{3}{8em}{B2 + Hernquist profile}
        & $M_d$ & $6.16^{+8.25}_{-4.51}$ & $8.95^{+2.24}_{-2.33}$ & $0.49^{+0.76}_{-0.37}$ & $2.7^{+4.07}_{-1.99}$ \\[4pt]
        & $\rho_0$ & $5.54^{+1.82}_{-1.87}$ & $5.22^{+0.71}_{-0.61}$ & $7.72^{+0.19}_{-0.29}$ & $7.73^{+1.25}_{-1.36}$ \\[4pt]
        & $r_c$ & $10.25^{+1.78}_{-1.12}$ & $10.7^{+0.54}_{-0.52}$ & $9.2^{+0.14}_{-0.1}$ & $8.94^{+0.7}_{-0.53}$ \\[4pt]
        \hline
    \rule{0pt}{3ex}\multirow{4}{8em}{B0 + Einasto profile}
        & $M_d$ & $23.75^{+16.78}_{-15.74}$ & $26.98^{+2.83}_{-3.07}$ & $45.69^{+1.12}_{-1.24}$ & $14.04^{+13.66}_{-9.76}$ \\[4pt]
        & $\rho_0$ & $0.8^{+0.4}_{-0.25}$ & $0.83^{+0.11}_{-0.08}$ & $0.55^{+0.01}_{-0.01}$ & $1.0^{+0.31}_{-0.3}$ \\[4pt]
        & $r_c$ & $15.29^{+1.76}_{-1.79}$ & $14.84^{+0.59}_{-0.69}$ & $16.69^{+0.12}_{-0.13}$ & $14.29^{+1.65}_{-1.2}$ \\[4pt]
        & $\alpha$ & $1.01^{+0.59}_{-0.32}$ & $0.45^{+0.05}_{-0.04}$ & $1.35^{+0.1}_{-0.1}$ & $0.74^{+0.24}_{-0.14}$ \\[4pt]
        \hline
    \rule{0pt}{3ex}\multirow{4}{8em}{B1 + Einasto profile}
        & $M_d$ & $30.06^{+15.16}_{-18.93}$ & $35.5^{+2.85}_{-3.12}$ & $51.28^{+0.83}_{-0.93}$ & $19.72^{+15.03}_{-13.41}$ \\[4pt]
        & $\rho_0$ & $0.65^{+0.53}_{-0.27}$ & $0.57^{+0.11}_{-0.08}$ & $0.35^{+0.01}_{-0.01}$ & $0.88^{+0.46}_{-0.35}$ \\[4pt]
        & $r_c$ & $15.57^{+1.97}_{-2.14}$ & $15.97^{+0.8}_{-0.88}$ & $17.24^{+0.15}_{-0.15}$ & $14.52^{+1.95}_{-1.53}$ \\[4pt]
        & $\alpha$ & $1.07^{+0.84}_{-0.4}$ & $0.43^{+0.05}_{-0.04}$ & $1.69^{+0.17}_{-0.16}$ & $0.75^{+0.32}_{-0.16}$ \\[4pt]
        \hline
    \rule{0pt}{3ex}\multirow{4}{8em}{B2 + Einasto profile}
        & $M_d$ & $24.4^{+17.41}_{-16.43}$ & $30.02^{+3.02}_{-3.25}$ & $48.62^{+0.99}_{-1.06}$ & $15.04^{+14.25}_{-10.42}$ \\[4pt]
        & $\rho_0$ & $0.8^{+0.4}_{-0.26}$ & $0.75^{+0.1}_{-0.08}$ & $0.52^{+0.01}_{-0.01}$ & $0.99^{+0.32}_{-0.3}$ \\[4pt]
        & $r_c$ & $15.29^{+1.79}_{-1.8}$ & $15.41^{+0.61}_{-0.71}$ & $16.85^{+0.11}_{-0.11}$ & $14.35^{+1.69}_{-1.28}$ \\[4pt]
        & $\alpha$ & $1.01^{+0.6}_{-0.32}$ & $0.47^{+0.06}_{-0.05}$ & $1.47^{+0.1}_{-0.1}$ & $0.74^{+0.25}_{-0.14}$ \\[4pt]
        \hline
    \multicolumn{6}{|c|}{$a_0$ as free parameter}\\[2pt]
    \rule{0pt}{3ex}\multirow{2}{8em}{B0 + standard IF}
        & $M_d$ & $31.36^{+3.63}_{-3.75}$ & $22.62^{+0.26}_{-0.27}$ & $23.83^{+0.36}_{-0.36}$ & $38.41^{+2.07}_{-2.19}$ \\[4pt]
        &  $a_0$ & $0.77^{+0.11}_{-0.1}$ & $1.2^{+0.01}_{-0.01}$ & $1.11^{+0.02}_{-0.02}$ & $0.6^{+0.05}_{-0.04}$ \\[4pt]
        \hline
    \rule{0pt}{3ex}\multirow{2}{8em}{B1 + standard IF}
        & $M_d$ & $38.81^{+2.16}_{-2.39}$ & $34.98^{+0.17}_{-0.17}$ & $35.52^{+0.22}_{-0.23}$ & $42.7^{+1.2}_{-1.3}$ \\[4pt]
        & $a_0$ & $1.64^{+0.26}_{-0.21}$ & $2.75^{+0.03}_{-0.03}$ & $2.52^{+0.04}_{-0.04}$ & $1.25^{+0.1}_{-0.09}$ \\[4pt]
        \hline
    \rule{0pt}{3ex}\multirow{2}{8em}{B2 + standard IF}
        & $M_d$ & $32.52^{+3.73}_{-3.81}$ & $24.21^{+0.27}_{-0.27}$ & $25.28^{+0.38}_{-0.37}$ & $39.41^{+2.1}_{-2.12}$ \\[4pt]
        & $a_0$ & $1.58^{+0.25}_{-0.2}$ & $2.59^{+0.03}_{-0.03}$ & $2.39^{+0.04}_{-0.04}$ & $1.22^{+0.09}_{-0.09}$ \\[4pt]
        \hline
    \rule{0pt}{3ex}\multirow{2}{8em}{B0 + simple IF}
        & $M_d$ & $37.94^{+4.26}_{-4.39}$ & $28.1^{+0.31}_{-0.3}$ & $29.25^{+0.41}_{-0.41}$ & $46.71^{+2.72}_{-2.63}$ \\[4pt]
        & $a_0$ & $0.6^{+0.11}_{-0.09}$ & $0.94^{+0.01}_{-0.01}$ & $0.87^{+0.02}_{-0.02}$ & $0.44^{+0.05}_{-0.04}$ \\[4pt]
        \hline
    \rule{0pt}{3ex}\multirow{2}{8em}{B1 + simple IF}
        & $M_d$ & $49.34^{+3.23}_{-3.45}$ & $44.61^{+0.25}_{-0.24}$ & $45.14^{+0.32}_{-0.33}$ & $55.18^{+1.92}_{-1.99}$ \\[4pt]
        & $a_0$ & $1.52^{+0.24}_{-0.2}$ & $2.53^{+0.03}_{-0.03}$ & $2.34^{+0.04}_{-0.04}$ & $1.15^{+0.09}_{-0.09}$ \\[4pt]
        \hline
    \rule{0pt}{3ex}\multirow{2}{8em}{B2 + simple IF}
        & $M_d$ & $39.28^{+4.4}_{-4.36}$ & $30.06^{+0.31}_{-0.32}$ & $31.03^{+0.42}_{-0.42}$ & $48.15^{+2.72}_{-2.75}$ \\[4pt]
        & $a_0$ & $1.47^{+0.23}_{-0.19}$ & $2.38^{+0.03}_{-0.03}$ & $2.22^{+0.04}_{-0.04}$ & $1.11^{+0.09}_{-0.09}$ \\[4pt]
        \hline
    \rule{0pt}{3ex}\multirow{2}{8em}{B0 + RAR}
        & $M_d$ & $40.65^{+4.63}_{-4.79}$ & $30.31^{+0.33}_{-0.33}$ & $31.44^{+0.46}_{-0.45}$ & $50.14^{+2.89}_{-2.91}$ \\[4pt]
        & $a_0$ & $1.44^{+0.23}_{-0.19}$ & $2.37^{+0.03}_{-0.03}$ & $2.21^{+0.04}_{-0.04}$ & $1.08^{+0.09}_{-0.08}$ \\[4pt]
        \hline
    \rule{0pt}{3ex}\multirow{2}{8em}{B1 + RAR}
        & $M_d$ & $53.32^{+3.38}_{-3.64}$ & $48.62^{+0.27}_{-0.26}$ & $49.08^{+0.36}_{-0.36}$ & $59.22^{+1.91}_{-2.0}$ \\[4pt]
        & $a_0$ & $0.54^{+0.1}_{-0.08}$ & $0.84^{+0.01}_{-0.02}$ & $0.48^{+0.02}_{-0.08}$ & $0.4^{+0.04}_{-0.03}$ \\[4pt]
        \hline
    \rule{0pt}{3ex}\multirow{2}{8em}{B2 + RAR}
        & $M_d$ & $41.98^{+4.72}_{-4.88}$ & $32.49^{+0.34}_{-0.34}$ & $33.42^{+0.46}_{-0.47}$ & $51.69^{+2.97}_{-2.94}$ \\[4pt]
        & $a_0$ & $1.39^{+0.23}_{-0.19}$ & $2.22^{+0.03}_{-0.03}$ & $2.09^{+0.04}_{-0.03}$ & $1.04^{+0.09}_{-0.08}$ \\[4pt]
        \hline
    \multicolumn{6}{|c|}{constant $a_0$; standard MOND paradigm}\\[2pt]
        B0 + standard IF & $M_d$ & $39.39^{+0.77}_{-0.75}$ & $40.45^{+0.03}_{-0.04}$ & $40.17^{+0.05}_{-0.05}$ & $39.62^{+0.5}_{-0.47}$ \\[4pt]
        \hline
        B1 + standard IF & $M_d$ & $30.95^{+0.54}_{-0.53}$ & $34.96^{+0.03}_{-0.03}$ & $34.27^{+0.04}_{-0.04}$ & $30.07^{+0.33}_{-0.31}$ \\[4pt]
        \hline
        B2 + standard IF & $M_d$ & $39.73^{+0.8}_{-0.78}$ & $41.05^{+0.04}_{-0.04}$ & $40.73^{+0.06}_{-0.06}$ & $39.91^{+0.48}_{-0.5}$ \\[4pt]
        \hline
        B0 + simple IF & $M_d$ & $46.45^{+0.89}_{-0.89}$ & $48.78^{+0.04}_{-0.04}$ & $48.27^{+0.06}_{-0.07}$ & $46.41^{+0.58}_{-0.57}$ \\[4pt]
        \hline
        B1 + simple IF & $M_d$ & $34.88^{+0.59}_{-0.59}$ & $39.94^{+0.03}_{-0.03}$ & $39.05^{+0.05}_{-0.05}$ & $33.75^{+0.37}_{-0.36}$ \\[4pt]
        \hline
        B2 + simple IF & $M_d$ & $45.53^{+0.89}_{-0.87}$ & $47.65^{+0.04}_{-0.04}$ & $47.18^{+0.06}_{-0.06}$ & $45.61^{+0.54}_{-0.55}$ \\[4pt]
        \hline
        B0 + RAR & $M_d$ & $46.45^{+0.89}_{-0.89}$ & $48.78^{+0.04}_{-0.04}$ & $48.27^{+0.06}_{-0.07}$ & $46.41^{+0.58}_{-0.57}$ \\[4pt]
        \hline
        B1 + RAR & $M_d$ & $35.69^{+0.62}_{-0.64}$ & $41.3^{+0.03}_{-0.03}$ & $40.32^{+0.05}_{-0.05}$ & $34.44^{+0.37}_{-0.37}$ \\[4pt]
        \hline
        B2 + RAR & $M_d$ & $46.83^{+0.91}_{-0.89}$ & $49.47^{+0.04}_{-0.04}$ & $48.92^{+0.07}_{-0.07}$ & $46.72^{+0.57}_{-0.55}$ \\[4pt]
        \hline
\end{longtable}

\newpage
\vspace{2cm}
\subsection{Corner plots of the model parameters}\label{appendixb}
\begin{figure}[h]
    \centering
    \caption{The corner plots of the baryon models {\tt baryon + dark matter} showing 68\% and 95\% credible intervals, analyzed with the four datasets. The plots display the marginalized distributions of parameters along the diagonal and the scatter plots of parameter pairs across the four datasets.}
    \includegraphics[width=1\textwidth]{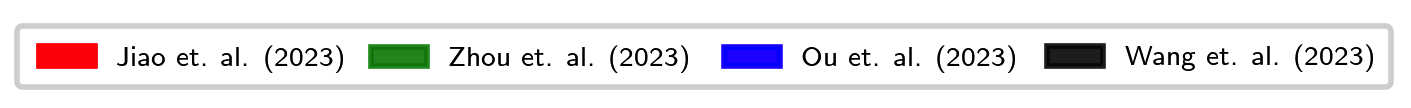}\label{fig:placeholder}
\end{figure}

\begin{figure}[h]
    \centering
    \subfloat[B0 + NFW profile]{\label{fig:b0nfw}\includegraphics[width=0.53\textwidth]{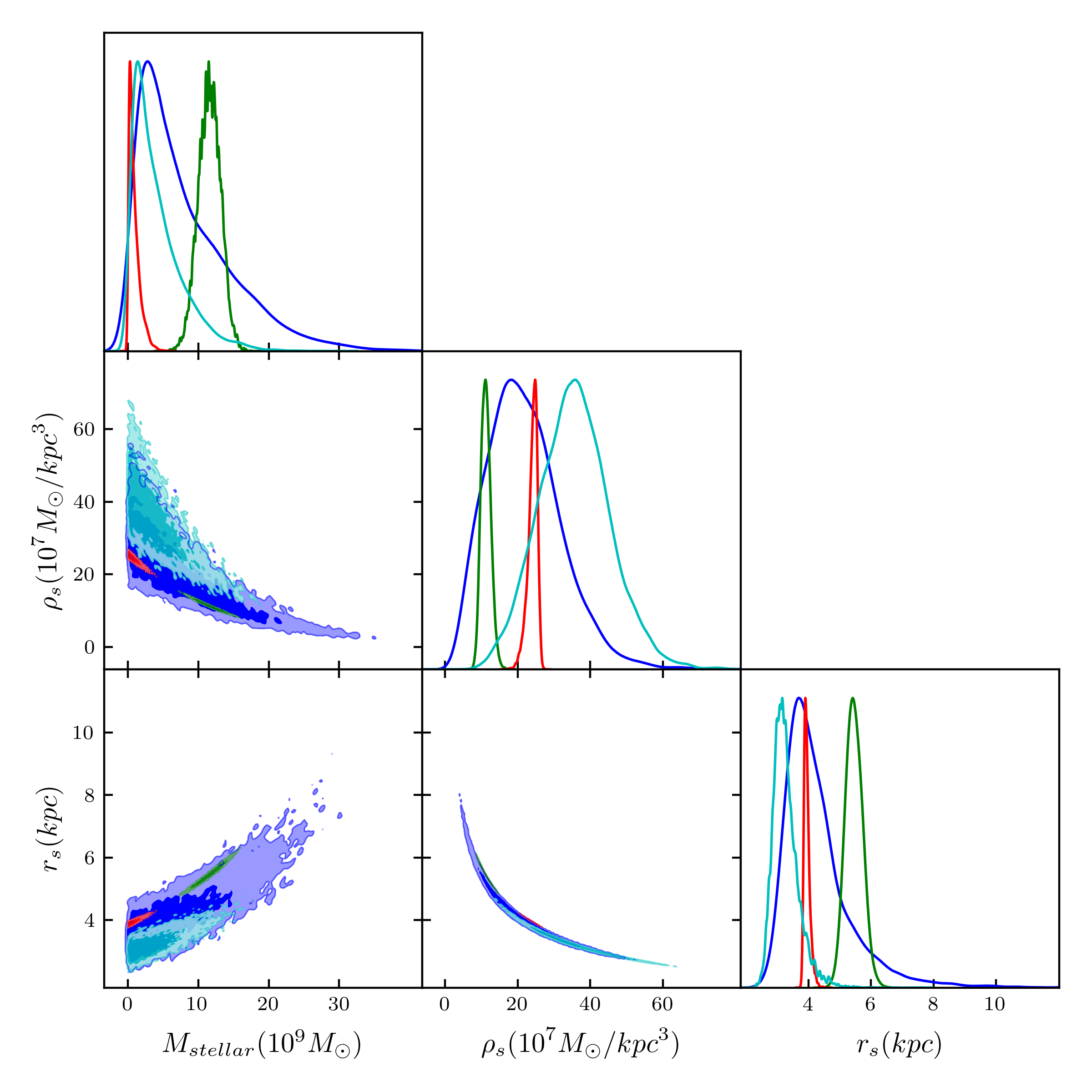}}
    \subfloat[B1 + NFW profile]{\label{fig:b1nfw}\includegraphics[width=0.53\textwidth]{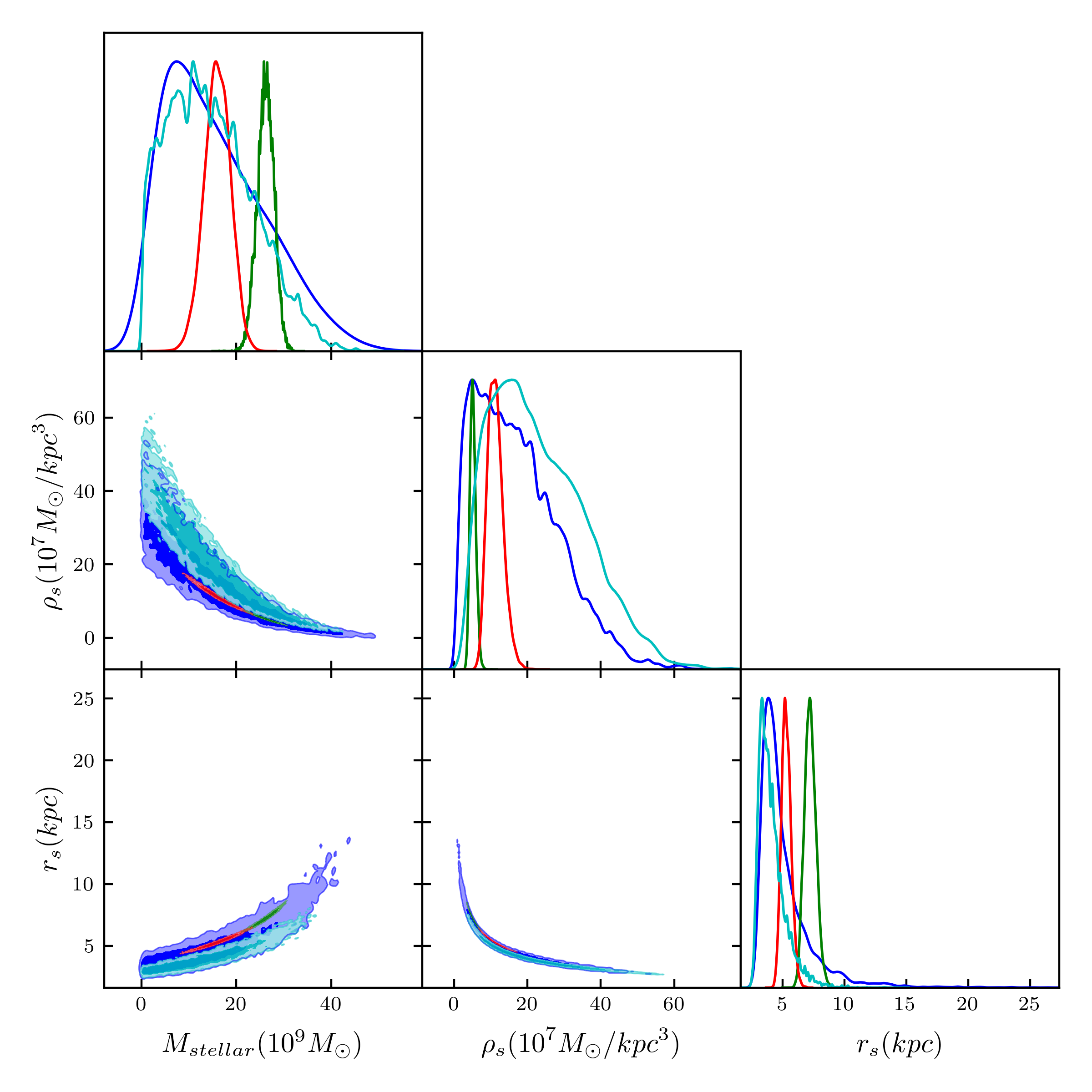}}
    \\
\end{figure}

\begin{figure}[h]
    \ContinuedFloat
    \centering
    \subfloat[B2 + NFW profile]{\label{fig:b2nfw}\includegraphics[width=0.53\textwidth]{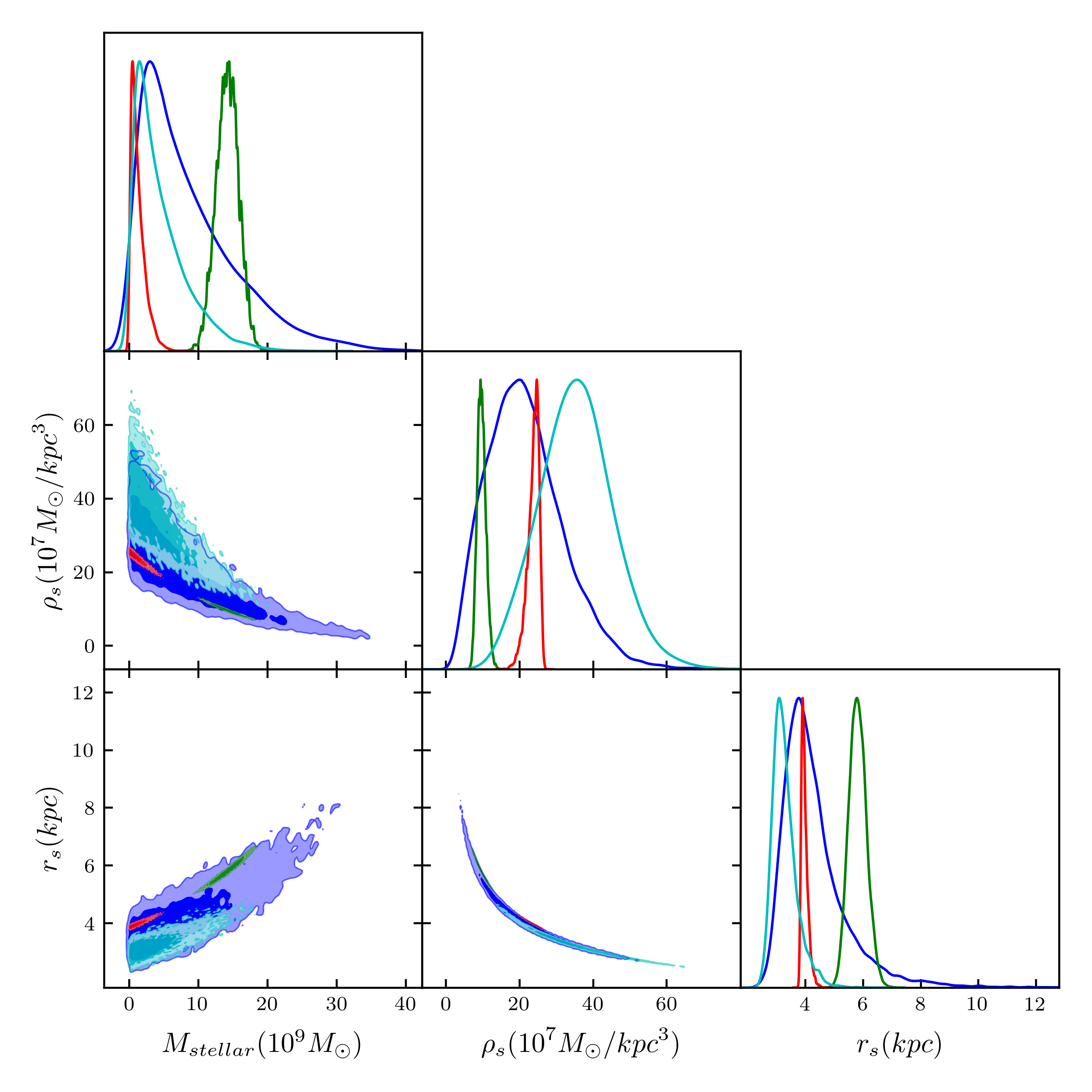}}
    \subfloat[B0 + Isothermal profile]{\label{fig:b0iso}\includegraphics[width=0.53\textwidth]{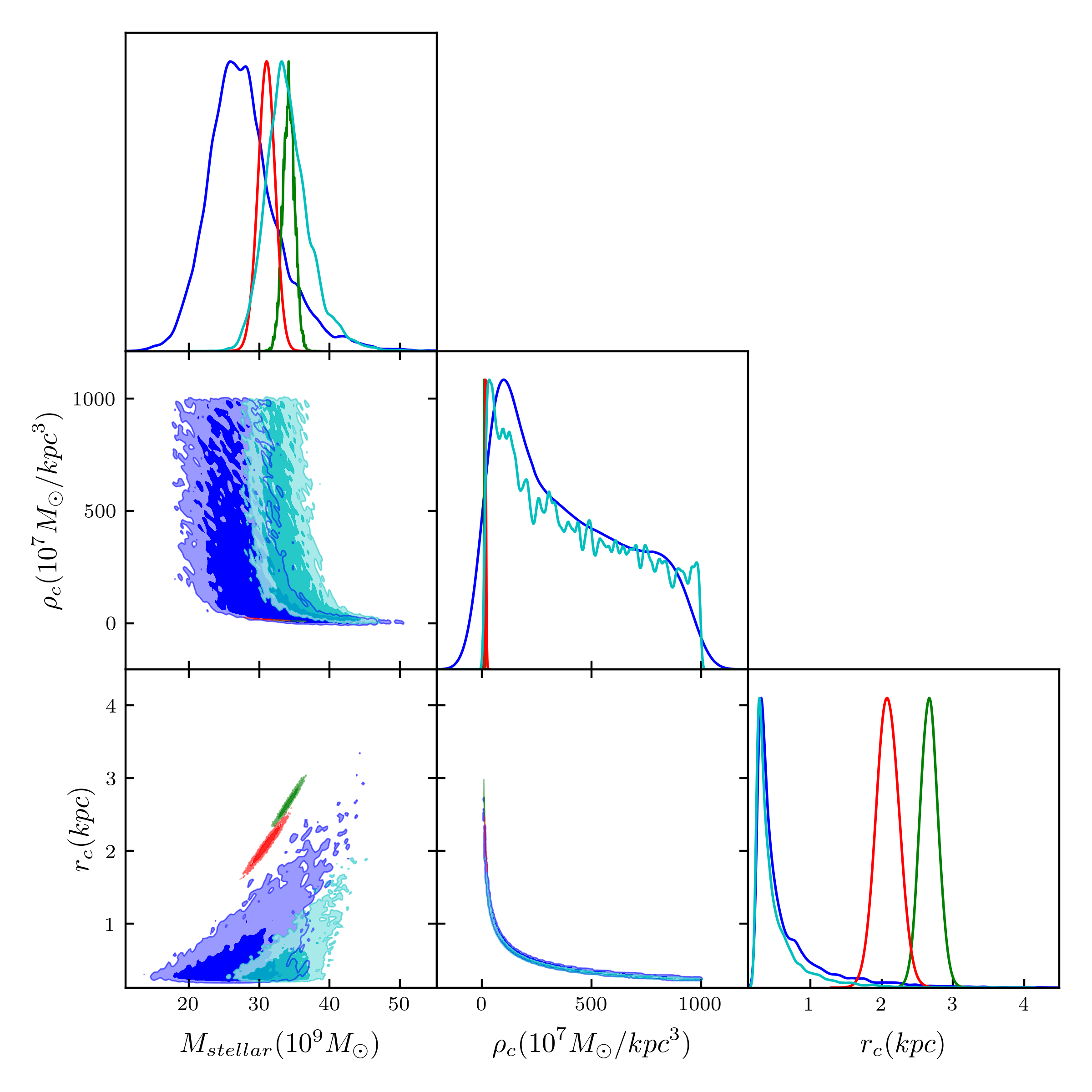}}
    \\
    \subfloat[B1 + Isothermal profile]{\label{fig:b1iso}\includegraphics[width=0.53\textwidth]{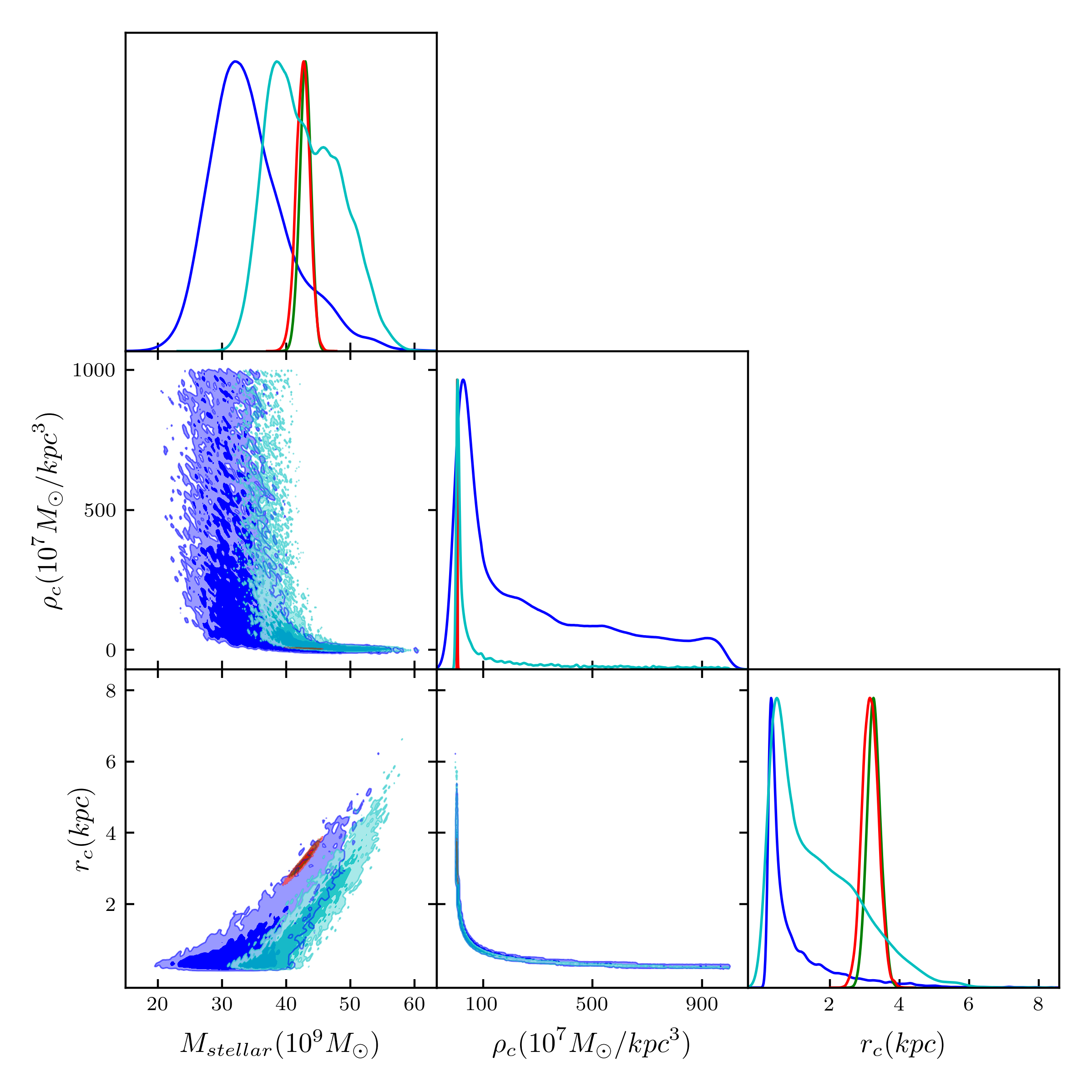}}
    \subfloat[B2 + Isothermal profile]{\label{fig:b2iso}\includegraphics[width=0.53\textwidth]{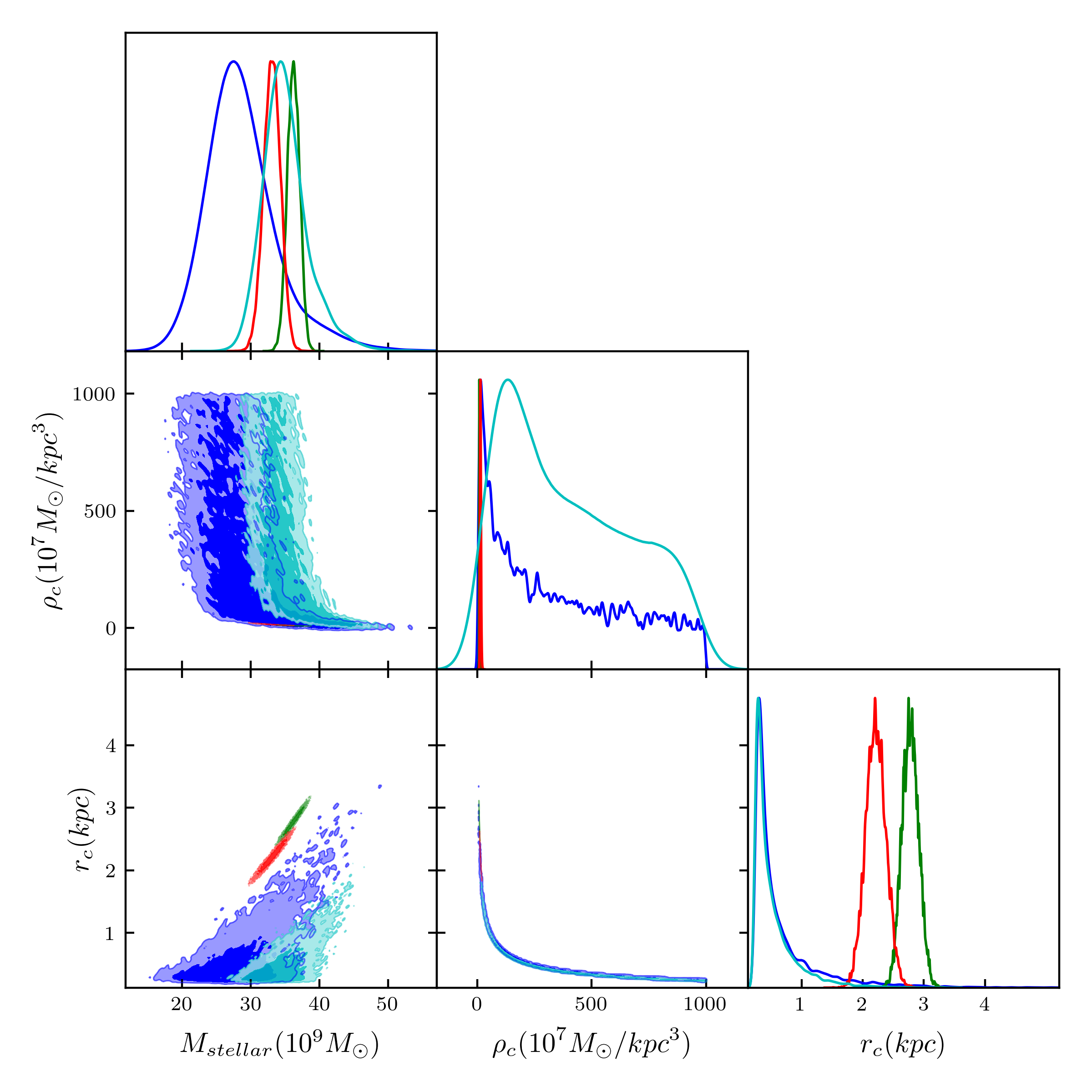}}
    \\
\end{figure}

\begin{figure}[h]
    \ContinuedFloat
    \centering
    \subfloat[B0 + Burkert profile]{\label{fig:b0burkert}\includegraphics[width=0.53\textwidth]{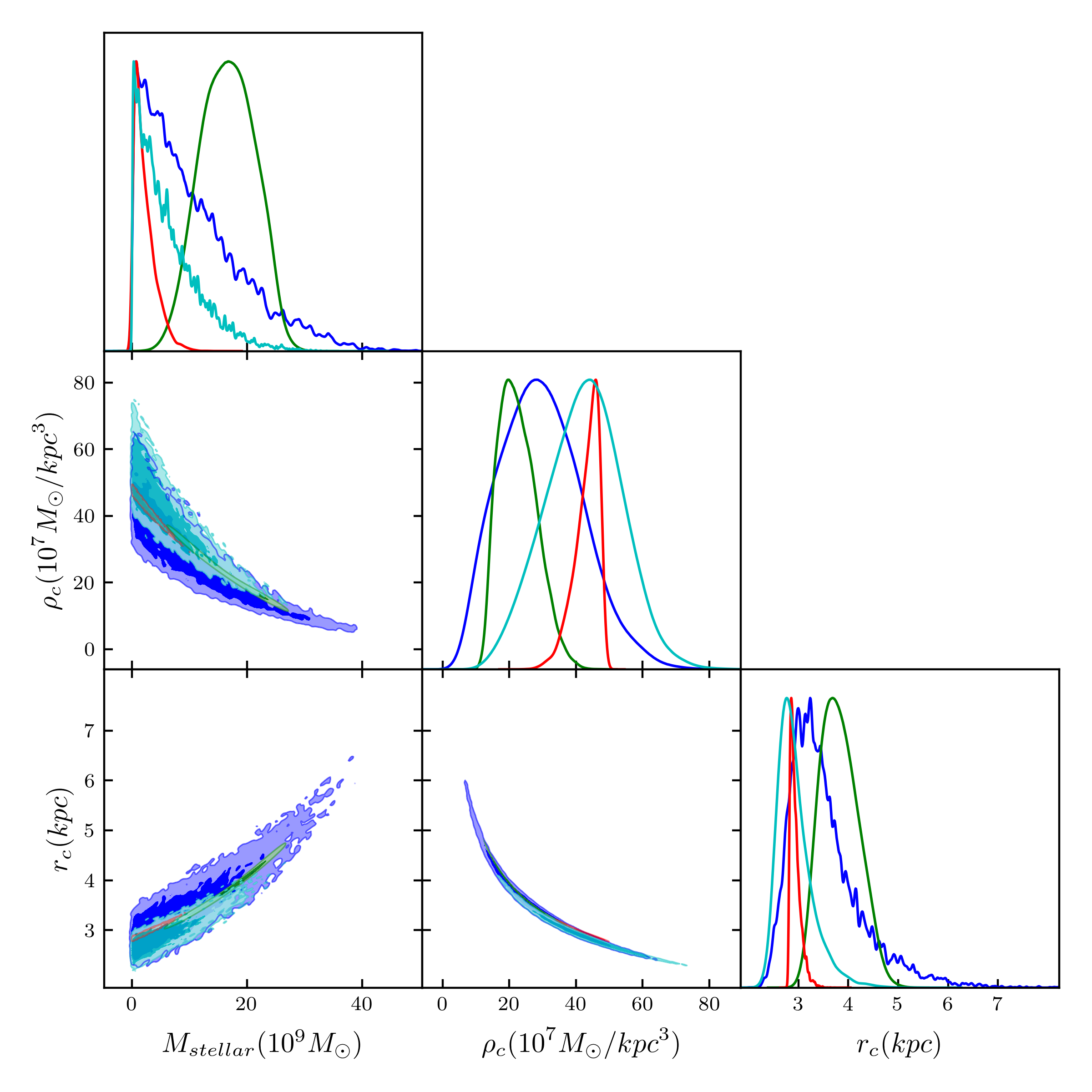}}
    \subfloat[B1 + Burkert profile]{\label{fig:b1burkert}\includegraphics[width=0.53\textwidth]{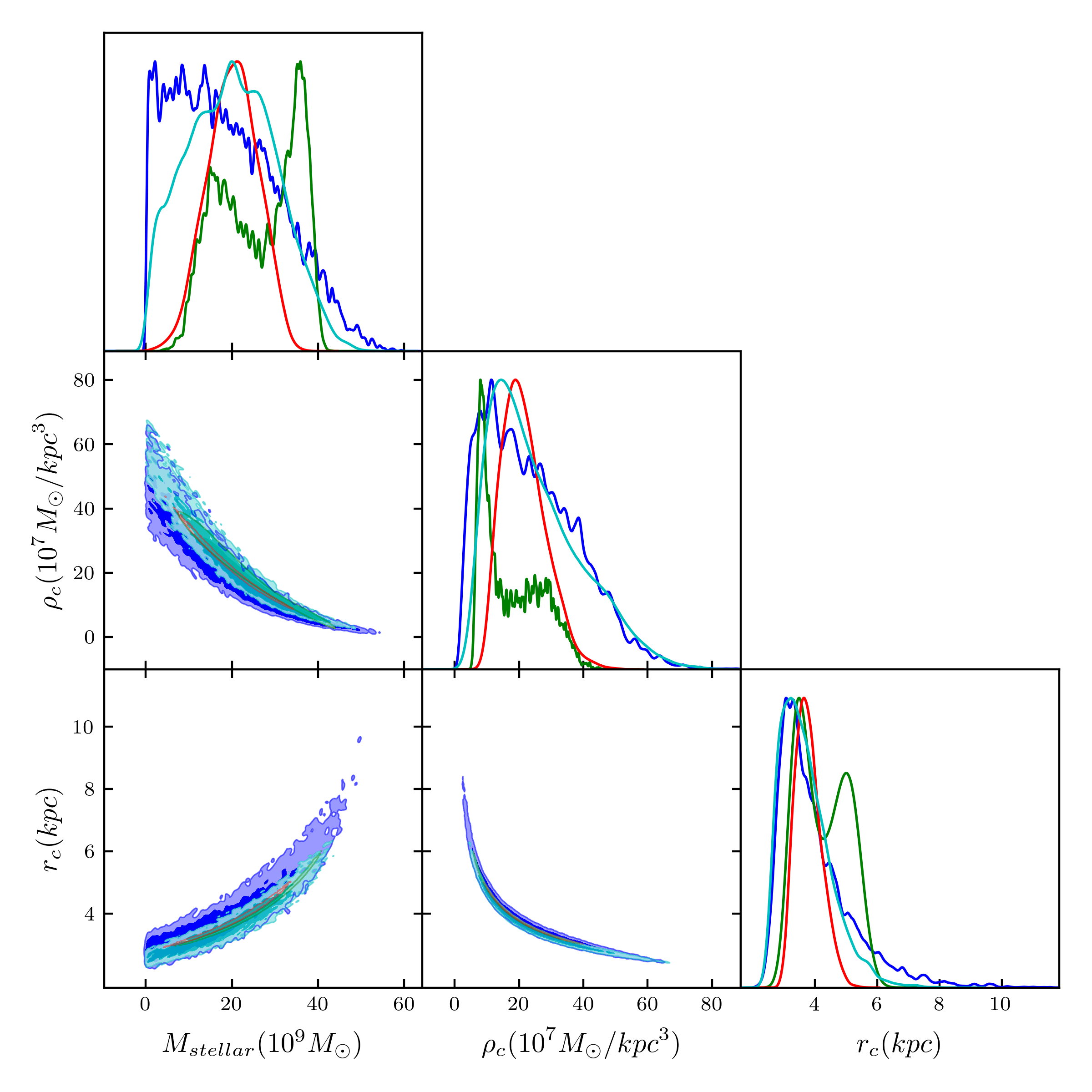}}
    \\
    \subfloat[B2 + Burkert profile]{\label{fig:b2burkert}\includegraphics[width=0.53\textwidth]{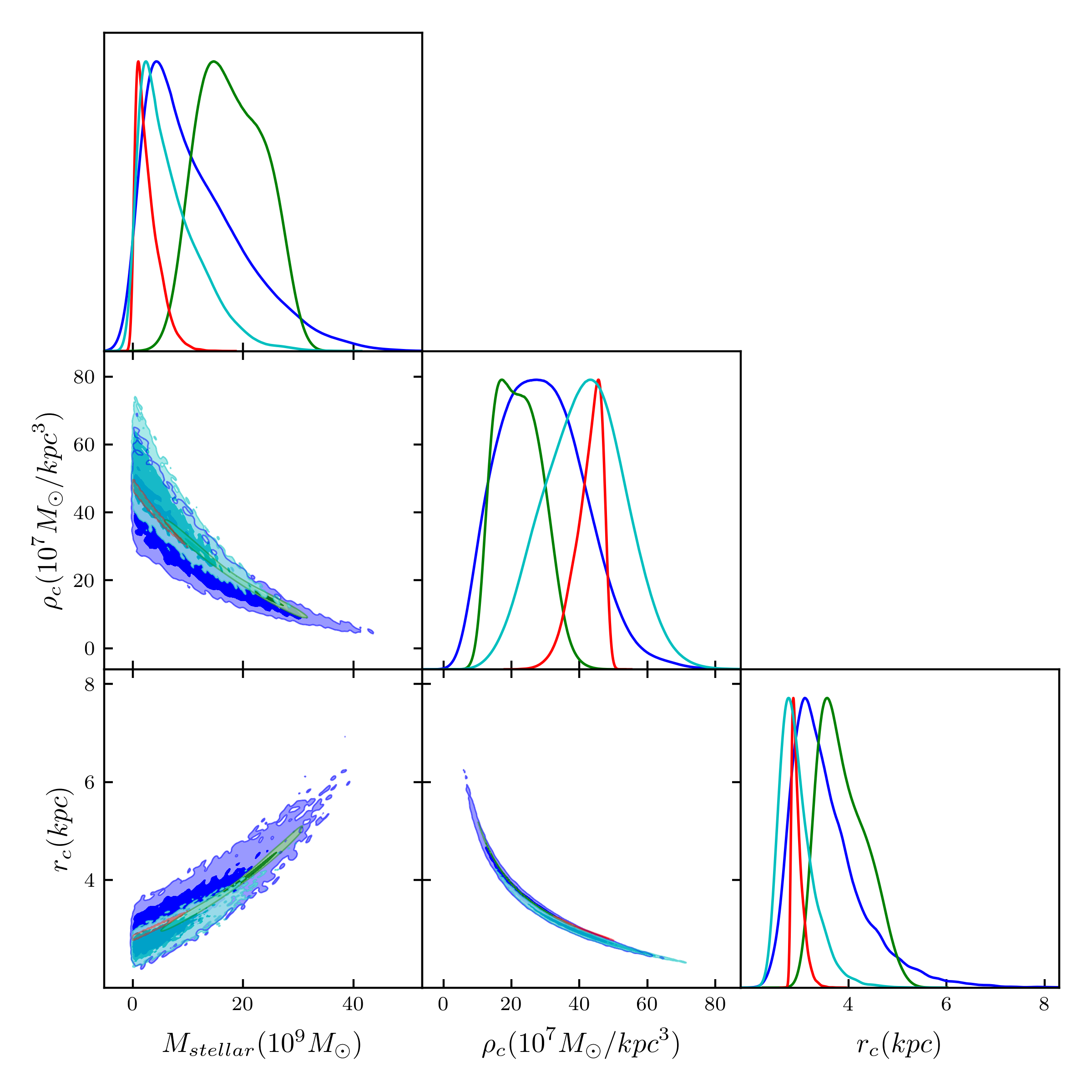}}
    \subfloat[B0 + gNFW profile]{\label{fig:b0gnfw}\includegraphics[width=0.53\textwidth]{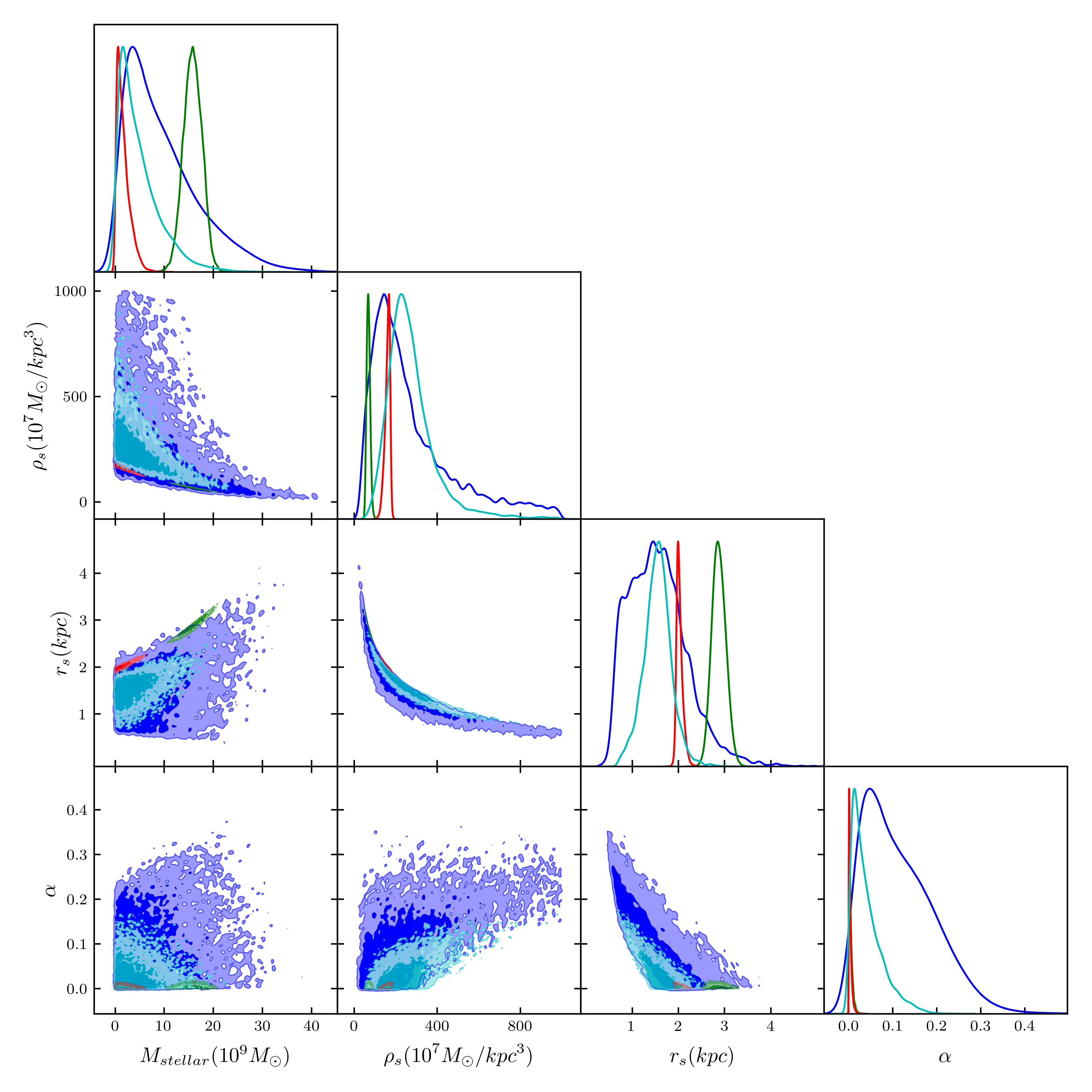}}
    \\
\end{figure}

\begin{figure}[h]
    \ContinuedFloat
    \centering
    \subfloat[B1 + gNFW profile]{\label{fig:b1gnfw}\includegraphics[width=0.53\textwidth]{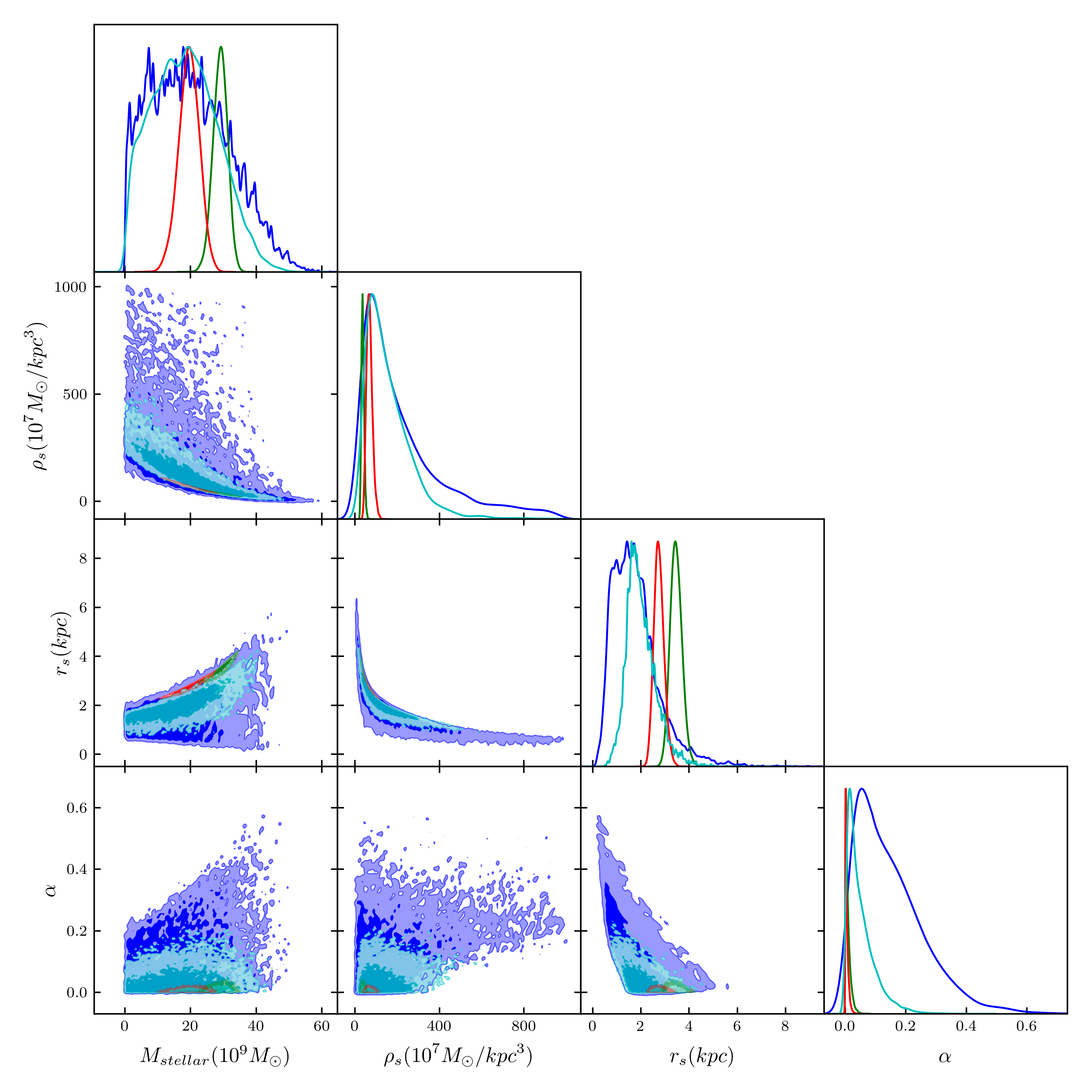}}
    \subfloat[B2 + gNFW profile]{\label{fig:b2gnfw}\includegraphics[width=0.53\textwidth]{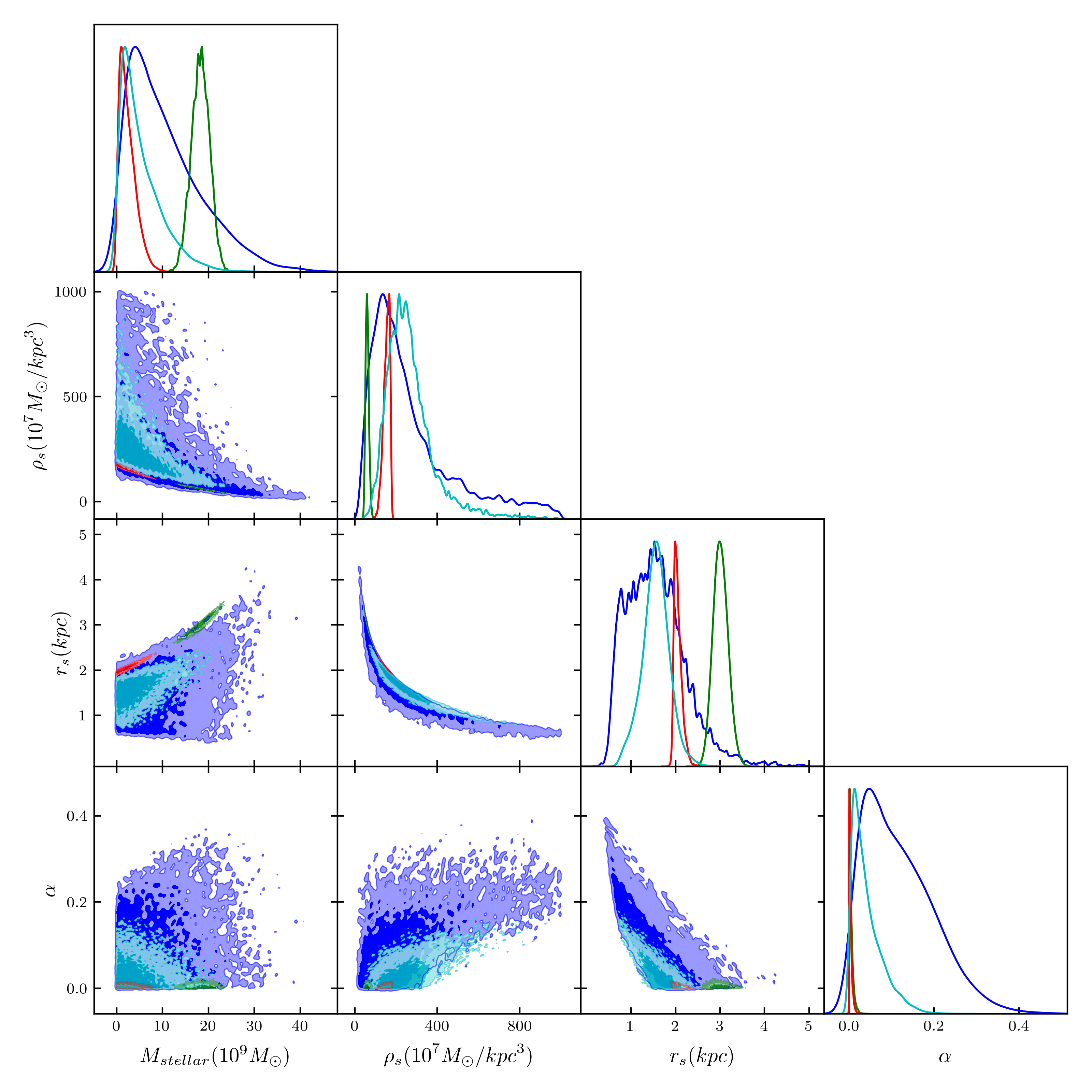}}
    \\
    \subfloat[B0 + $\beta$ profile]{\label{fig:b0beta}\includegraphics[width=0.53\textwidth]{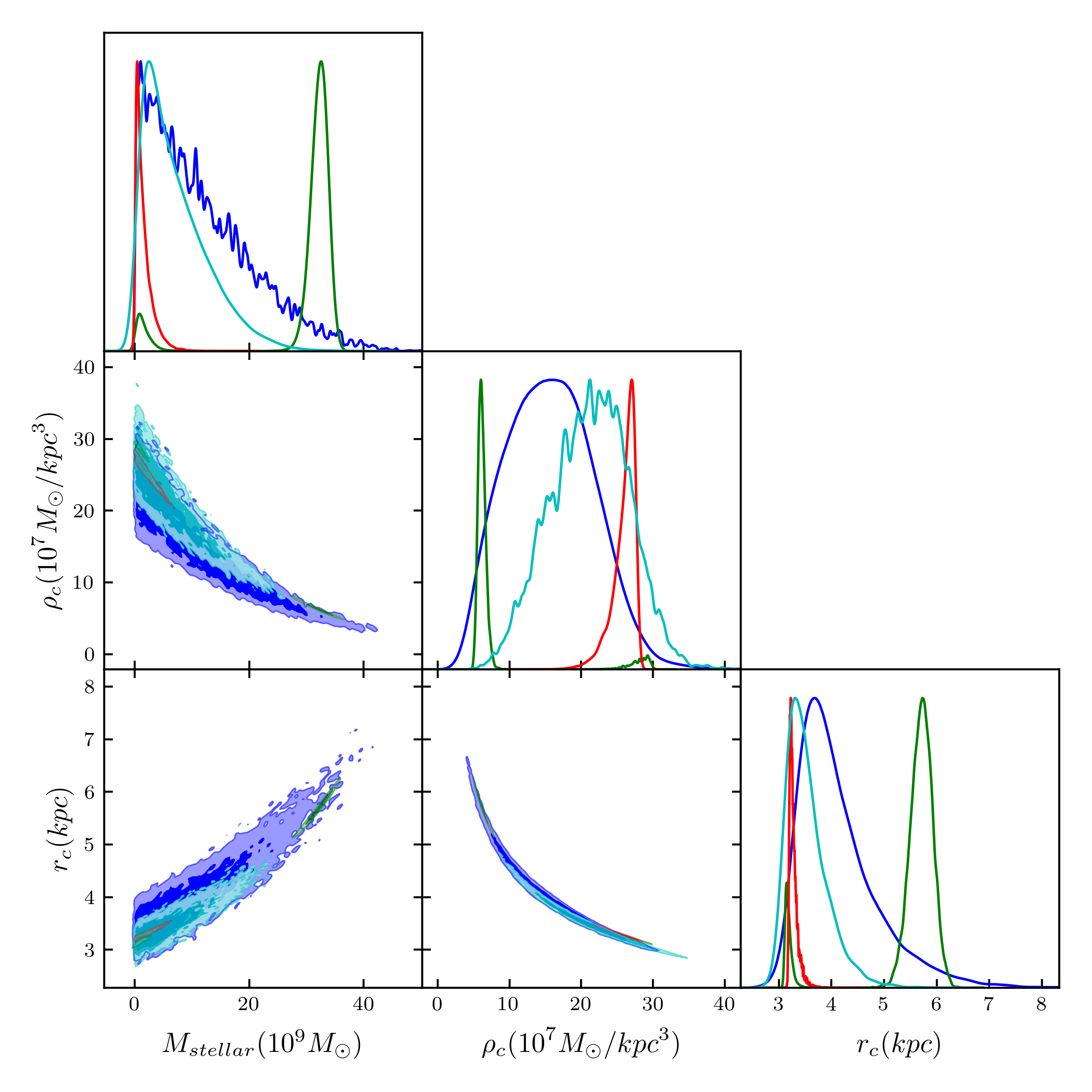}}
    \subfloat[B1 + $\beta$ profile]{\label{fig:b1beta}\includegraphics[width=0.53\textwidth]{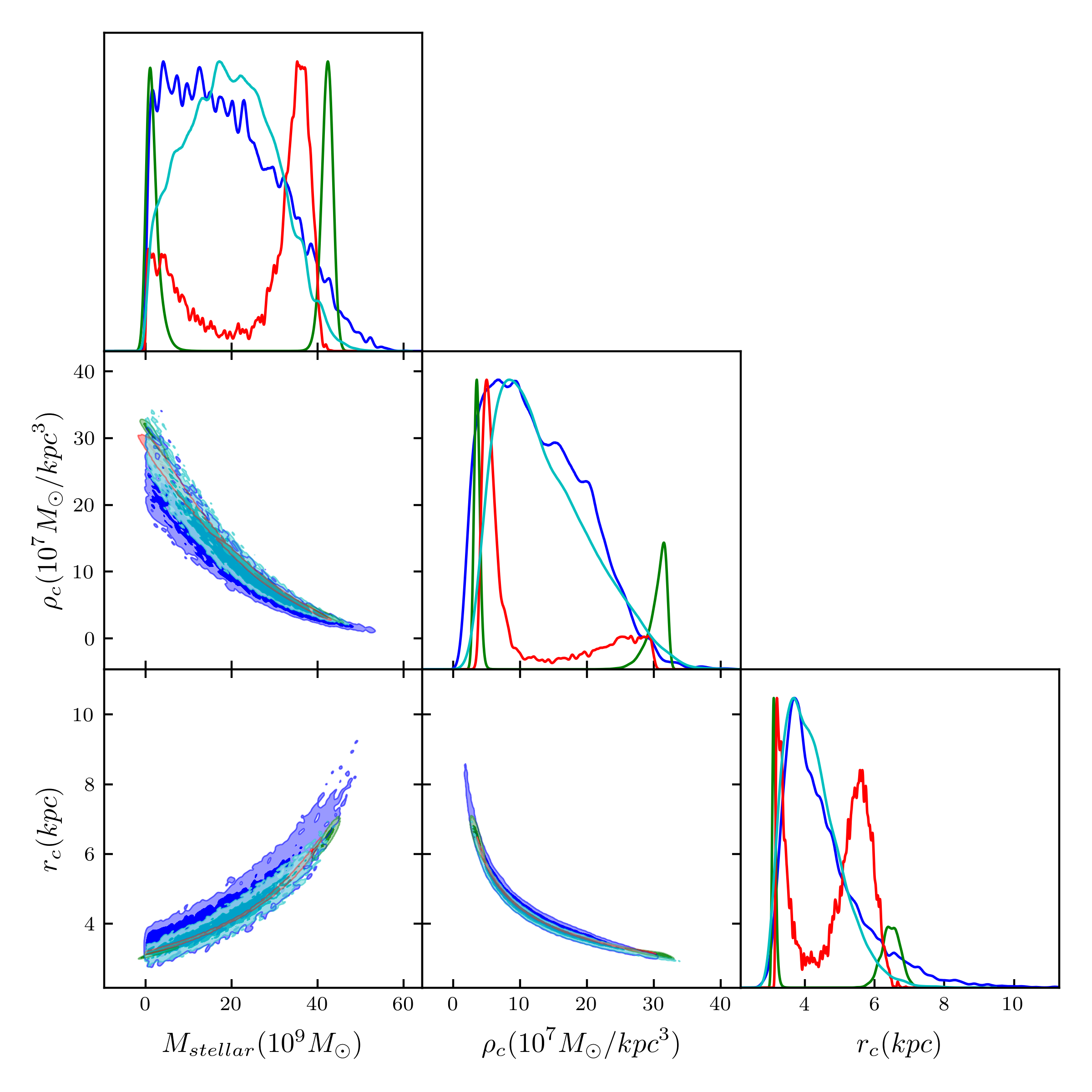}}
    \\
\end{figure}

\begin{figure}[h]
    \ContinuedFloat
    \centering
    \subfloat[B2 + $\beta$ profile]{\label{fig:b2beta}\includegraphics[width=0.53\textwidth]{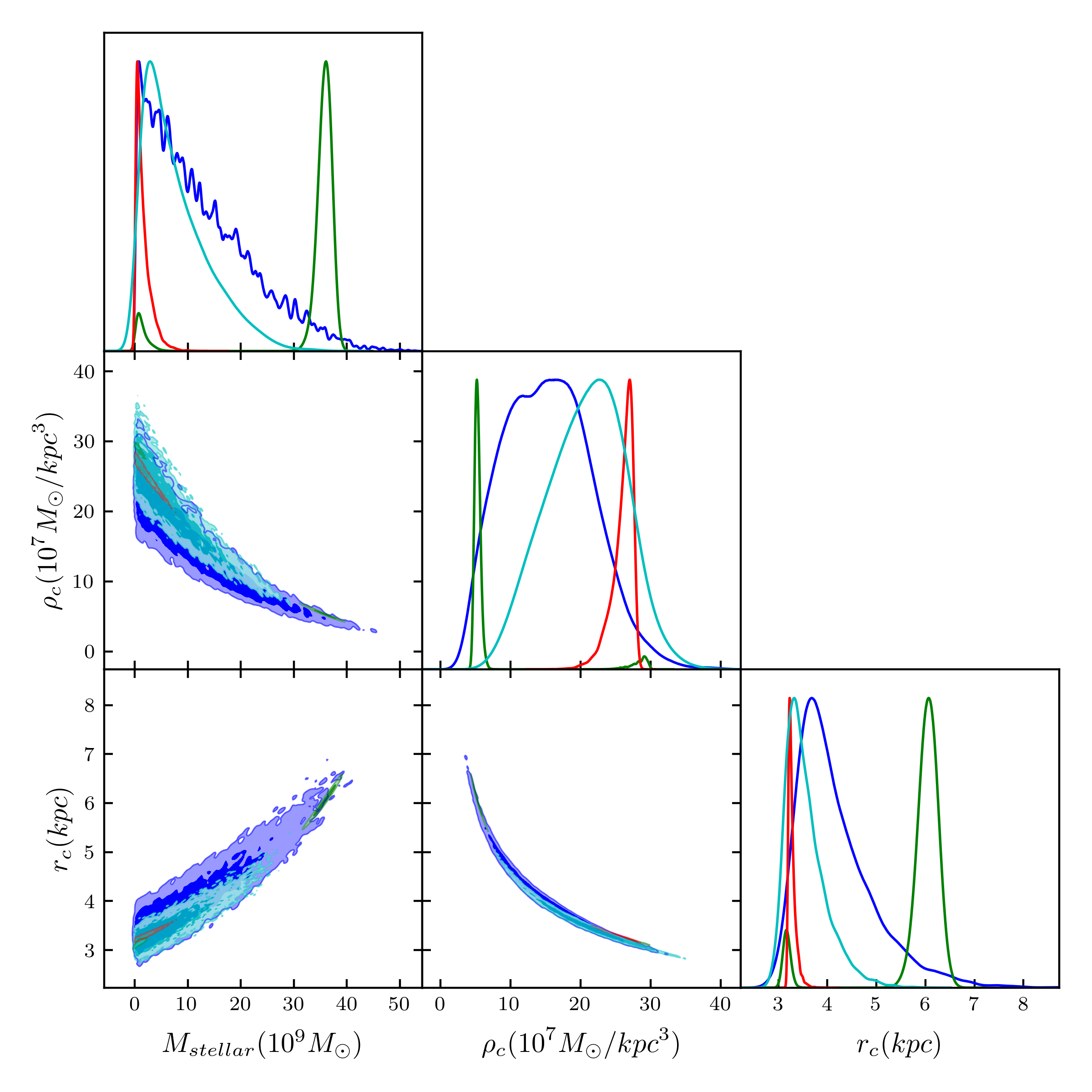}}
    \subfloat[B0 + Hernquist profile]{\label{fig:b0hern}\includegraphics[width=0.53\textwidth]{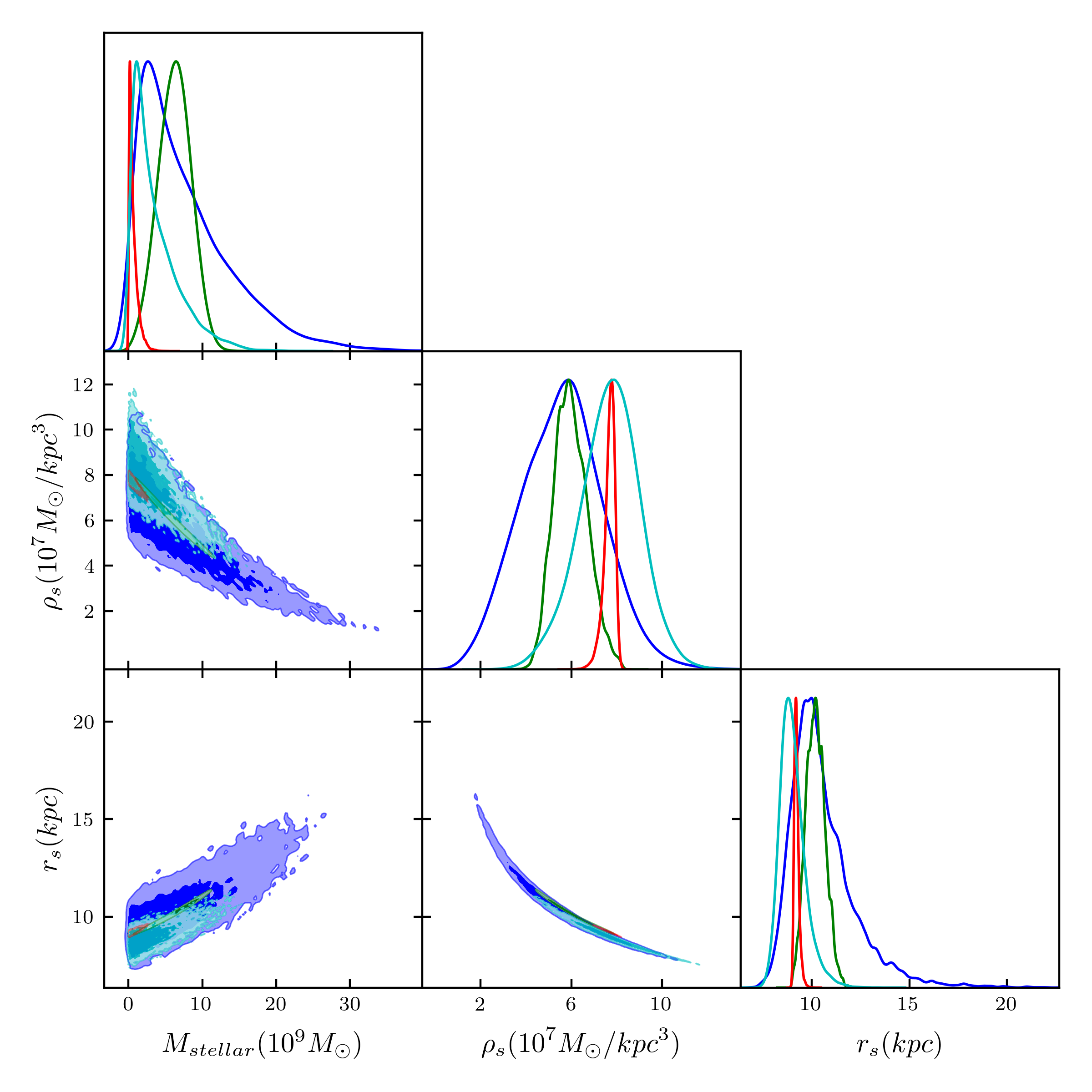}}
    \\
    \subfloat[B1 + Hernquist profile]{\label{fig:b1hern}\includegraphics[width=0.53\textwidth]{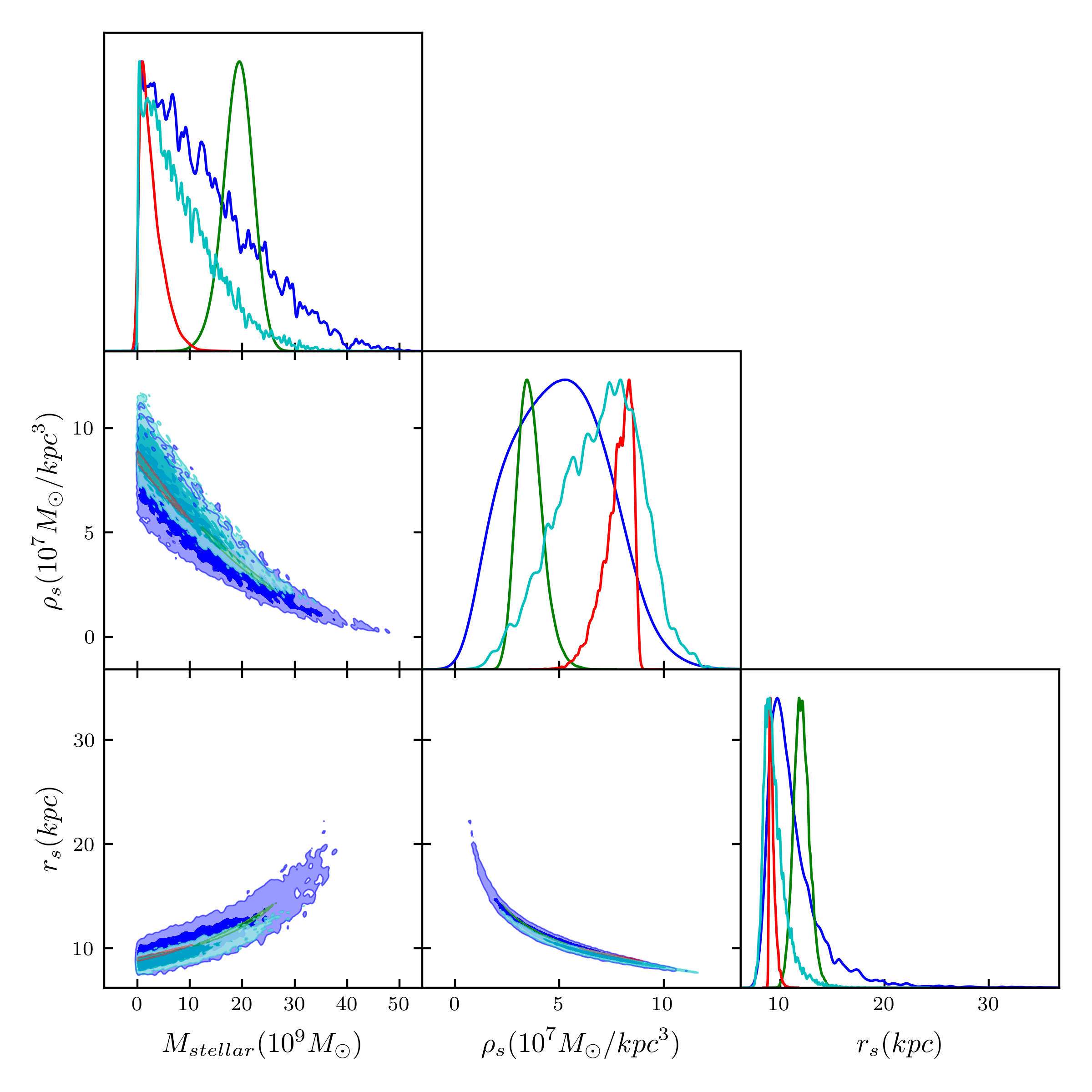}}
    \subfloat[B2 + Hernquist profile]{\label{fig:b2hern}\includegraphics[width=0.53\textwidth]{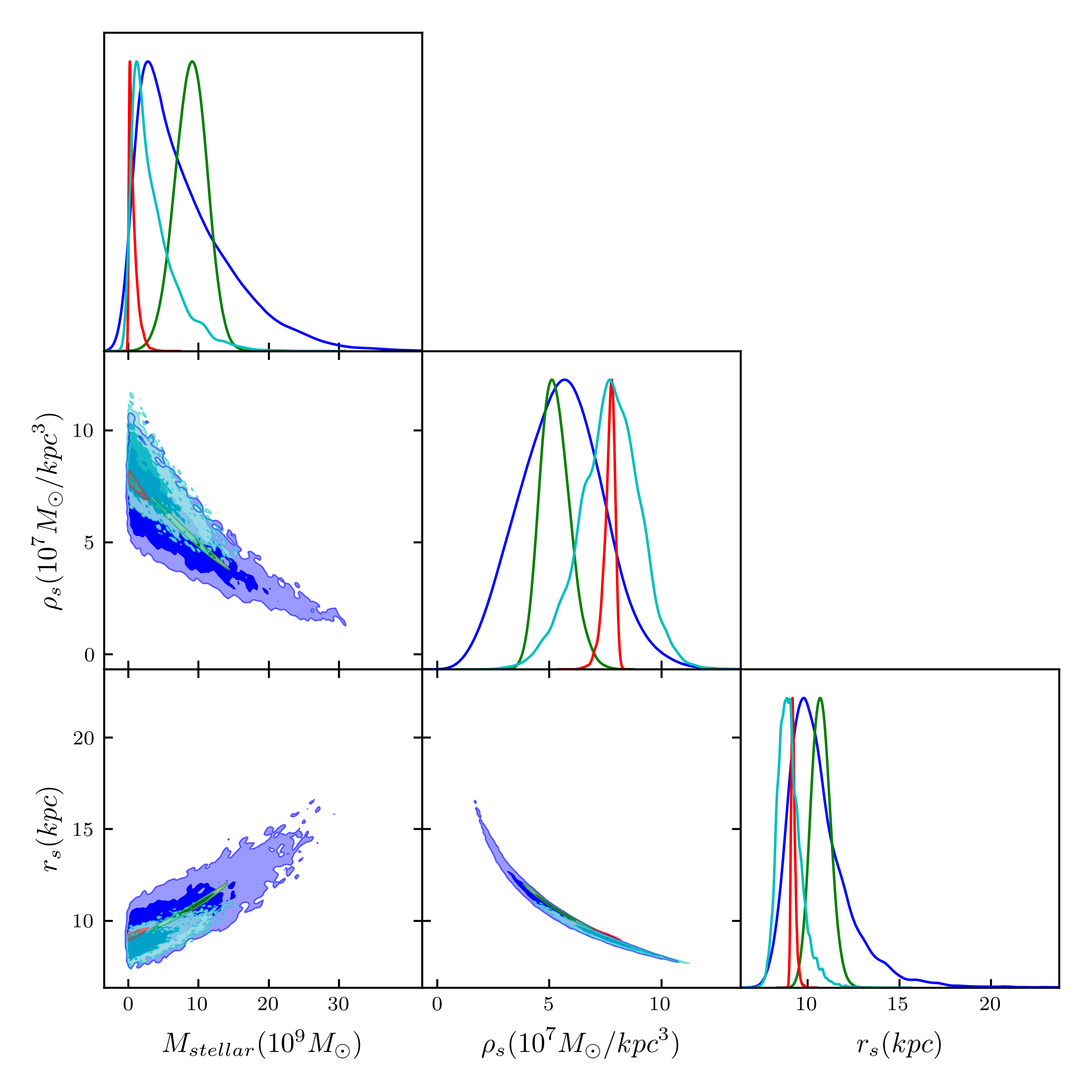}}
    \\
\end{figure}

\begin{figure}[h]
    \ContinuedFloat
    \centering
    \subfloat[B0 + Einasto profile]{\label{fig:b0einasto}\includegraphics[width=0.53\textwidth]{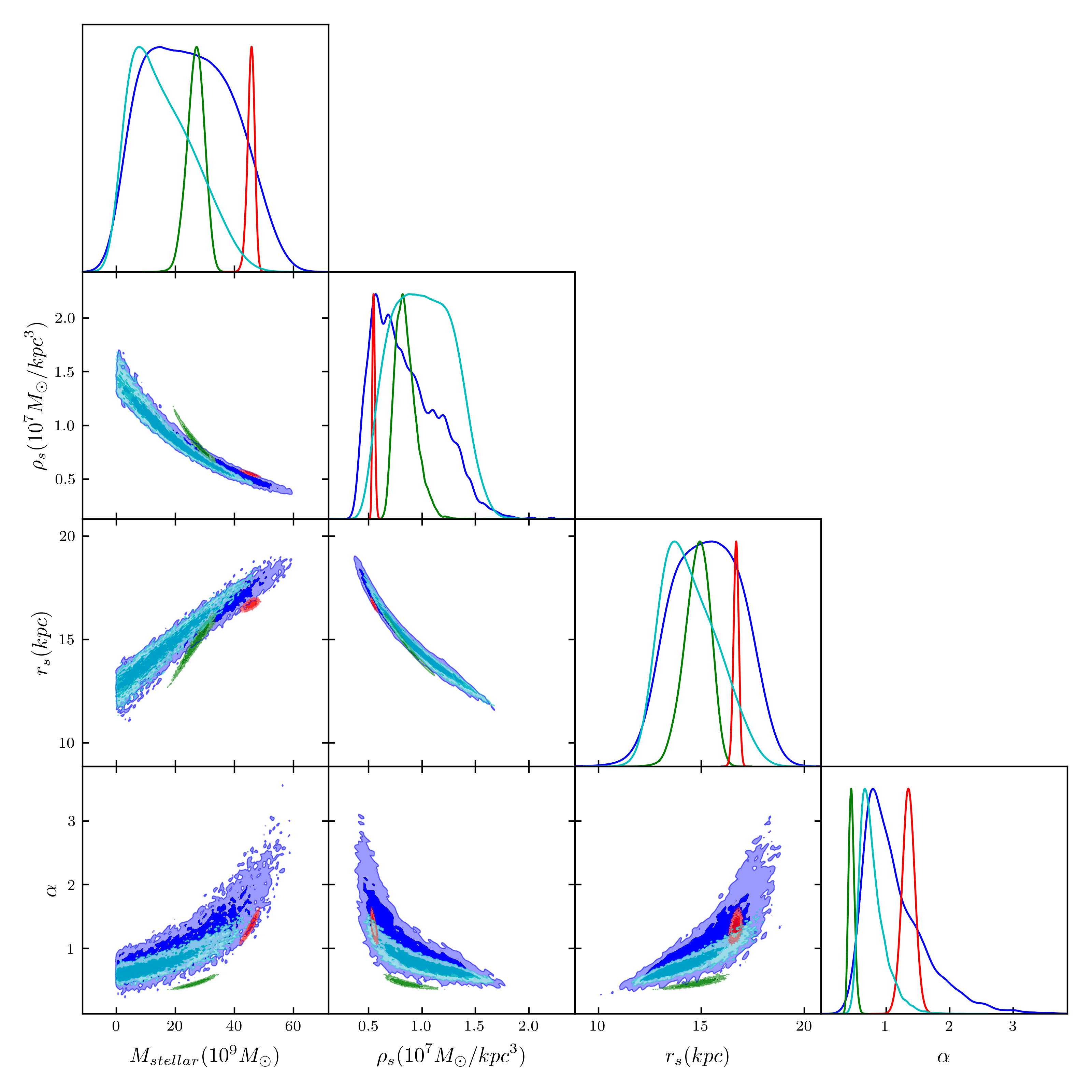}}
    \subfloat[B1 + Einasto profile]{\label{fig:b1einasto}\includegraphics[width=0.53\textwidth]{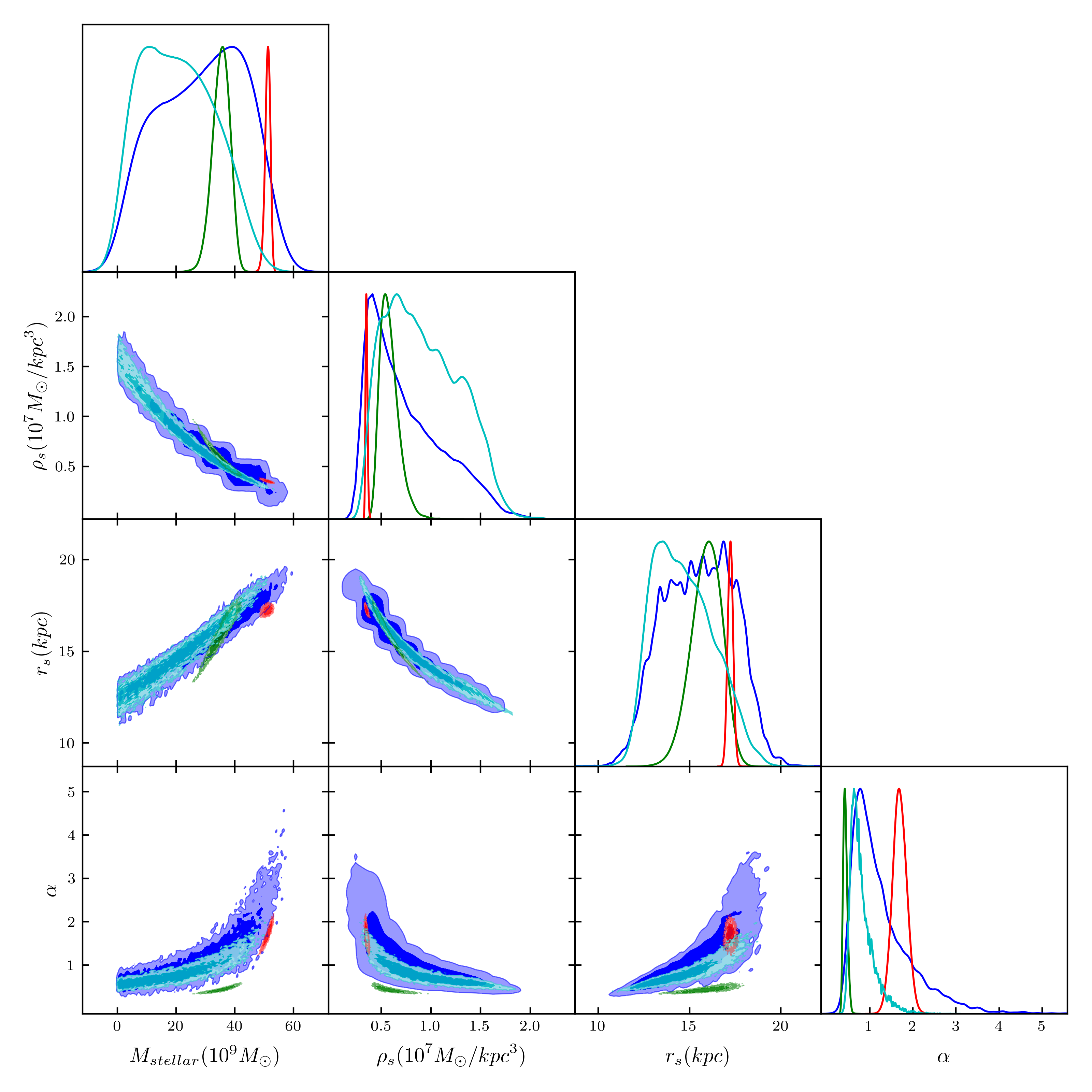}}
    \\
    \subfloat[B2 + Einasto profile]{\label{fig:b2einasto}\includegraphics[width=0.53\textwidth]{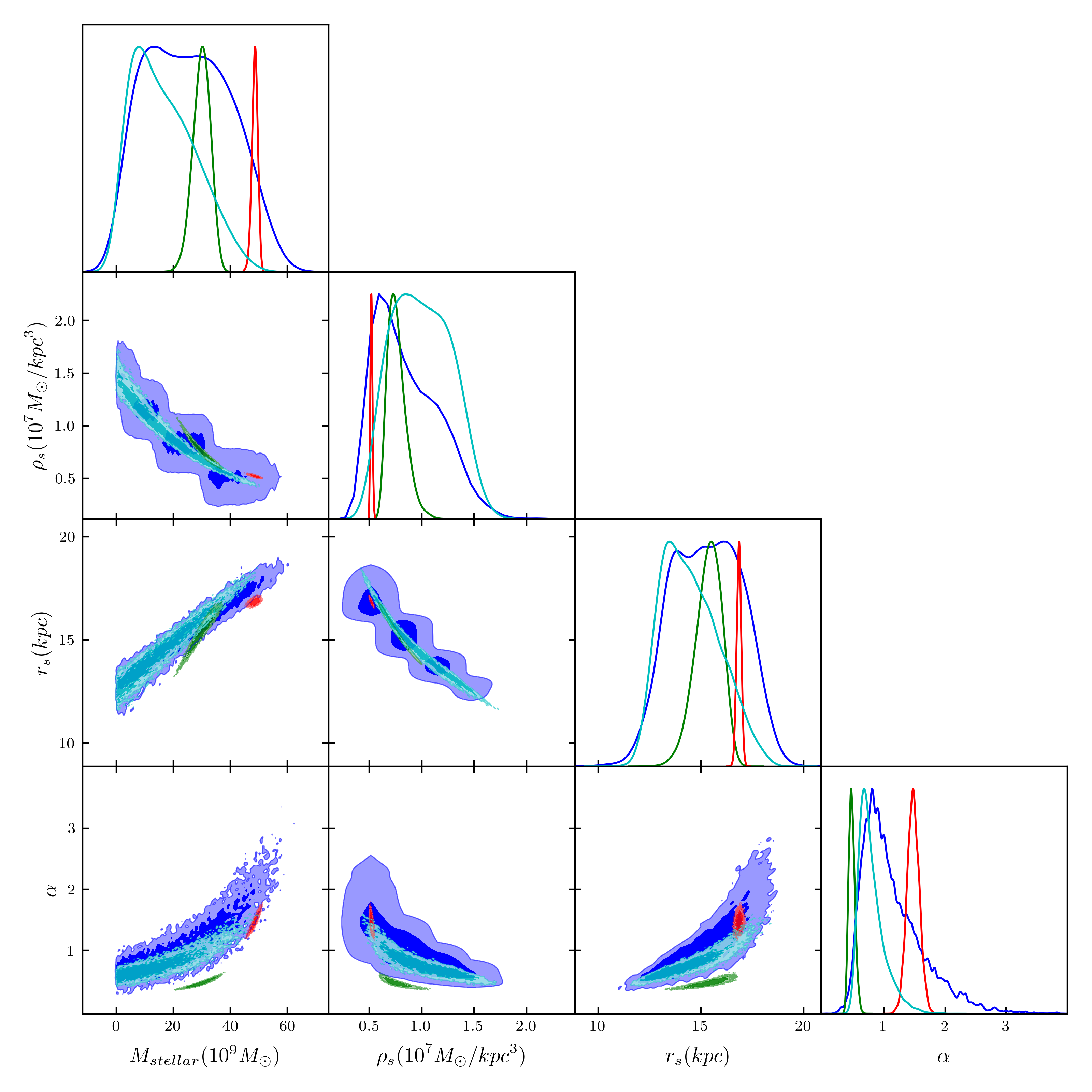}}
    \\
\end{figure}

\begin{figure}[h]
    \centering
    \subfloat[B0 + standard IF (free $a_0$)]{\label{fig:b0std}\includegraphics[width=0.53\textwidth]{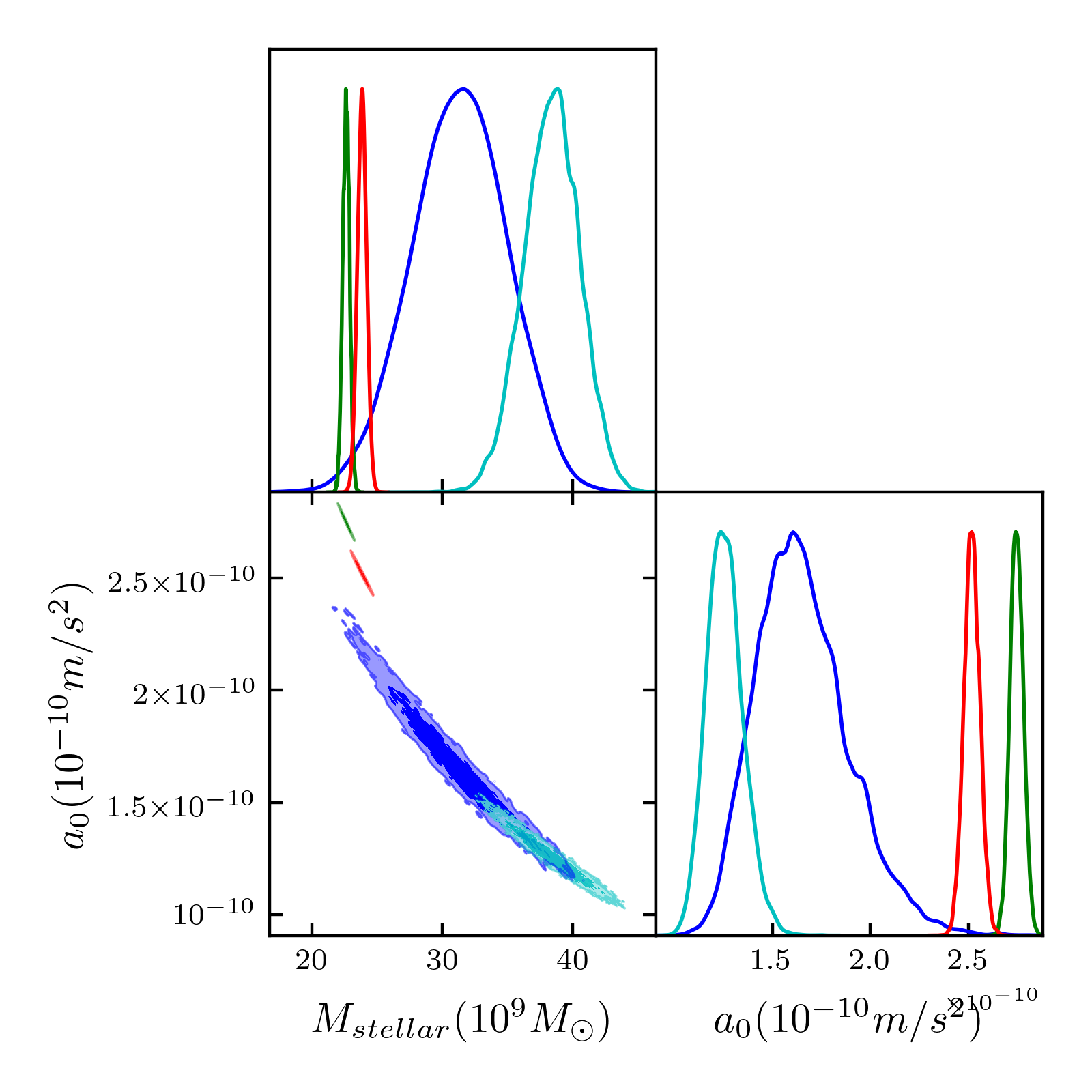}}
    \subfloat[B1 + standard IF (free $a_0$)]{\label{fig:b1std}\includegraphics[width=0.53\textwidth]{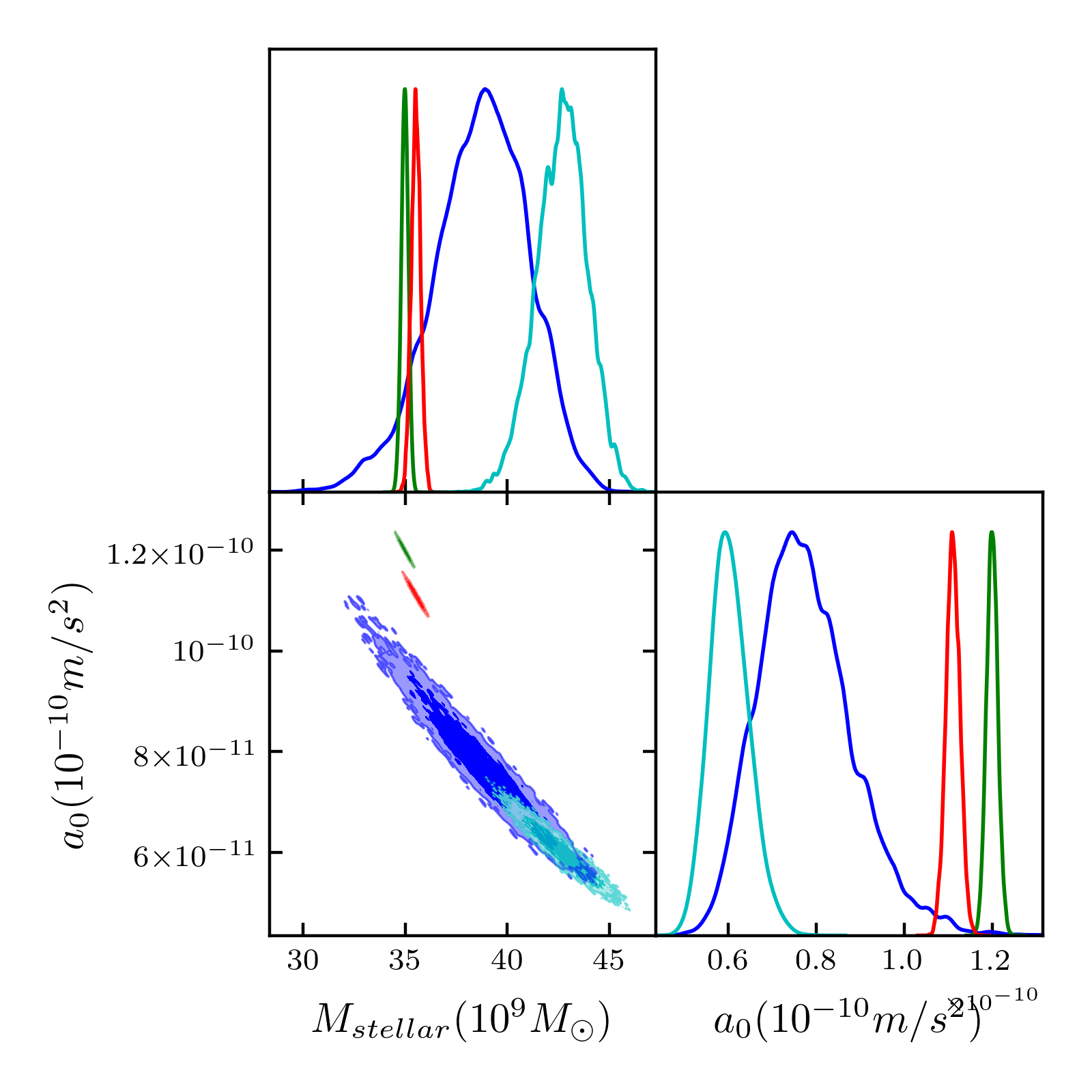}}
    \\
    \subfloat[B2 + standard IF (free $a_0$)]{\label{fig:b2std}\includegraphics[width=0.53\textwidth]{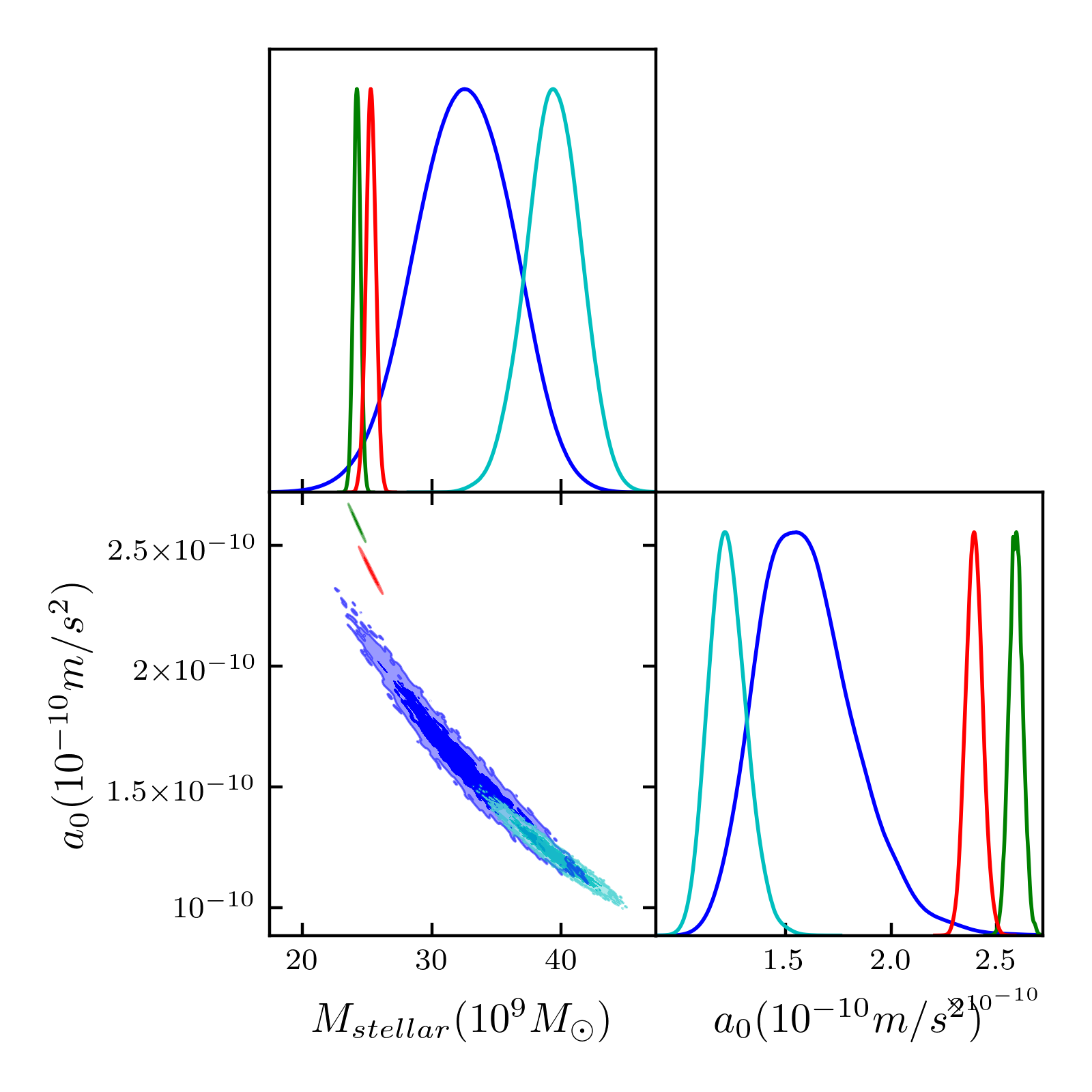}}
    \subfloat[B0 + simple IF (free $a_0$)]{\label{fig:b0simple}\includegraphics[width=0.53\textwidth]{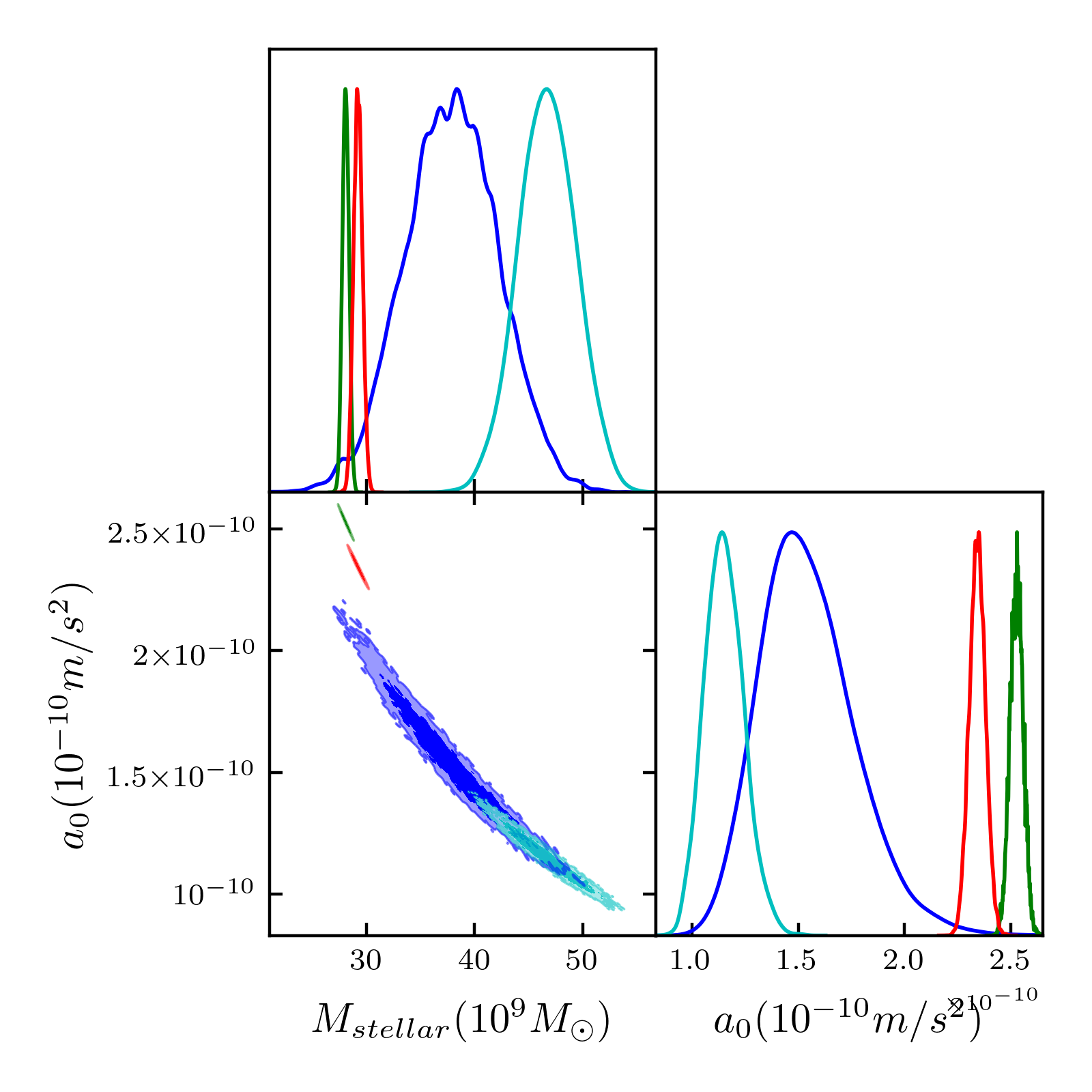}}
    \\
\end{figure}

\begin{figure}[h]
   \ContinuedFloat
    \centering
    \subfloat[B1 + simple IF (free $a_0$)]{\label{fig:b1simple}\includegraphics[width=0.53\textwidth]{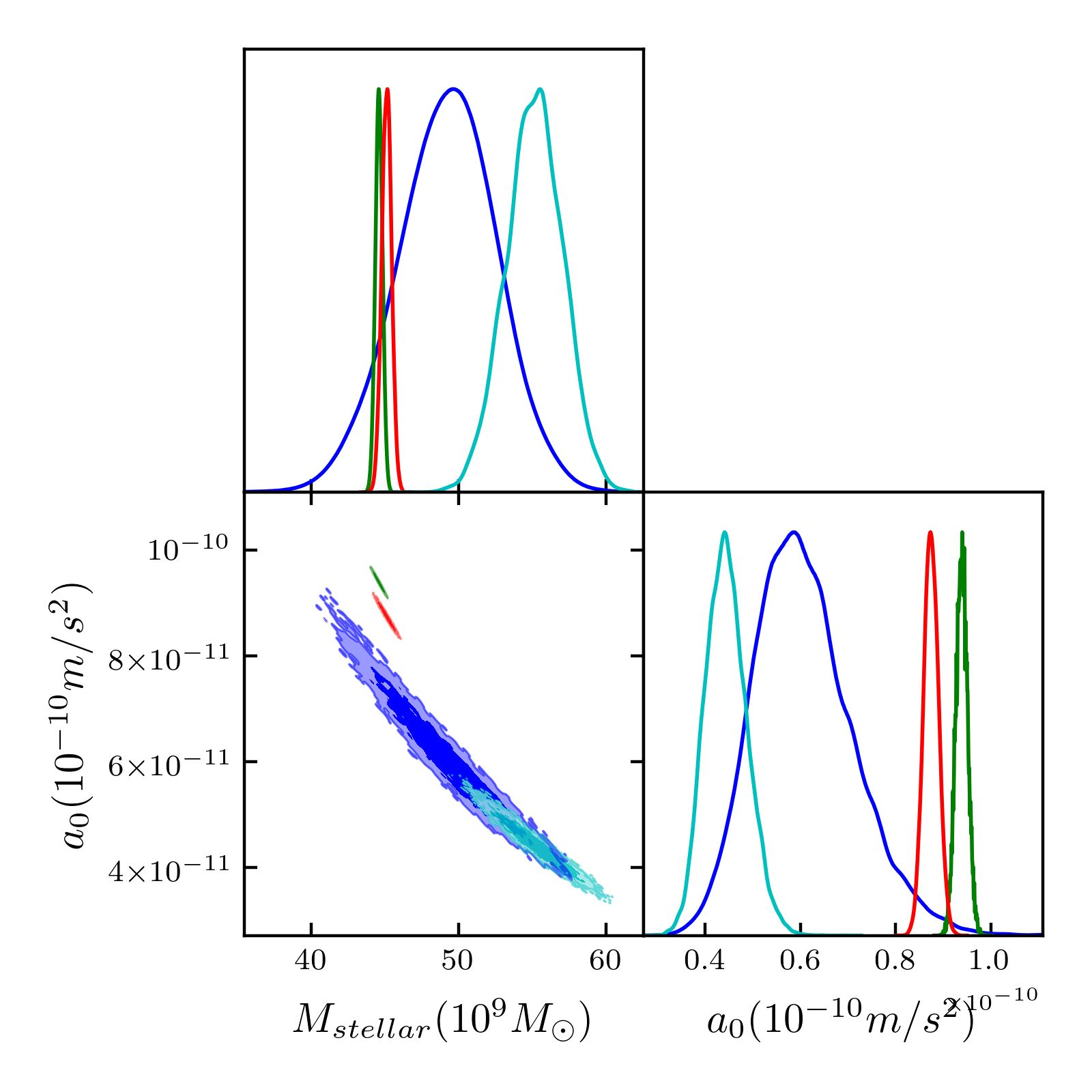}}
    \subfloat[B2 + simple IF (free $a_0$)]{\label{fig:b2simple}\includegraphics[width=0.53\textwidth]{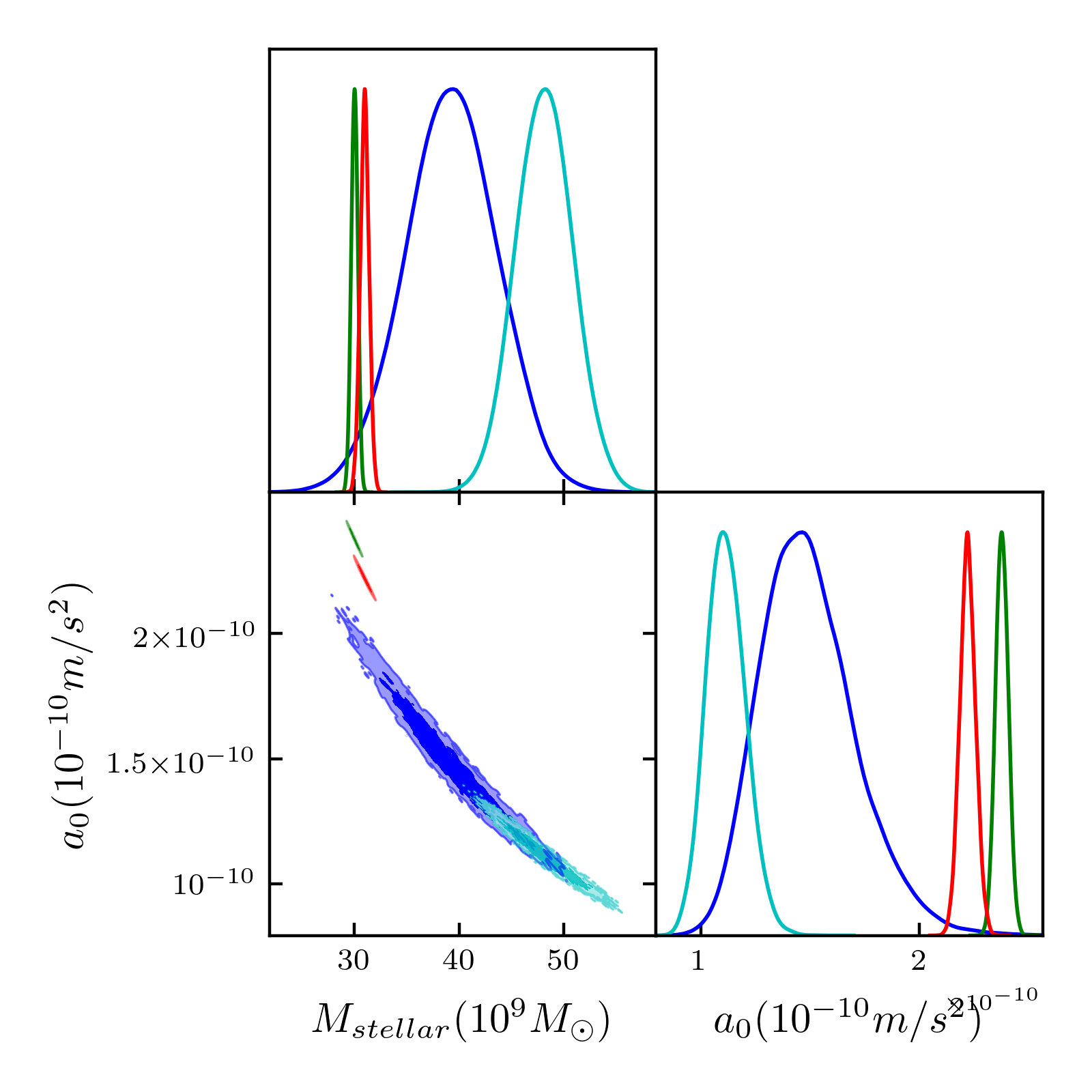}}
    \\
    \subfloat[B0 + RAR (free $a_0$)]{\label{fig:b0rar}\includegraphics[width=0.53\textwidth]{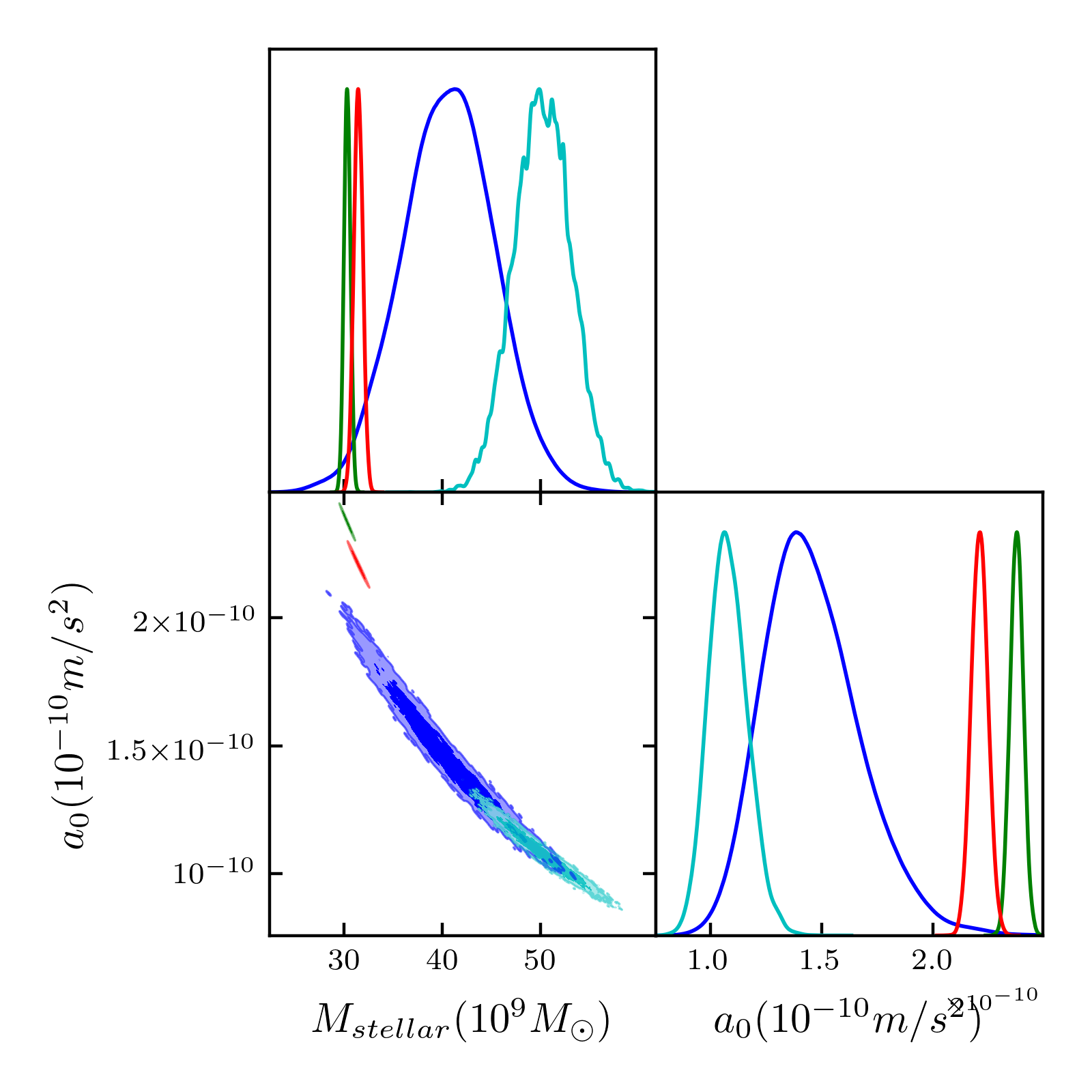}}
    \subfloat[B1 + RAR (free $a_0$)]{\label{fig:b1rar}\includegraphics[width=0.53\textwidth]{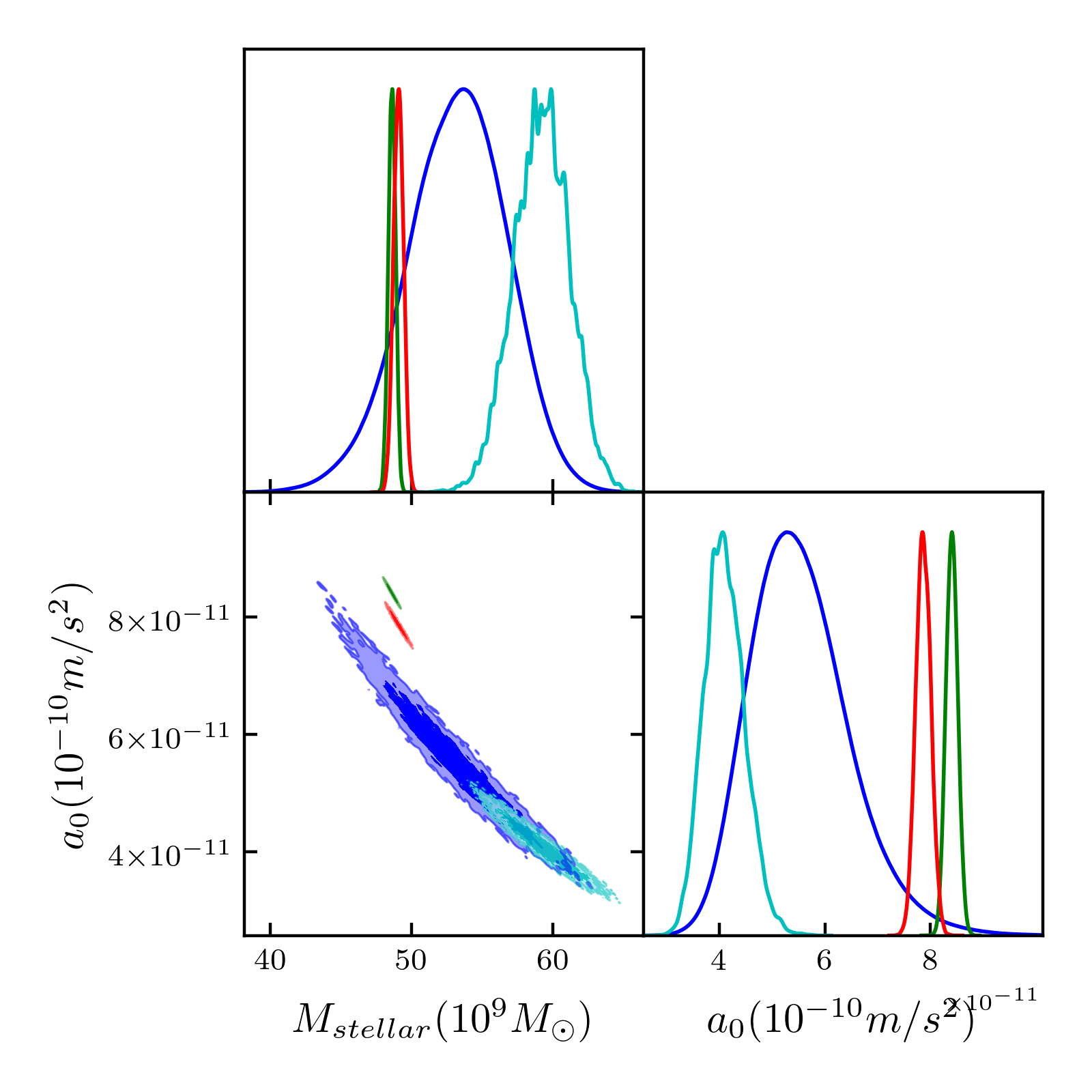}}
    \\
\end{figure}

\begin{figure}[h]
    \ContinuedFloat
    \centering
    \subfloat[B2 + RAR (free $a_0$)]{\label{fig:b2rar}\includegraphics[width=0.53\textwidth]{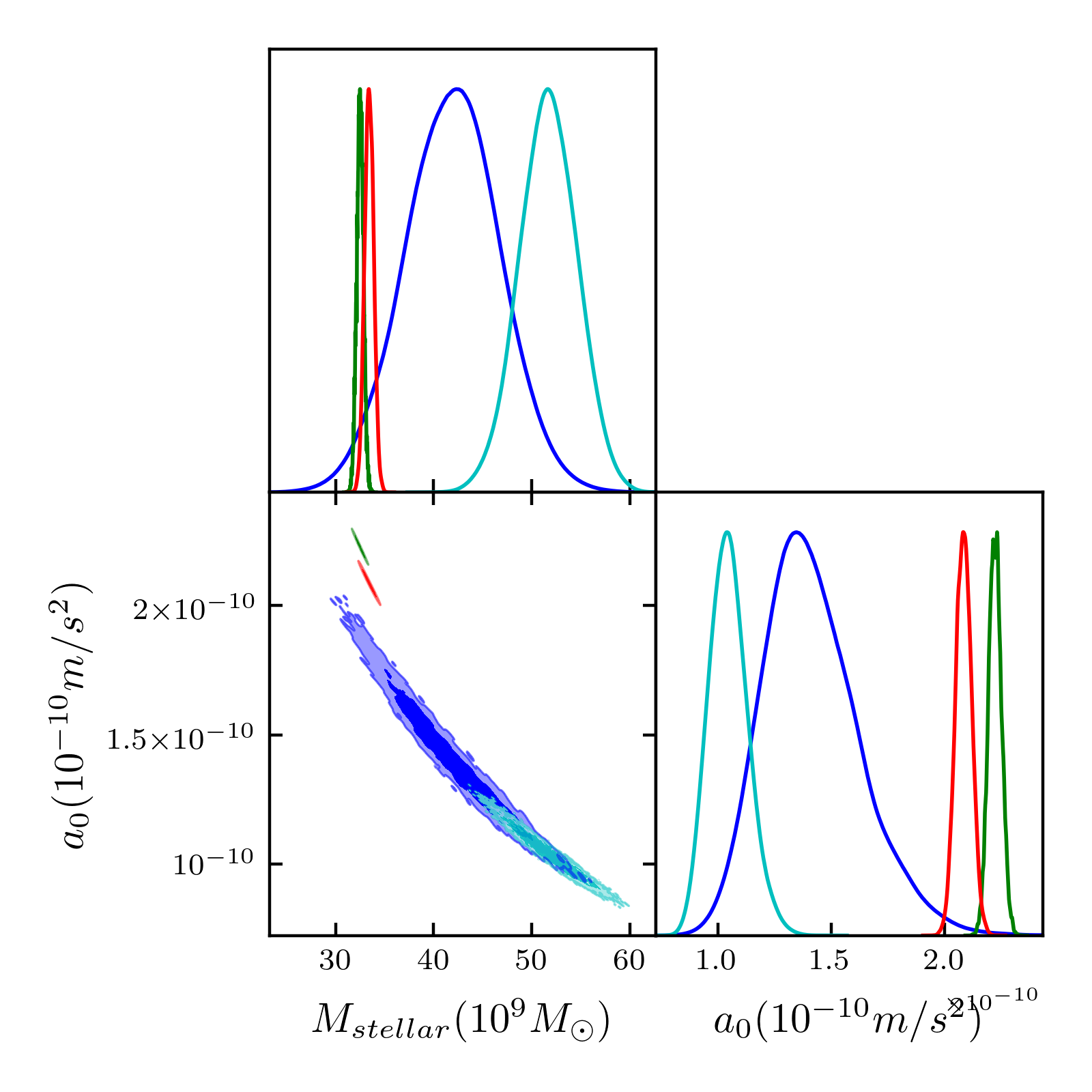}}
    \subfloat[B0 + standard IF (constant ($a_0$)]{\label{fig:b0stdmond}\includegraphics[width=0.53\textwidth]{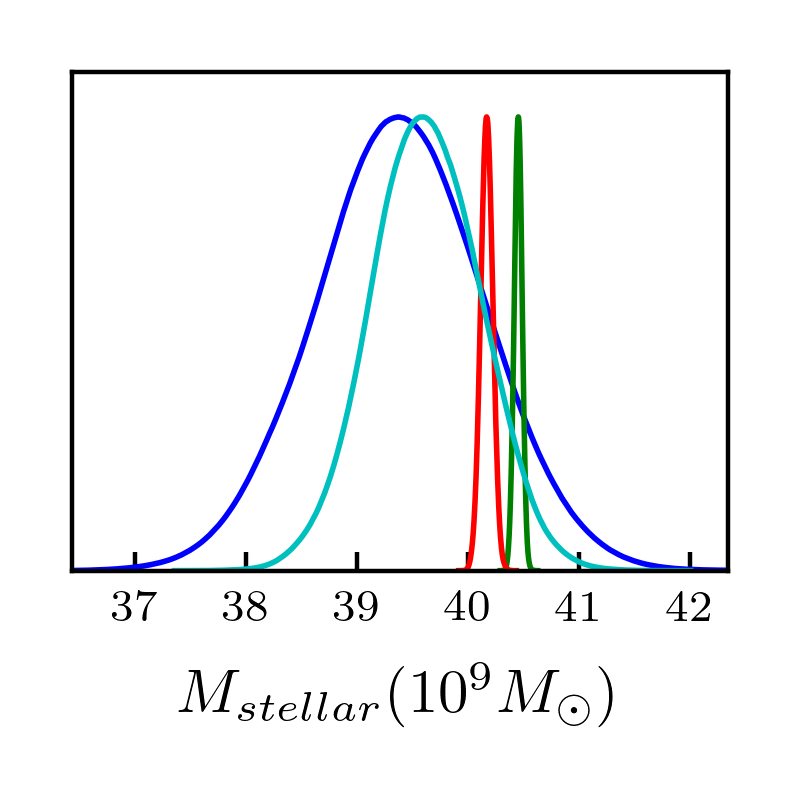}}
    \\
    \subfloat[B1 + standard IF (constant ($a_0$)]{\label{fig:b1stdmond}\includegraphics[width=0.53\textwidth]{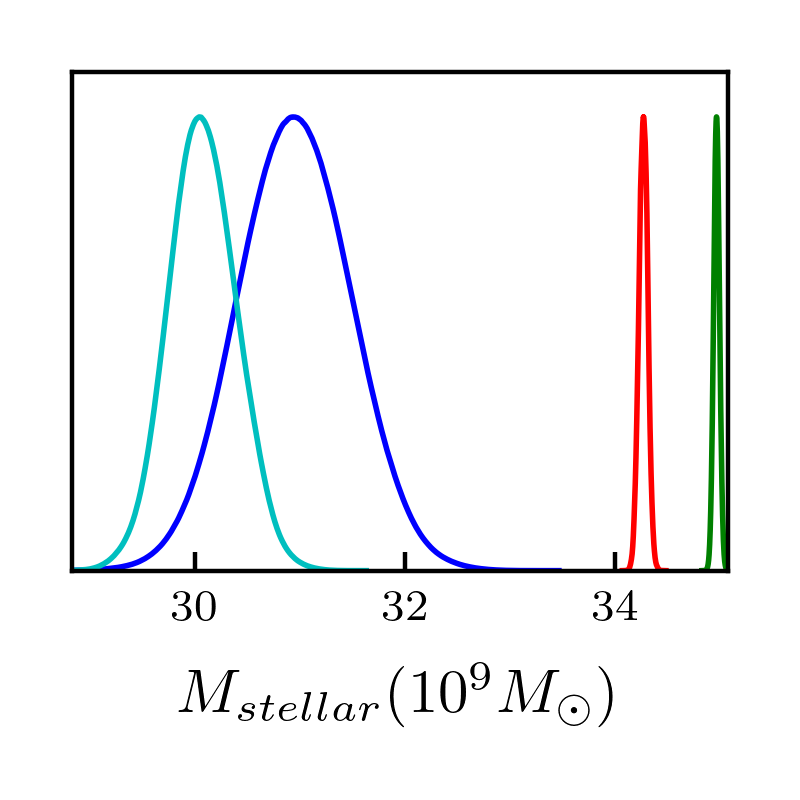}}
    \subfloat[B2 + standard IF (constant ($a_0$)]{\label{fig:b2stdmond}\includegraphics[width=0.53\textwidth]{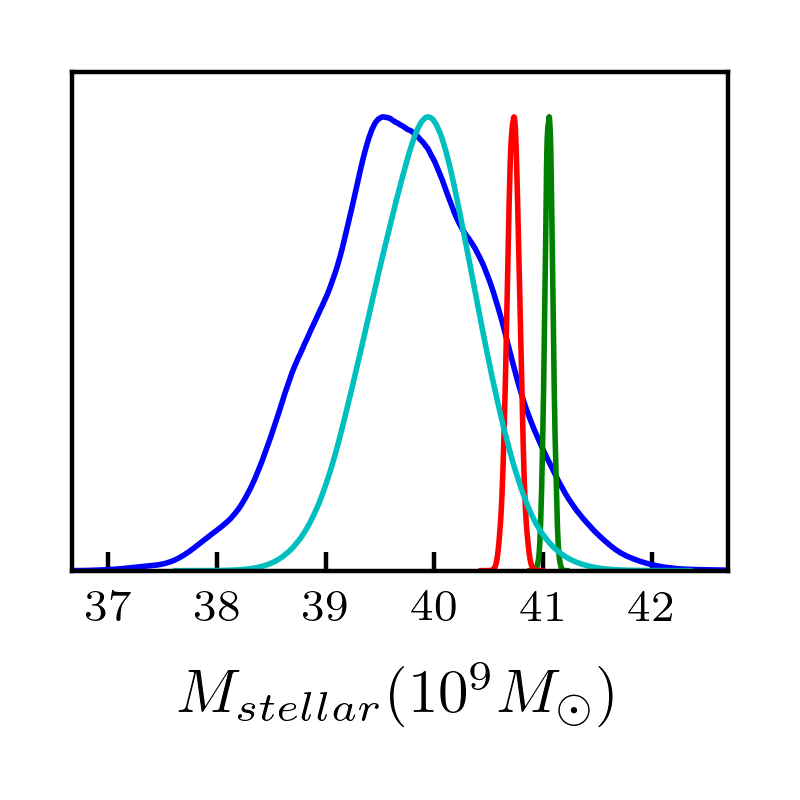}}
    \\
\end{figure}

\begin{figure}[h]
    \ContinuedFloat
    \centering
    \subfloat[B0 + simple IF (constant $a_0$)]{\label{fig:b0simplemond}\includegraphics[width=0.53\textwidth]{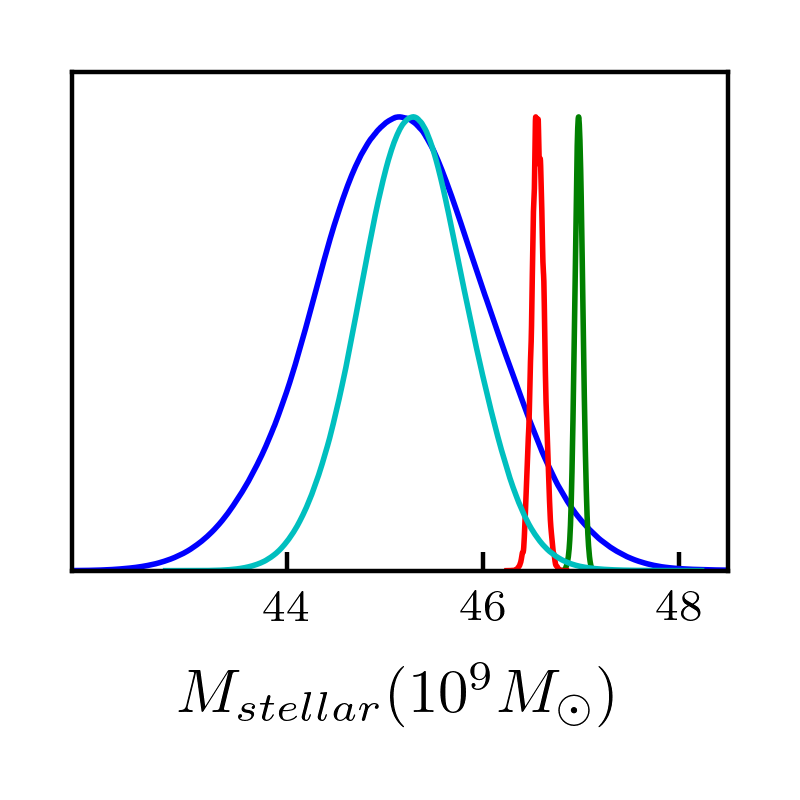}}
    \subfloat[B1 + simple IF (constant ($a_0$)]{\label{fig:b1simplemond}\includegraphics[width=0.53\textwidth]{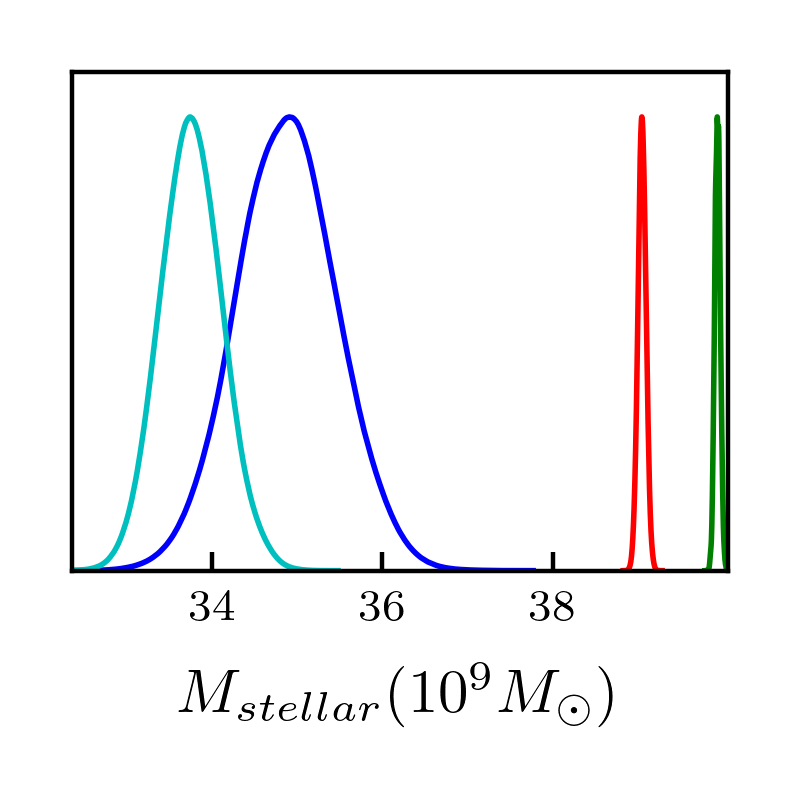}}
    \\
    \subfloat[B2 + simple IF (constant ($a_0$)]{\label{fig:b2simplemond}\includegraphics[width=0.53\textwidth]{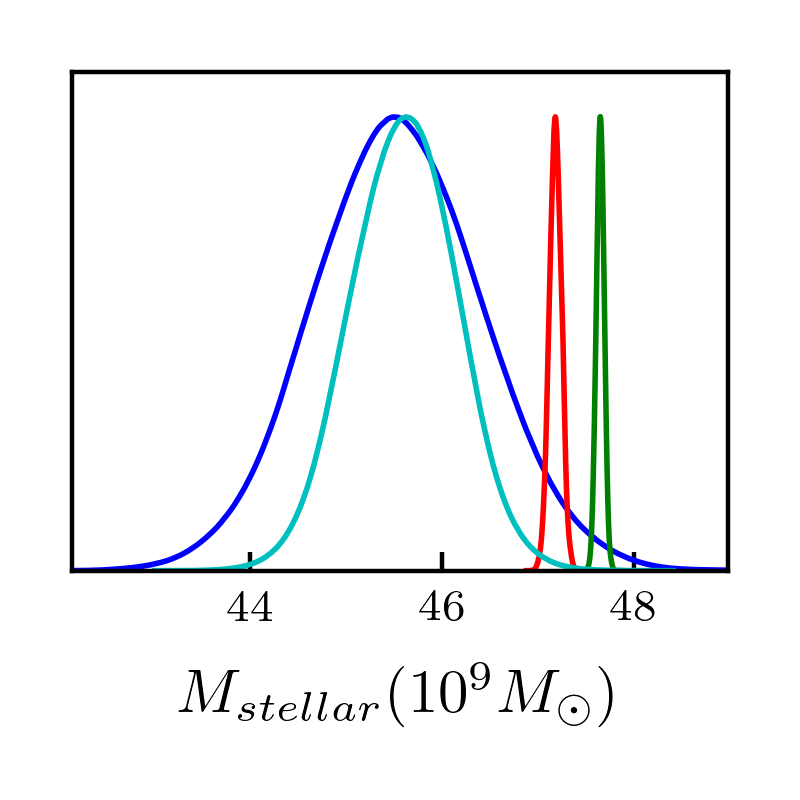}}
    \subfloat[B0 + RAR (constant ($a_0$)]{\label{fig:b0rarMOND}\includegraphics[width=0.53\textwidth]{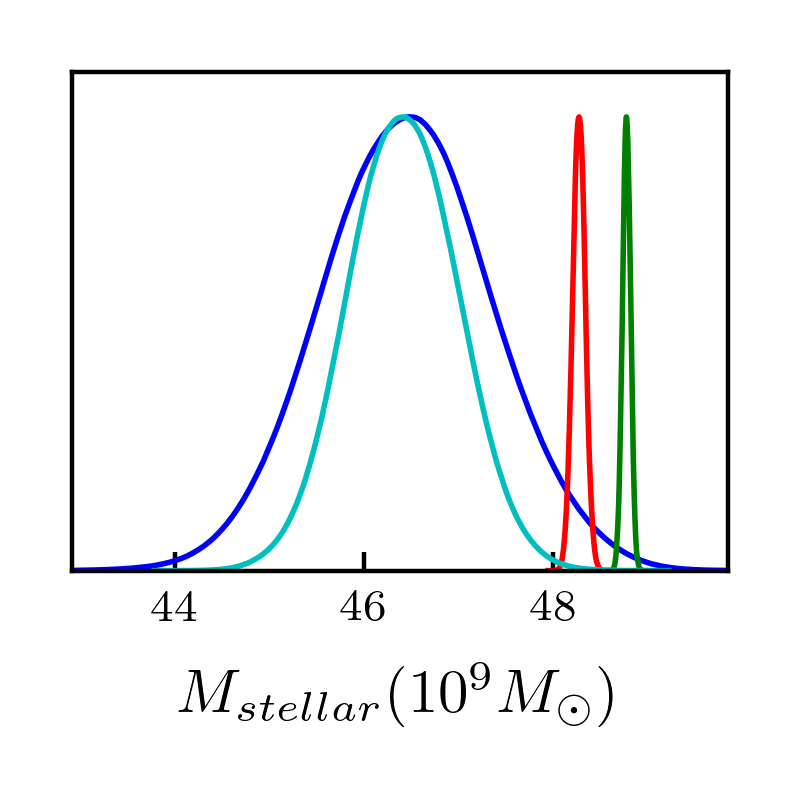}}
    \\
\end{figure}

\begin{figure}[h]
    \ContinuedFloat
    \centering
    \subfloat[B1 + RAR (constant $a_0$)]{\label{fig:b1rarMOND}\includegraphics[width=0.53\textwidth]{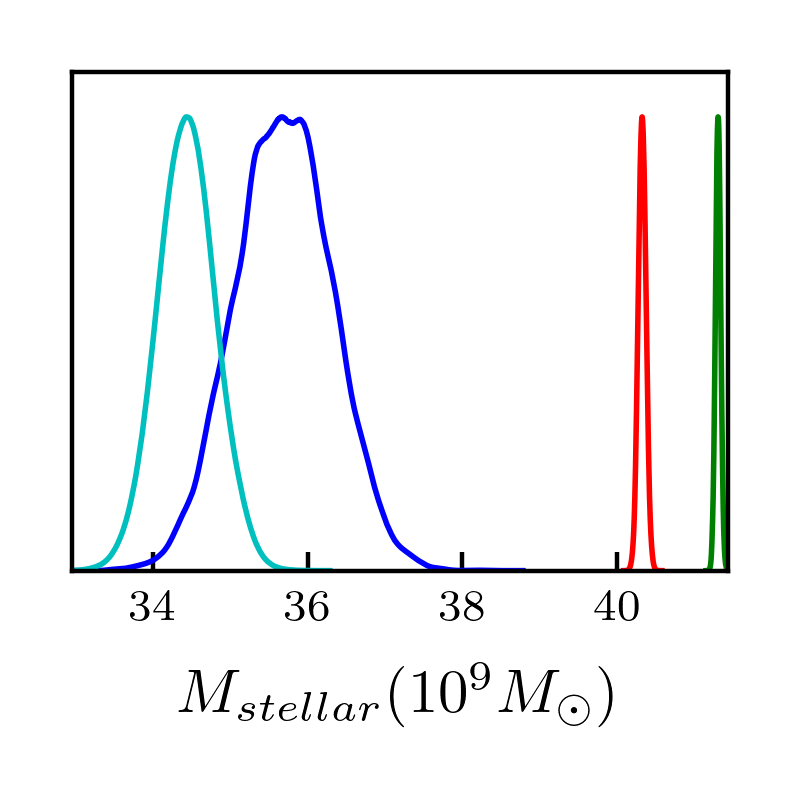}}
    \subfloat[B2 + RAR (constant ($a_0$)]{\label{fig:b2rarMOND}\includegraphics[width=0.53\textwidth]{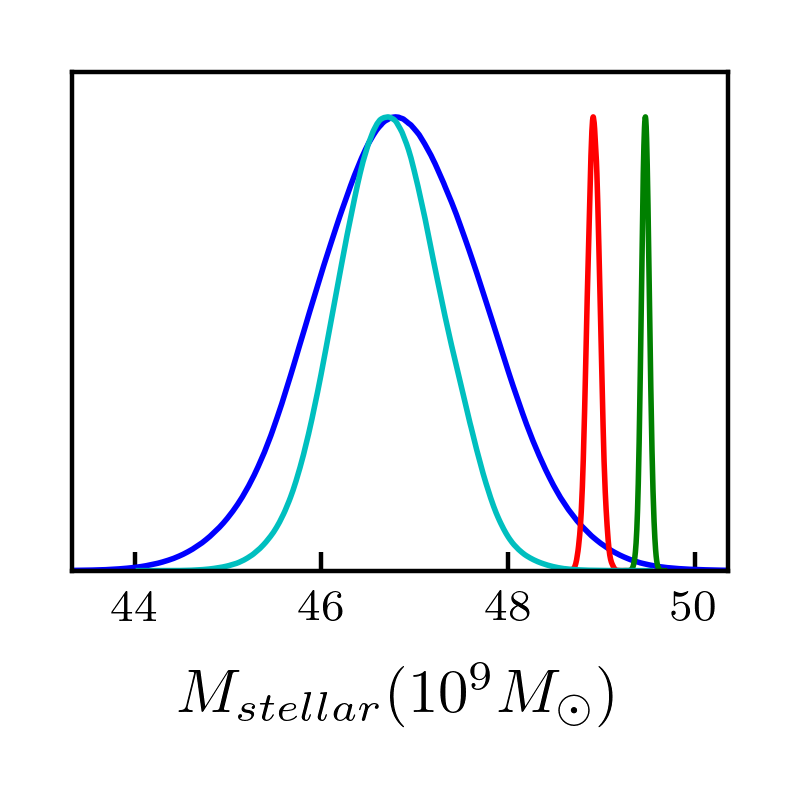}}
    \\
\end{figure}

\end{document}